\documentclass[11pt]{article}
\pdfoutput=1

\usepackage{jheppub}
\usepackage{url,comment}
\usepackage{times}
\usepackage{mathrsfs}
\usepackage{dsfont}
\usepackage{latexsym}
\usepackage{graphicx, graphics, hyperref, amsmath, amssymb, slashed, color, bbm,amsthm}
\usepackage{array}

\let\Re\relax
\let\Im\relax
\DeclareMathOperator{\Tr}{Tr}
\DeclareMathOperator{\Re}{Re}
\DeclareMathOperator{\Im}{Im}
\newcommand{\nc}{\newcommand}
\nc{\nn}{\nonumber}

\nc{\beq}{\begin{equation}}
\nc{\eeq}{\end{equation}}
\nc{\barray}{\begin{eqnarray}}
\nc{\earray}{\end{eqnarray}}
\nc{\barrayn}{\begin{eqnarray*}}
\nc{\earrayn}{\end{eqnarray*}}
\nc{\bcenter}{\begin{center}}
\nc{\ecenter}{\end{center}}
\nc{\mc}{\mathcal}
\nc{\er}[1]{(\ref{eq:#1})} 
\nc{\onehalf}{\frac{1}{2}}
\nc{\partialbar}{\bar{\partial}}
\nc{\psit}{\widetilde{\psi}} 
\nc{\hc}{\mbox{H.c.}}
\nc{\ev}{\;\mathrm{eV}}
\nc{\mev}{\;\mathrm{MeV}}
\nc{\gev}{\;\mathrm{GeV}}
\nc{\tev}{\;\mathrm{TeV}}
\nc{\f}{\frac}

\def\chii0{\chi_i^0}
\def\chij0{\chi_j^0}

\newcommand{\gsim}{\lower.7ex\hbox{$\;\stackrel{\textstyle>}{\sim}\;$}}
\newcommand{\lsim}{\lower.7ex\hbox{$\;\stackrel{\textstyle<}{\sim}\;$}}
\nc{\ttbar}{t\bar t}

\def\eeq{\end{equation}}
\newenvironment{Eqnarray}%
     {\arraycolsep 0.14em\begin{eqnarray}}{\end{eqnarray}}
\def\beqa{\begin{Eqnarray}}
\def\eeqa{\end{Eqnarray}}
\def\lsim{\mathrel{\raise.3ex\hbox{$<$\kern-.75em\lower1ex\hbox{$\sim$}}}}
\def\gsim{\mathrel{\raise.3ex\hbox{$>$\kern-.75em\lower1ex\hbox{$\sim$}}}}

\def\anti{\overline}
\def\wtil{\widetilde}
\def\ur{U_R}
\def\dr{D_R}

\def\abar{{\bar a}}
\def\bbar{{\bar b}}

\def\qlo{Q^0_L}
\def\uro{U^0_R}
\def\dro{D^0_R}

\def\ur{U_R}
\def\dr{D_R}

\def\eiuoa{\eta_a^{U,0}}
\def\eidoa{\eta_a^{D,0}}

\def\eidoab{\eta_{\abar}^{D,0}}

\def\eiuob{\eta_b^{U,0}}
\def\eidob{\eta_b^{D,0}}
\def\eiuobb{\eta_{\bbar}^{U,0}}
\def\eidobb{\eta_{\bbar}^{D,0}}

\def\elo{E^0_L}

\def\ero{E^0_R}

\def\er{E_R}

\def\eieoa{\eta_a^{E,0}}
\def\eieoab{\eta_{\abar}^{E,0}}

\def\eieob{\eta_b^{E,0}}
\def\eieobb{\eta_{\bbar}^{E,0}}

\def\cba{\cos(\beta-\alpha)}
\def\sba{\sin(\beta-\alpha)}
\def\cbma{c_{\beta-\alpha}}
\def\sbma{s_{\beta-\alpha}}

\def\mc{m_{H^\pm}}

\def\eq#1{Eq.~(\ref{#1})}
\def\eqs#1#2{Eqs.~(\ref{#1}) and (\ref{#2})}

\def\eqst#1#2{Eqs.~(\ref{#1})--(\ref{#2})}

\def\Eq#1{Eq.~(\ref{#1})}

\def\Eqs#1#2{Eqs.~(\ref{#1}) and (\ref{#2})}

\def\ifmath#1{\relax\ifmmode #1\else $#1$\fi}
\def\ls#1{\ifmath{_{\lower1.5pt\hbox{$\scriptstyle #1$}}}}
\def\lss#1{\ifmath{^{\,\lower2.5pt\hbox{$\scriptstyle #1$}}}}
\def\lsup#1{^{\lower 6pt\hbox{$\scriptstyle#1$}}}
\def\llsup#1{^{\lower 3pt\hbox{$\scriptstyle#1$}}}
\def\lasup#1{^{\lower 2pt\hbox{$\scriptstyle#1$}}}

\def\half{\tfrac{1}{2}}

\def\vev#1{\langle #1 \rangle}
\def\lsub#1{\ifmath{_{\lower1.5pt\hbox{$\scriptstyle #1$}}}}
\def\lsup#1{^{\lower 6pt\hbox{$\scriptstyle#1$}}}
\def\phaa{\phantom{AA}}

\def\ddel{\!\!\mathrel{\raise1.5ex\hbox{$\leftrightarrow$\kern-.85em
\lower1.7ex\hbox{$\partial$}}}}

\title{High scale flavor alignment in two-Higgs doublet models and its phenomenology}

\author[a]{Stefania Gori,}
\author[b]{Howard E. Haber,}
\author[b]{Edward Santos}
\affiliation[a]{Department of Physics, University of Cincinnati, Cincinnati, Ohio 45221, USA}
\affiliation[b]{Santa Cruz Institute for Particle Physics, University of California, Santa Cruz, CA 95064}

\emailAdd{stefania.gori@uc.edu}
\emailAdd{haber@scipp.ucsc.edu}
\emailAdd{eddie.santos.3@gmail.com}

\abstract{
The most general two-Higgs doublet model (2HDM) includes potentially large sources of flavor changing neutral currents (FCNCs) that must be suppressed in order to achieve a phenomenologically viable model. The flavor alignment ansatz postulates that all Yukawa coupling matrices are diagonal when expressed in the basis of mass-eigenstate fermion fields, in which case tree-level Higgs-mediated FCNCs are eliminated.  In this work, we explore models with the flavor alignment condition imposed at a very high energy scale, which results in the generation of Higgs-mediated FCNCs via renormalization group running from the  high energy
scale to the electroweak scale.  Using the current experimental bounds on flavor changing observables, constraints are derived on the aligned 2HDM parameter space.  In the favored parameter region, we analyze the implications for Higgs boson phenomenology.}

\begin{document}

\maketitle

\section{Introduction}\label{sec:intro}

With the discovery of a particle closely resembling the Standard Model (SM) Higgs boson at the Large Hadron Collider (LHC) \cite{Aad:2012tfa,Chatrchyan:2012ufa,Khachatryan:2016vau}, attention now turns to elucidating the dynamics of electroweak symmetry breaking.  Many critical question still remain unanswered.  What is the origin of the electroweak scale, and what mechanism ensures its stability?  In light of the existence of multiple generations of fermions, are there also multiple copies of the scalar multiplets, implying the existence of additional Higgs scalars? If yes, how are the Higgs-fermion Yukawa interactions 
compatible with the apparent Minimal Flavor Violation (MFV), which is responsible for suppressed flavor changing neutral currents (FCNCs)?  

Motivations for extending the Higgs sector beyond its minimal form  have appeared often in the literature.  For example, the minimal supersymmetric extension of the Standard Model, which is invoked to explain the stability of the electroweak symmetry breaking scale with respect to very high mass scales (such as the grand unification or Planck scales), requires a second Higgs doublet~\cite{Fayet:1974pd,Inoue:1982ej,Flores:1982pr,Gunion:1984yn}
to avoid anomalies due to the Higgsino partners of the Higgs bosons.  More complicated scalar sectors may also be required for a realistic model of baryogenesis~\cite{Fromme:2006cm}.  Finally, the metastability of the SM Higgs vacuum~\cite{Bezrukov:2012sa,Degrassi:2012ry,Buttazzo:2013uya} can be rendered stable up to the Planck scale in models of extended Higgs sectors~\cite{EliasMiro:2012ay,Lebedev:2012zw,Pruna:2013bma,Costa:2014qga,Chakrabarty:2014aya,Das:2015mwa,Ferreira:2015rha,Chowdhury:2015yja}.
Even in the absence of a specific model of new physics beyond the Standard Model, an enlarged scalar sector can provide a rich phenomenology that can be probed by experimental searches now underway at the LHC.

One of the simplest extensions of the SM Higgs sector is the two-Higgs doublet model (2HDM).\footnote{For a review with a comprehensive list of references, see Ref.~\cite{Branco:2011iw}.}  In its most general form, the 2HDM is incompatible with experimental data due to the existence of unsuppressed tree-level Higgs-mediated FCNCs, in contrast to the SM where
tree-level Higgs-mediated FCNCs are absent.  To see why this is so,
consider the Higgs-fermion Yukawa interactions expressed in terms of interaction eigenstate fermion fields.   Due to the non-zero vacuum expectation value (vev) of the neutral Higgs field, fermion mass matrices are generated.   Redefining the left and right-handed fermion fields by separate unitary transformations, the fermion mass matrices are diagonalized.  In the SM, this diagonalization procedure also diagonalizes the neutral Higgs-fermion couplings, and consequently no tree-level Higgs-mediated FCNCs are present.  In contrast, in a generic 2HDM, the diagonalization of the fermion mass matrices implies the diagonalization of one linear combination of Higgs-fermion Yukawa coupling matrices.  As a result, tree-level Higgs-mediated FCNCs remain in the 2HDM Lagrangian when expressed in terms of mass-eigenstate fermion fields. 
If it were possible in the 2HDM to realize flavor-diagonal neutral Higgs couplings at tree-level (thereby eliminating all tree-level Higgs-mediated FCNCs), then all FCNC processes arising in the model would be generated at the loop-level, with magnitudes more easily in agreement with experimental constraints.\footnote{Even in models with flavor-diagonal neutral Higgs couplings, one-loop processes mediated by the charged Higgs boson can generate significant FCNC effects involving third generation quarks.  Such models, in order to be consistent with experimental data, will produce constraints in the $[m_{H^\pm}$\,,\,$\tan\beta$] plane.   The most stringent constraint of this type, obtained in Ref.~\cite{Misiak:2017bgg} in the analysis of the Type-II 2HDM prediction for $b\to s\gamma$, yields $m_{H^\pm}\gtrsim 580$~GeV at $95\%$ CL. \label{fn2}} 

A natural mechanism for eliminating the tree-level Higgs-mediated FCNCs was proposed by Glashow and Weinberg \cite{Glashow:1976nt} and by Paschos~\cite{Paschos:1976ay} [GWP].  One can implement the GWP mechanism in the 2HDM by introducing a $\mathbb{Z}_2$ symmetry to eliminate half of the Higgs-fermion Yukawa coupling terms.  In this case, the fermion mass matrices and the non-zero Higgs-fermion Yukawa coupling matrices (which are consistent with the $\mathbb{Z}_2$ symmetry) are simultaneously diagonalized.  Indeed, there are a number of inequivalent implementations of the GWP mechanism, resulting in the so-called Types I~\cite{Haber:1978jt,Hall:1981bc}, and II~\cite{Donoghue:1978cj,Hall:1981bc}, and Types X and Y~\cite{Barger:1989fj,Aoki:2009ha} versions of the 2HDM.\footnote{However, if additional degrees of freedom exist at the TeV scale, then the GWP mechanism is in general not sufficient to protect the theory from FCNCs that are incompatible with the experimental data. These TeV-scale degrees of freedom, when integrated out, can generate higher-dimensional operators of the type $(c_1/\Lambda^2)\bar Q_LY_{u1}^{(6)}U_RH_2|H_1|^2+\cdots$, which break the proportionality relation between quark masses and effective Yukawa interactions with the neutral scalars.  As a result, such models generically generate FCNC processes that are not sufficiently suppressed~\cite{Buras:2010mh}.}

Another strategy for eliminating tree-level Higgs-mediated FCNCs is by fiat.
The flavor alignment ansatz proposed in Ref.~\cite{Pich:2009sp} asserts a proportionality between the two sets of Yukawa matrices.  If this flavor-alignment condition is implemented at the electroweak scale, then the diagonalization of the fermion mass matrices simultaneously yields flavor-diagonal neutral Higgs couplings.  Moreover, this flavor-aligned 2HDM (henceforth denoted as the A2HDM) preserves the relative hierarchy in the quark mass matrices, and provides additional sources of CP-violation in the Yukawa Lagrangian via the introduction of three complex alignment parameters. 
Unfortunately, apart from the special cases enumerated in Ref.~\cite{Botella:2015yfa},
there are no symmetries within the 2HDM that guarantee the stability of the flavor alignment ansatz with respect to radiative corrections.  As such, flavor alignment at the electroweak scale must be generically regarded as an unnatural fine-tuning of the Higgs-fermion Yukawa matrix parameters. 
Indeed, the Types I, II, X and Y 2HDMs are the \textit{unique} special cases of flavor alignment that are radiatively stable after imposing the observed fermion masses and mixing~\cite{Ferreira:2010xe}.

In this paper, we consider the possibility that flavor alignment arises from New Physics beyond the 2HDM.  Without a specific ultraviolet completion in mind, we shall assert that flavor alignment is imposed at some high energy scale, $\Lambda$, perhaps as large as a grand unification scale or the Planck scale, where new dynamics can emerge (e.g., see Ref.~\cite{Knapen:2015hia} for a viable model).  Once we impose the flavor alignment ansatz at the scale $\Lambda$, the effective field theory below this scale corresponds to a 2HDM with both Higgs doublets coupling to up type and down type quarks and leptons.\footnote{In practice, one should also append to the 2HDM some mechanism for generating neutrino masses.  An example of incorporating the effects of neutrino masses and mixing in the context of a 2HDM with flavor changing neutral Higgs couplings can be found in Ref.~\cite{Botella:2015hoa}.
In this paper, we shall simply put all neutrino masses to zero for the sake of simplicity.  The extension of the results of this paper to models that incorporate a mechanism for neutrino mass generation will be considered in a future publication.\label{fn4}}   We then employ renormalization group (RG) evolution to determine the structure of the 2HDM Yukawa couplings at the electroweak scale.  For a generic flavor alignment ansatz at the scale $\Lambda$, flavor alignment in the Higgs-fermion Yukawa couplings at the electroweak scale is violated, thereby generating Higgs-mediated FCNCs.  However, these FCNCs will be of Minimal Flavor Violation \cite{D'Ambrosio:2002ex} type and therefore may be small enough to be consistent with experimental constraints, depending on the choice of the initial alignment parameters at the scale $\Lambda$.

We therefore examine the phenomenology of Higgs-mediated FCNCs that arise from the assumption of flavor alignment at some high energy scale, $\Lambda$, that, for the purpose of our analyses, is fixed to be the Planck scale ($M_{\rm P}$).
We note that similar work was performed in \cite{Braeuninger:2010td}, where meson mixing and $B$ decays were used to constrain the A2HDM parameter space with flavor alignment at the Planck scale. Numerical results were obtained analytically in \cite{Braeuninger:2010td}, using the leading logarithmic approximation.
The results of this paper are first obtained in the leading log approximation, and then numerically by evolving the full one-loop renormalization group equations (RGEs) down from the Planck scale to the electroweak scale.  In our work, we discuss the validity of the leading log approximation and examine additional FCNC processes at high energy (top and Higgs decays) and at low energy ($B$~meson decays) to place bounds on the A2HDM parameters.  

This paper is organized as follows.  In section \ref{sec:2hdm}, we review the theoretical framework of the general 2HDM.  It is convenient to make use of the Higgs basis, which is unique up to a phase degree of freedom.  All physical observables must be independent of this phase.  In particular, we examine in detail the structure of the Higgs-fermion Yukawa couplings and exhibit its flavor structure.   In the formalism presented in  section \ref{sec:2hdm}, we initially allow for the most general form of the Higgs scalar potential and the Yukawa coupling matrices.  In particular, new sources of CP-violation beyond the SM can arise due to unremovable complex phases in both the scalar potential parameters and the Yukawa couplings.
For simplicity, we subsequently choose to analyze the case of a CP-conserving Higgs scalar potential and vacuum, in which case the neutral mass-eigenstates consist of two CP-even and one CP-odd neutral Higgs bosons.  
 We then introduce the flavor-aligned 2HDM, in which the Yukawa coupling matrices are diagonal in the basis of quark and lepton mass-eigenstates.  However, alignment is not stable under renormalization group running.   Following the framework for flavor discussed above, we impose the alignment condition at the Planck scale and then evaluate the Yukawa coupling matrices of the Higgs basis at the electroweak scale as determined by renormalization group running, subject to the observed quark and lepton masses and the CKM mixing matrix.   The renormalization group running is performed numerically and checked in the leading log approximation, where simple analytic expressions can be obtained. In this context, a comparison with general Minimal Flavor Violating 2HDMs is performed.

In  section \ref{sec:3hdm}, we discuss the implications of high-scale flavor alignment for high energy processes.   We focus on flavor-changing decays of the top quark and on the phenomenology of the heavy neutral CP-even and CP-odd Higgs bosons.  In  section \ref{sec:4hdm}, we discuss the implications of high-scale flavor alignment for low energy processes.   Here we consider constraints arising from neutral meson mixing observables and from $B_s\to \ell^+\ell^-$, which receive contributions at tree-level from neutral Higgs exchange, and from the charged Higgs mediated $B\to\tau\nu$ decay.  By comparing
theoretical predictions to experimental data, one can already probe certain regions of the A2HDM parameter space. 
Additional parameter regions will be probed by future searches for heavy Higgs bosons and measurements of $B$-physics observables. Conclusions of this work are presented in section \ref{sec:conclusions}. Finally, in
Appendix~\ref{appA} we review the derivation of the Yukawa sector of our model in the fermion mass-eigenstate basis, and in Appendix \ref{sec:RGEs} we exhibit the one-loop matrix Yukawa coupling RGEs used in this analysis.


\section{The flavor-aligned 2HDM}\label{sec:2hdm}

\subsection{Theoretical framework for the 2HDM}

Consider a generic 2HDM consisting of two complex, hypercharge-one scalar doublets, $\Phi_1$ and $\Phi_2$.
The most general renormalizable scalar potential that is invariant under local SU(2)$\times$U(1) gauge transformations can be written as
\beqa
\mathcal{V}&=& m_{11}^2 \Phi_1^\dagger \Phi_1+ m_{22}^2 \Phi_2^\dagger \Phi_2 -[m_{12}^2
\Phi_1^\dagger \Phi_2+{\rm h.c.}]
+\half \lambda_1(\Phi_1^\dagger \Phi_1)^2+\half \lambda_2(\Phi_2^\dagger \Phi_2)^2
+\lambda_3(\Phi_1^\dagger \Phi_1)(\Phi_2^\dagger \Phi_2)
\nn\\
&&\quad
+\lambda_4( \Phi_1^\dagger \Phi_2)(\Phi_2^\dagger \Phi_1)
 +\left\{\half \lambda_5 (\Phi_1^\dagger \Phi_2)^2 +\big[\lambda_6 (\Phi_1^\dagger
\Phi_1) +\lambda_7 (\Phi_2^\dagger \Phi_2)\big] \Phi_1^\dagger \Phi_2+{\rm
h.c.}\right\}\,.\label{genpot}
\eeqa
The parameters of the scalar potential can be chosen so that the minimum of the scalar potential is achieved when the neutral components of the two scalar doublet fields acquire non-zero vacuum expectation vales, $\langle \Phi_1^0 \rangle= v_1/\sqrt{2}$ and $\langle \Phi_2^0\rangle = v_2/\sqrt{2}$, where the (potentially complex) vevs satisfy
\beq \label{vev2}
v^2 \equiv  |v_1|^2 + |v_2|^2 \simeq (246~{\rm GeV})^2\,,
\eeq
as required by the observed $W$ boson mass, $m_W=\half gv$.
The SU(2)$\times$U(1) gauge symmetry is then spontaneously broken, leaving an unbroken U(1)$_{\rm EM}$ gauge group.  

In the most general 2HDM, the fields $\Phi_1$ and $\Phi_2$ are indistinguishable.
Thus, it is always possible to define two orthonormal linear combinations of the two
doublet fields without modifying any prediction of the model.  Performing such a
redefinition of fields leads to a new scalar potential with the same form as \eq{genpot}
but with modified coefficients.  This implies that the coefficients that parameterize
the scalar potential in \eq{genpot} are not directly physical~\cite{Davidson:2005cw}.

To obtain a scalar potential that is more closely related to physical observables, one
can introduce the so-called \textit{Higgs basis} in which the redefined doublet fields (denoted below
by $H_1$ and $H_2$) have the property that $H_1$ has a non-zero vev whereas $H_2$ has a zero
vev~\cite{Branco:1999fs,Davidson:2005cw}. In particular, we define the new Higgs doublet fields:
\beq \label{higgsbasispot}
H_1=\begin{pmatrix}H_1^+\\ H_1^0\end{pmatrix}\equiv \frac{v_1^* \Phi_1+v_2^*\Phi_2}{v}\,,
\qquad\quad H_2=\begin{pmatrix} H_2^+\\ H_2^0\end{pmatrix}\equiv\frac{-v_2 \Phi_1+v_1\Phi_2}{v}
 \,.
\eeq
It follows that $\vev{H_1^0}=v/\sqrt{2}$ and $\vev{H_2^0}=0$.
The Higgs basis is uniquely defined
up to an overall rephasing, $H_2\to e^{i\chi} H_2$ (which does not alter the fact that
$\vev{H_2^0}=0$).  In the Higgs basis, the scalar potential is
given by~\cite{Branco:1999fs,Davidson:2005cw}:
\beqa \mathcal{V}&=& Y_1 H_1^\dagger H_1+ Y_2 H_2^\dagger H_2 +[Y_3
H_1^\dagger H_2+{\rm h.c.}]
+\half Z_1(H_1^\dagger H_1)^2+\half Z_2(H_2^\dagger H_2)^2
+Z_3(H_1^\dagger H_1)(H_2^\dagger H_2)
\nn\\
&&\quad
+Z_4( H_1^\dagger H_2)(H_2^\dagger H_1)
+\left\{\half Z_5 (H_1^\dagger H_2)^2 +\big[Z_6 (H_1^\dagger
H_1) +Z_7 (H_2^\dagger H_2)\big] H_1^\dagger H_2+{\rm
h.c.}\right\}\,,
\eeqa
where $Y_1$, $Y_2$ and $Z_1,\ldots,Z_4$ are real and uniquely defined,
whereas $Y_3$, $Z_5$, $Z_6$ and $Z_7$ are potentially complex and transform under
the rephasing of $H_2\to e^{i\chi} H_2$ as
\beq \label{rephase}
[Y_3, Z_6, Z_7]\to e^{-i\chi}[Y_3, Z_6, Z_7] \quad{\rm and}\quad
Z_5\to  e^{-2i\chi} Z_5\,,
\eeq
since $\mathcal{V}$ must be independent of $\chi$.  After minimizing the scalar potential, 
\beq
Y_1=-\half Z_1 v^2\,,\qquad\quad
Y_3=-\half Z_6 v^2\,.
\eeq
 This leaves 11 free parameters:
1 vev, 8 real parameters, $Y_2$, $Z_{1,2,3,4}$, $|Z_{5,6,7}|$, and two relative phases.

In the general 2HDM,
the physical charged Higgs boson is the charged component of the Higgs-basis doublet $H_2$, and its mass
is given by
\beq \label{chhiggsmass}
m_{H^\pm}^2=Y_{2}+\half Z_3 v^2\,.
\eeq
The three physical neutral Higgs boson mass-eigenstates
are determined by diagonalizing a $3\times 3$ real symmetric squared-mass
matrix that is defined in the Higgs basis~\cite{Branco:1999fs,Haber:2006ue}
\beq   \label{mtwo}
\mathcal{M}^2=v^2\left( \begin{array}{ccc}
Z_1&\,\, \Re Z_6 &\,\, -\Im Z_6\\
\Re Z_6 &\,\, \half (Z_{345}+Y_2/v^2) & \,\,
- \half \Im Z_5\\ -\Im Z_6 &\,\, - \half \Im Z_5 &\,\,
 \half (Z_{345}+Y_2/v^2)-\Re Z_5\end{array}\right),
\eeq
where $Z_{345}\equiv Z_3+Z_4+\Re Z_5$.
\clearpage

To identify the neutral Higgs mass-eigenstates, we diagonalize the squared-mass matrix $\mathcal{M}^2$.
The diagonalization matrix is a $3\times 3$
real orthogonal matrix that depends on three angles:
$\theta_{12}$, $\theta_{13}$ and~$\theta_{23}$.  Following Ref.~\cite{Haber:2006ue},
\beq \label{mixingmatrix}
\begin{pmatrix} h_1\\ h_2 \\ h_3\end{pmatrix}=\begin{pmatrix} c_{12} c_{13} & \quad -s_{12}c_{23}-c_{12}s_{13}s_{23} & \quad -c_{12}s_{13}c_{23}+s_{12}s_{23} \\
s_{12} c_{13} & \quad c_{12}c_{23}-s_{12}s_{13}s_{23} & \quad -s_{12}s_{13}c_{23}-c_{12}s_{23} \\
s_{13} & \quad c_{13}s_{23} & c_{13}c_{23}\end{pmatrix}\begin{pmatrix}
\sqrt{2}\Re H_1^0 -v \\ \sqrt{2}\Re H_2^0\\ \sqrt{2}\Im H_2^0 \end{pmatrix}\,,
\eeq 
where the $h_i$ are the mass-eigenstate neutral Higgs fields, $c_{ij}\equiv\cos\theta_{ij}$ and $s_{ij}\equiv\sin\theta_{ij}$.
Under the rephasing $H_2\to e^{i\chi}H_2$,
\beq \label{rephasing}
\theta_{12}\,,\, \theta_{13}~{\hbox{\text{are invariant, and}}}\quad
\theta_{23}\to  \theta_{23}-\chi\,.
\eeq
Assuming that $Z_6\equiv |Z_6|e^{i\theta_6}\neq 0$,\footnote{If $Z_6=0$, then one can always rephase the Higgs basis field $H_2$ such that $Z_5$ is real.  In this basis, the neutral Higgs boson squared-mass matrix, $\mathcal{M}^2$, is diagonal, and the identification of the neutral Higgs boson mass-eigenstates is trivial.} 
it is convenient to define the invariant mixing angle,
\beq \label{sinphi}
\phi\equiv \theta_{23}-\theta_6\,.
\eeq
In light of the freedom to define the mass-eigenstate Higgs fields up to an overall sign, the invariant mixing angles $\theta_{12}$, $\theta_{13}$ and $\phi$ can be determined modulo $\pi$.  By convention, we choose
\beq \label{range}
-\half\pi\leq\theta_{12}\,,\,\theta_{13}<\half\pi\,,\,\quad \text{and}\quad 0\leq\phi<\pi\,.
\eeq
The physical neutral Higgs states ($h_{1,2,3}$) are then given by:
\beq \label{hsubk}
h_k=\frac{1}{\sqrt{2}}\biggl\{q_{k1}^*\left(H_1^0-\frac{v}{\sqrt{2}}\right)+q_{k2}^*H_2^0 e^{i\theta_{23}}+{\rm h.c.}\biggr\}\,,
\eeq
where the $q_{k1}$ and $q_{k2}$ are invariant
combinations of  $\theta_{12}$ and $\theta_{13}$, which are exhibited
in Table~\ref{tabinv}~\cite{Haber:2006ue}.
The masses of the neutral Higgs bosons $h_i$ will be denoted by $m_i$, respectively.
It is convenient to define the physical charged Higgs states by 
\beq \label{chiggs}
H^\pm\equiv e^{\pm i\theta_{23}}H_2^\pm\,, 
\eeq
so that all the Higgs mass-eigenstate
fields ($h_1$, $h_2$, $h_3$ and $H^\pm$) are invariant under $H_2\to e^{i\chi}H_2$.
\begin{table}[t!]
\centering
\begin{tabular}{|c||c|c|}\hline
$\phaa k\phaa $ &\phaa $q_{k1}\phaa $ & \phaa $q_{k2} \phaa $ \\
\hline
$1$ & $c_{12} c_{13}$ & $-s_{12}-ic_{12}s_{13}$ \\
$2$ & $s_{12} c_{13}$ & $c_{12}-is_{12}s_{13}$ \\
$3$ & $s_{13}$ & $ic_{13}$ \\ \hline
\end{tabular}
\caption{Invariant combinations of the neutral Higgs boson mixing angles $\theta_{12}$ and $\theta_{13}$,
where $c_{ij}\equiv\cos\theta_{ij}$ and $s_{ij}\equiv\sin\theta_{ij}$.          \label{tabinv}}
\end{table}

Although the explicit formulae for the neutral  Higgs boson masses and mixing
angles are quite complicated, there are numerous relations among them which take on rather simple forms.  The
following results are noteworthy~\cite{Haber:2006ue,Haber:2010bw}:
\beqa
Z_1 v^2&=&m_1^2 c_{12}^2 c_{13}^2+m_2^2 s_{12}^2 c_{13}^2 + m_3^2
s_{13}^2\,,\label{z1v} \\[5pt]
\Re(Z_6\,e^{-i\theta_{23}})\,v^2 &=& c_{13}s_{12}c_{12}(m_2^2-m_1^2)\,,
\label{z6rv} \\[5pt]
\Im(Z_6\,e^{-i\theta_{23}})\,v^2 &=& s_{13}c_{13}(c_{12}^2 m_1^2+s_{12}^2
m_2^2-m_3^2) \,, \label{z6iv}   \\[5pt]
\Re(Z_5\,e^{-2i\theta_{23}})\,v^2 &=& m_1^2(s_{12}^2-c_{12}^2 s_{13}^2)+m_2^2(c_{12}^2-s_{12}^2 s_{13}^2)-m_3^2 c_{13}^2\,,
\label{z5rv} \\[5pt]
\Im(Z_5\,e^{-2i\theta_{23}})\,v^2 &=& 2s_{12}c_{12}s_{13}(m_2^2-m_1^2)\,. \label{z5iv}
\eeqa

We next turn to the Higgs-fermion Yukawa couplings.  As reviewed in Appendix \ref{appA}, one starts out initially with a Lagrangian
expressed in terms of the scalar doublet fields $\Phi_i$ ($i=1,2$) and
interaction--eigenstate quark and lepton fields.  After electroweak symmetry breaking,
one can re-express the scalar doublet fields in terms of the Higgs basis fields $H_1$ and $H_2$.
At the same time,
one can identify the $3\times 3$ quark and lepton mass matrices.  By redefining the left
and right-handed quark and lepton fields appropriately, the quark and lepton
mass matrices are transformed into diagonal form, where the diagonal elements are real
and non-negative. The resulting Higgs--fermion Yukawa Lagrangian is given by in \eq{yukhbasis2} and is repeated here for the convenience of the reader~\cite{Haber:2010bw},
\beqa
-\mathscr{L}_{\rm Y}&=&\overline U_L (\kappa^U H_1^{0\,\dagger}
+\rho^U H_2^{0\,\dagger})\ur
-\anti D_L K^\dagger(\kappa^U H_1^{-}+\rho^U H_2^{-})\ur \nonumber \\
&& +\anti U_L K (\kappa^{D\,\dagger}H_1^++\rho^{D\,\dagger}H_2^+)\dr
+\anti D_L (\kappa^{D\,\dagger}H_1^0+\rho^{D\,\dagger}H_2^0)\dr \nonumber \\
&& +\anti N_L (\kappa^{E\,\dagger}H_1^++\rho^{E\,\dagger}H_2^+)\er
+\anti E_L (\kappa^{E\,\dagger}H_1^0+\rho^{D\,\dagger}H_2^0)\er 
+{\rm h.c.},
\label{lyuk}
\eeqa
where $U=(u,c,t)$ and $D=(d,s,b)$ are the mass-eigenstate quark fields, $K$ is the
CKM
mixing matrix, $N=(\nu_e,\nu_\mu,\nu_\tau)$ and $E=(e,\mu,\tau)$ are the mass-eigenstate lepton fields,
and $\kappa$ and $\rho$ are $3\times 3$
Yukawa coupling matrices.  Note that $F_{R,L}\equiv P_{R,L}F$,
where $F=U$, $D$, $N$ and $E$, and $P_{R,L}\equiv\half(1\pm\gamma\lsub{5})$ are
the right and left-handed projection operators, respectively.   At this stage, the neutrinos are exactly massless,
so we are free to define the physical left-handed neutrino fields, $N_L$, such that their charged current interactions are generation-diagonal.\footnote{To incorporate the neutrino masses, one can employ a seesaw mechanism~\cite{Minkowski:1977sc,GellMann:1980vs,Yanagida:1980xy,Mohapatra:1979ia,Mohapatra:1980yp}
and introduce three right-handed neutrino fields along with an explicit SU(2)$\times$ U(1) conserving mass term.  See footnote~\ref{fn4}.}

By setting $H_1^0=v/\sqrt{2}$ and $H_2^0=0$, one can relate
$\kappa^U$, $\kappa^D$, and $\kappa^E$ to the diagonal (up-type and down-type)
quark and charged lepton mass matrices $M_U$, $M_D$, and $M_E$,
respectively,
\beqa
M_U&=&\frac{v}{\sqrt{2}}\kappa^U={\rm diag}(m_u\,,\,m_c\,,\,m_t)\,,\qquad
M_D=\frac{v}{\sqrt{2}}\kappa^{D\,\dagger}={\rm
diag}(m_d\,,\,m_s\,,\,m_b) \,,   \nn \\
M_E&=&\frac{v}{\sqrt{2}}\kappa^{E\,\dagger}={\rm
diag}(m_e\,,\,m_\mu\,,\,m_\tau) \,.\label{mumd}
\eeqa
However, the complex matrices $\rho^F$ ($F=U,D,E$) are unconstrained.   Moreover,
under the rephasing $H_2\to e^{i\chi}H_2$, the Yukawa matrix acquires an overall phase, $\rho^F\to e^{i\chi}\rho^F$, since
$\mathscr{L}_{\rm Y}$ must be independent of $\chi$.

To obtain the physical Yukawa couplings of the Higgs boson, one must relate the
Higgs basis scalar fields to the Higgs mass-eigenstate fields.  Using \eqs{hsubk}{chiggs}, the
Higgs--fermion Yukawa couplings are given by,
\beqa
\!\!\!\!\!\!\!\!\!\!\!\!\!\!
 -\mathscr{L}_Y &=&  \overline U \sum_{k=1}^3\biggl\{q_{k1}\frac{M_U}{v}+\frac{1}{\sqrt{2}}\left[
q^*_{k2}\,e^{i\theta_{23}}\rho^U P_R+
q_{k2}\,[e^{i\theta_{23}}\rho^U]^\dagger P_L\right]\biggr\}U h_k
\nonumber \\
&& +\overline D\sum_{k=1}^3
\biggl\{q_{k1}\frac{M_D}{v} +\frac{1}{\sqrt{2}}\left[
q_{k2}\,[e^{i\theta_{23}}\rho^D]^\dagger P_R+
q^*_{k2}\,e^{i\theta_{23}}\rho^D P_L\right]\biggr\}Dh_k \nonumber \\
&& +\overline E\sum_{k=1}^3
\biggl\{q_{k1}\frac{M_E}{v} +\frac{1}{\sqrt{2}}\left[
q_{k2}\,[e^{i\theta_{23}}\rho^E]^\dagger P_R+
q^*_{k2}\,e^{i\theta_{23}}\rho^E P_L\right]\biggr\}Eh_k \nonumber \\
&& +\biggl\{\overline U\bigl[K[e^{i\theta_{23}}\rho^D]^\dagger
P_R-[e^{i\theta_{23}}\rho^U]^\dagger KP_L \bigr]DH^+ +\overline N[e^{i\theta_{23}}\rho^E]^\dagger
P_R EH^+  +{\rm
h.c.}\biggr\}\,. \label{YUK}
\eeqa
The
combinations $e^{i\theta_{23}}\rho^U$, $e^{i\theta_{23}}\rho^D$ and $e^{i\theta_{23}}\rho^E$ that appear in these interactions are invariant under
the rephasing of $H_2$.
It is convenient to rewrite the Higgs-fermion Yukawa couplings in terms of
the following three $3\times 3$ hermitian matrices that are invariant
with respect to the rephasing of $H_2$,
\beqa
\rho^F_R &\equiv& \frac{v}{2\sqrt{2}}\,M^{-1/2}_F
\biggl\{e^{i\theta_{23}}\rho^F +
[e^{i\theta_{23}}\rho^F]^\dagger\biggr\}M^{-1/2}_F\,,
\qquad \text{for $F=U,D,E$}\,,\label{rhoR} \\[6pt]
\rho^F_I &\equiv& \frac{v}{2\sqrt{2}\,i}M^{-1/2}_F
\biggl\{e^{i\theta_{23}}\rho^F -
[e^{i\theta_{23}}\rho^F]^\dagger\biggr\}M^{-1/2}_F\,,
\qquad \text{for $F=U,D,E$}\,,\label{rhoI}
\eeqa
where the $M_F$ are the diagonal 
fermion mass matrices [cf.~\eq{mumd}] and the Yukawa
coupling matrices are introduced in \eq{lyuk}.
Then, the Yukawa couplings take the following form:
\beqa
\!\!\!\!\!\!\!\!\!\!
-\mathscr{L}_Y &=& \frac{1}{v}\,\overline U \sum_{k=1}^3 M_U^{1/2}\biggl\{q_{k1}\mathds{1}
+ \Re(q_{k2})\bigl[\rho^U_R+i\gamma\ls{5}\rho^U_I\bigr]+\Im(q_{k2})\bigl[\rho^U_I-i\gamma\ls{5}\rho^U_R\bigr]\biggr\} M_U^{1/2}Uh_k \nonumber \\
&&  +\frac{1}{v}\,\overline D\sum_{k=1}^3 M_D^{1/2}
\biggl\{q_{k1}\mathds{1}  +\Re(q_{k2})\bigl[\rho^D_R-i\gamma\ls{5}\rho^D_I\bigr]+\Im(q_{k2})\bigl[\rho^D_I+i\gamma\ls{5}\rho^D_R\bigr]\biggr\} M_D^{1/2}Dh_k  \nonumber \\
&&  +\frac{1}{v}\,\overline E\sum_{k=1}^3 M_E^{1/2}
\biggl\{q_{k1}\mathds{1}  + \Re(q_{k2})\bigl[\rho^E_R-i\gamma\ls{5}\rho^E_I\bigr]+\Im(q_{k2})\bigl[\rho^E_I+i\gamma\ls{5}\rho^E_R\bigr]\biggr\} M_E^{1/2}Eh_k  \nonumber \\
&& +\frac{\sqrt{2}}{v}\biggl\{\overline U\bigl[KM_D^{1/2}(\rho^D_R-i\rho^D_I)
M_D^{1/2}P_R-M_U^{1/2}(\rho^U_R-i\rho^U_I) M_U^{1/2}KP_L\bigr] DH^+ \nonumber \\
&&\qquad\qquad  \overline{N}M_E^{1/2}(\rho^E_R-i\rho^E_I)M_E^{1/2}P_R EH^+
+{\rm
h.c.}\biggr\}, \label{YUK2}
\eeqa
where $\mathds{1}$ is the $3\times 3$ identity matrix.
The appearance of unconstrained hermitian $3\times 3$ Yukawa matrices
$\rho^F_{R,I}$ in \eq{YUK2} indicates the presence of potential flavor-changing neutral Higgs--quark and lepton interactions.
If the off-diagonal elements of $\rho^F_{R,I}$ are unsuppressed, they will generate tree-level Higgs-mediated FCNCs
that are incompatible with the strong suppression of FCNCs observed in nature.

\subsection{The limit of a SM-like Higgs boson}

Current LHC data suggest that the properties of the observed Higgs boson are consistent with the predictions of the Standard Model.  In this paper, we shall identify $h_1$ as the SM-like Higgs boson. In light of the expression for the $h_1$ coupling to a pair of vector bosons $VV=W^+ W^-$ or $ZZ$,
\beq \label{hvv}
\frac{g_{h_1VV}}{g_{h_{\rm SM}VV}}=c_{12} c_{13}\simeq 1\,,\qquad \text{where $V=W$ or $Z$}\,,
\eeq
it follows that $|s_{12}|$, $|s_{13}|\ll 1$.  Thus, in the
limit of a SM-like Higgs boson, \eqs{z6rv}{z6iv} yield~\cite{Haber:2006ue}:
\beqa
|s_{12}|&\simeq &
\left|\frac{\Re(Z_6 e^{-i\theta_{23}})v^2}{m_2^2-m_1^2}\right|\ll 1\,, \label{done}\\
|s_{13}|&\simeq&
\left|\frac{\Im(Z_6 e^{-i\theta_{23}})v^2}{m_3^2-m_1^2}\right|\ll 1\,.\label{dtwo}
\eeqa
In addition, Eq.~(\ref{z5iv}) implies that one additional small quantity characterizes the
limit of a SM-like Higgs boson,
\beq \label{dthree}
|\Im(Z_5 e^{-2i\theta_{23}})|\simeq \left|\frac{2(m_2^2-m_1^2) s_{12}s_{13}}{v^2}\right|
\simeq \left|\frac{\Im(Z_6^2 e^{-2i\theta_{23}})v^2}{m_3^2-m_1^2}\right|\ll 1\,.
\eeq
Moreover, in the limit of a SM-like Higgs boson, \eq{z5rv} yields
\beq \label{m23}
m_2^2-m_3^2\simeq \Re(Z_5 e^{-2i\theta_{23}})v^2\,.
\eeq

As a consequence of \eqs{done}{dtwo},
the limit of a SM-like Higgs boson\footnote{In the literature, this is often referred to as the alignment limit~\cite{Gunion:2002zf,Craig:2013hca,Asner:2013psa,Carena:2013ooa,Haber:2013mia,Dev:2014yca,Pilaftsis:2016erj}.  We do not use this nomenclature here in order to avoid confusion with flavor alignment, which is the focus of this paper.}
can be achieved if either $|Z_6|\ll 1$ and/or if $m_2$, $m_3\gg v$.  The latter corresponds to the well-known decoupling limit of the 2HDM~\cite{Haber:1989xc,Gunion:2002zf,Haber:2006ue}.\footnote{Note that \eq{m23} implies that in the decoupling limit, 
$m_2\gg v$ implies that $m_3\gg v$ and vice versa.}
In this paper, we will focus on the decoupling regime of the 2HDM to ensure that $h_1$ is sufficiently SM-like, in light of the current LHC Higgs data~\cite{Khachatryan:2016vau}. 

\subsection{Neutral scalars of definite CP}\label{Sec:neutralCP}

In the exact SM-Higgs boson limit, the couplings of $h_1$ are precisely those of the SM Higgs boson.  In this case, we can identify $h_1$ as a CP-even scalar.  In general, the heavier neutral Higgs bosons, $h_2$ and $h_3$ can be mixed CP states.   The limit in which  $h_2$ and $h_3$ are approximate eigenstates of CP is noteworthy.  This limit is achieved assuming that $|s_{13}|\ll |s_{12}|$.  That is,
\beq \label{sfrac}
\left|\frac{s_{13}}{s_{12}}\right|\simeq \left|\left(\frac{m_2^2-m_1^2}{m_3^2-m_1}\right)\frac{\Im(Z_6 e^{-i\theta_{23}})}{\Re(Z_6 e^{-i\theta_{23}})}\right|\ll 1\,.
\eeq
In the decoupling limit, the ratio of squared-mass differences in \eq{sfrac} is of $\mathcal{O}(1)$.
Moreover, unitarity and perturbativity constraints suggest that $\Re(Z_6 e^{-i\theta_{23}})$ cannot be significantly larger than $\mathcal{O}(1)$.  Hence, it follows that
\beq \label{imz6}
|\Im(Z_6 e^{-i\theta_{23}})|\ll 1\,.
\eeq

In light of \eq{rephasing}, we can rephase $H_2 \to e^{i\chi} H_2$ such that $\theta_{23}=0$~(mod~$\pi$), i.e.~$c_{23}=\pm 1$.  \Eqs{dthree}{imz6} then yield
$
|\Im Z_5|\,,\,|\Im Z_6| \ll 1\,.
$
For simplicity in the subsequent analysis, we henceforth assume that a \textit{real Higgs basis} exists in which $Z_5$ and $Z_6$ are simultaneously real.  In this case, the scalar Higgs potential and the Higgs vacuum are CP-invariant, and
the squared-mass matrix of the neutral Higgs bosons given in \eq{mtwo} simplifies,
\beq   \label{mtwocp}
\mathcal{M}^2=
\begin{pmatrix}
Z_1 v^2&\quad Z_6 v^2 &\,\, 0\\
Z_6 v^2  &\quad Y_2+\half (Z_3+Z_4+Z_5)v^2 & \,\, 0 \\
0 &\quad 0 &\,\,
 Y_2+\half (Z_3+Z_4-Z_5)v^2\end{pmatrix}\,,
\eeq
where $Z_5$ and $Z_6$ are real.  Moreover, $c_{13}=1$ and we can set $\theta_{23}=\theta_6=0$~(mod~$\pi$), or equivalently
\beq \label{theta6}
e^{i\theta_{23}}=c_{23}=\epsilon_6\,,
\eeq
where $\epsilon_6\equiv {\rm sgn}~Z_6$, in the real Higgs basis [cf.~\eqs{sinphi}{range}].
To maintain the reality of the Higgs basis, the only remaining
freedom in defining the Higgs basis fields is the overall sign of the field $H_2$.  In particular, under $H_2\to -H_2$, we see that $Z_5$ is invariant whereas $Z_6$ (and $Z_7$) and $c_{23}$ change sign.
We immediately identify the CP-odd Higgs boson $A=\sqrt{2}\,{\rm Im}~H_2^0$ with
squared mass,
\beq \label{ma}
m_A^2=Y_2+\half (Z_3+Z_4-Z_5)v^2\,.
\eeq
Note that the real Higgs mass-eigenstate field, $A$, is defined up to an overall sign change, which corresponds to the freedom to redefine $H_2\to -H_2$.  In contrast, the charged Higgs field $H^\pm$ defined (as a matter of convenience) by \eq{chiggs} is invariant with respect to $H_2\to -H_2$.  Indeed, by using \eq{theta6}, we can now write $H^\pm=\epsilon_6 H_2^\pm$.
In light of \eqs{chhiggsmass}{ma},
\beq \label{hpmmass}
m_{H^\pm}^2=m_A^2-\half(Z_4-Z_5)v^2\,.
\eeq

The upper $2\times 2$ matrix block given in \eq{mtwocp} is the CP-even Higgs squared-mass matrix,
\beq \label{Hmm}
\mathcal{M}_H^2=\begin{pmatrix} Z_1 v^2 & \quad Z_6 v^2 \\  Z_6 v^2 & \quad m_A^2+Z_5 v^2\end{pmatrix}\,,
\eeq
where we have used \eq{ma} to eliminate $Y_2$.
To diagonalize $\mathcal{M}^2_H$, we define
the CP-even mass-eigenstates, $h$ and $H$ (with $m_{h}\leq m_{H}$) by
\beq \label{hH}
\begin{pmatrix} H\\ h\end{pmatrix}=\begin{pmatrix} \cbma & \,\,\, -\sbma \\
\sbma & \,\,\,\phantom{-}\cbma\end{pmatrix}\,\begin{pmatrix} \sqrt{2}\,\,{\rm Re}~H_1^0-v \\ 
\sqrt{2}\,{\rm Re}~H_2^0
\end{pmatrix}\,,
\eeq
where $\cbma\equiv\cba$ and $\sbma\equiv\sba$ are defined in terms of the angle $\beta$ defined via $\tan\beta\equiv v_2/v_1$, and
the mixing angle $\alpha$ that diagonalizes the CP-even Higgs squared-mass matrix when expressed relative to the original basis of scalar fields, $\{\Phi_1\,,\,\Phi_2\}$, which is assumed here to be a \textit{real} basis.\footnote{Given the assumption [indicated above \eq{mtwocp}] that the scalar Higgs potential and the Higgs vacuum are CP-invariant, it follows that there must exist a \textit{real} basis of scalar fields in which all scalar potential parameters and the vacuum expectation values of the two neutral Higgs fields, $\langle\Phi_i^0\rangle\equiv v_i/\sqrt 2$ (for $i=1$, 2), are simultaneously real~\cite{Gunion:2005ja}.}    
 Since the
real Higgs mass-eigenstate fields $H$ and $h$ are defined up to an overall sign change,
it follows that $\beta-\alpha$ is determined modulo $\pi$.  To make contact with the notation of \eq{mixingmatrix}, we note that $c_{13}=1$ and $c_{23}=\epsilon_6$ [cf.~\eq{theta6}].
Assuming that $h_1$ is the lighter of the two neutral CP-even Higgs bosons, then
\eq{hH} implies the following identifications:
\beq\label{eq:epsilon6}
h=h_1\,,\qquad H=-\epsilon_{6} h_2\,,\qquad A=\epsilon_{6} h_3\,,
\eeq
and
\beq \label{anglelimit}
c_{12}=\sbma\,,\qquad\quad s_{12}=-\epsilon_{6}\cbma\,.
\eeq
This means that the signs of the fields $H$ and $A$ and the sign of $\cbma$ all flip under the redefinition of the Higgs basis field $H_2\to -H_2$.  

Note that $0\leq\sbma\leq 1$ in
the convention specified in \eq{range}.
Moreover, \eq{z6rv} yields
\beq \label{sc}
\sbma\cbma=-\frac{Z_6 v^2}{m_H^2-m_h^2}\,,
\eeq
and it therefore follows that $0\leq s_{12}\,,\,c_{12}\leq 1$ and $\cbma Z_6\leq 0$.  The decoupling limit corresponds to $m_H\gg m_h$ and $|\cbma|\ll 1$ [cf.~\eq{done}], in which case we can identify $h$ as the SM-like Higgs boson and $H$ as the heavier CP-even Higgs boson.  
Finally, \eqst{z1v}{z5iv} yield
\beqa
Z_1 v^2&=&m_h^2\ s^2_{\beta-\alpha}+ m_H^2\ c^2_{\beta-\alpha}\,,\label{z1v2} \\
Z_6 v^2&=&(m_h^2-m_H^2)s_{\beta-\alpha}c_{\beta-\alpha}\,, \\
Z_5 v^2 &=& m_H^2\ s^2_{\beta-\alpha}+ m_h^2\ c^2_{\beta-\alpha}-m_A^2\,.\label{z5v2}
\eeqa
In particular, $m_h^2\simeq Z_1 v^2$ in the limit of a SM-like Higgs boson $h$.
Applying \eq{anglelimit} to Table~\ref{tabinv}, 
\beqa
q_{11}=\sbma\,, \qquad && q_{12}=\epsilon_{6}\cbma\,,\label{q1} \\
q_{21}=-\epsilon_{6}\cbma\,, \qquad && q_{22}=\sbma\,,\label{q2} \\
q_{31}=0\,, \qquad && q_{32}=i\,.\label{q3}
\eeqa
Inserting these results into the general form of the Yukawa couplings given in \eq{YUK2}, we obtain the following Higgs-fermion couplings  in the case of a CP-conserving Higgs scalar potential and vacuum, 
\beqa
\!\!\!\!\!\!\!\!\!\!
-\mathscr{L}_Y &=& \frac{1}{v}\sum_{F=U,D,E}\overline F \biggl\{\sbma M_F
+\epsilon_{6}\cbma M_F^{1/2}\bigl[\rho^F_R +i\varepsilon_F\gamma\ls{5}\rho^F_I\bigr] M_F^{1/2} \biggr\}F h 
\nonumber \\[8pt]
&&
+\frac{1}{v}\sum_{F=U,D,E}\overline F\biggl\{\cbma M_F
-\epsilon_{6}\sbma  M_F^{1/2}\bigl[\rho^F_R +i\varepsilon_F\gamma\ls{5}\rho^F_I \bigr] M_F^{1/2}\biggr\} FH
\nonumber \\[8pt]
&& +\frac{1}{v}\sum_{F=U,D,E} \overline{F} \biggl\{M_F^{1/2} \epsilon_{6}\bigl(\rho_I^F-i\varepsilon_F\gamma\ls{5}\rho_R^F\bigr)
M_F^{1/2}\biggr\}FA
\nonumber \\[8pt]
&& +\frac{\sqrt{2}}{v}\biggl\{\overline U\bigl[KM_D^{1/2}(\rho^D_R-i\rho^D_I)
M_D^{1/2}P_R-M_U^{1/2}(\rho^U_R-i\rho^U_I) M_U^{1/2}KP_L\bigr] DH^+ \nonumber \\
&&\qquad\qquad  \overline{N}M_E^{1/2}(\rho^E_R-i\rho^E_I)M_E^{1/2}P_R EH^+
+{\rm
h.c.}\biggr\}, \label{YUK3}
\eeqa
where we have introduced the notation,
\beq
\varepsilon_F=\begin{cases} +1 & \quad \text{for $F=U$}\,,\\
-1 & \quad \text{for $F=D,E$}\,.\end{cases}
\eeq
Moreover, by employing \eq{theta6} in \eqs{rhoR}{rhoI}, the expressions for $\rho^F_R$ and $\rho^F_I$ in terms of the
Higgs Yukawa coupling matrices $\rho^F$ simplify,
\beqa
\epsilon_6 M_F^{1/2}\rho^F_R M_F^{1/2}&=&\frac{v}{2\sqrt{2}}\bigl(\rho^F+[\rho^F]^\dagger\bigr)\,,\\
i\epsilon_6 M_F^{1/2}\rho^F_I M_F^{1/2}&=&\frac{v}{2\sqrt{2}}\bigl(\rho^F-[\rho^F]^\dagger\bigr)\,.
\eeqa

The structure of the neutral Higgs couplings given in \eq{YUK3} is easily ascertained.
If $\rho_I^F\neq 0$, then the neutral Higgs fields will exhibit CP-violating Yukawa couplings.\footnote{Likewise, if $\Im Z_7\neq 0$ in a basis where $Z_5$ and $Z_6$ are real, then the neutral Higgs fields will also possess CP-violating trilinear and quadralinear scalar couplings.}
Moreover, the two sign choices, $\epsilon_6=\pm 1$ are physically indistinguishable, since the sign of $Z_6$ can always be flipped by redefining the Higgs basis field $H_2\to -H_2$.  Under this field redefinition, $\rho^F$, $\cbma$, $H$ and $A$ also flip sign, in which case $\mathscr{L}_Y$ is unchanged.

For completeness, we briefly consider the case where $h_1$ is the heavier of the two neutral CP-even 
Higgs bosons.  In this case, \eq{hH} implies the following identifications,
\beq
h=\epsilon_6 h_2\,,\qquad H=h_1\,,\qquad A=\epsilon_{6} h_3\,,
\eeq
and
\beq \label{anglelimit2}
c_{12}=\cbma\,,\qquad\quad s_{12}=\epsilon_6\sbma\,.
\eeq
This means that the signs of the fields $h$ and $A$ and the sign of $\sbma$ all flip under the redefinition of the Higgs basis field $H_2\to -H_2$. Note that \eqst{sc}{z5v2} are still valid.
Invoking the convention given by \eq{range} now implies that $0\leq\cbma\leq 1$ and $Z_6\sbma\leq 0$.  Moreover in light of \eq{hvv}, if $|\sbma|\ll 1$ then $H$ is SM-like and $m_H^2\simeq Z_1 v^2$, which is achieved in the limit of $|Z_6|\ll 1$.  No decoupling limit is possible in this case since
$m_h<m_H=125$~GeV.  Using \eq{anglelimit2}, one can check that  \eqst{q1}{q3} are modified by taking $\sbma\to\cbma$ and $\cbma\to -\sbma$.  As a result, \eq{YUK3} remains unchanged.

So far, the parameters $\alpha$ and $\beta$ have no separate significance.  Only the combination, $\beta-\alpha$ is meaningful.  Moreover the matrices $\rho^F_R$ and $\rho^F_I$ are generic complex matrices, which implies the existence of tree-level Higgs-mediated flavor changing neutral currents, as well as new sources of CP violation. However, experimental data suggest that such Higgs-mediated FCNCs must be highly suppressed.  One can eliminate these FCNCs by imposing a discrete $\mathbb{Z}_2$ symmetry 
$\Phi_1\to\Phi_1$ and $\Phi_2\to -\Phi_2$ on the quartic terms of the Higgs potential given in \eq{genpot}, which sets $\lambda_6=\lambda_7=0$ and gives physical significance to the $\Phi_1$--$\Phi_2$ basis choice.  This in turn promotes the CP-even Higgs mixing angle $\alpha$ in the real $\Phi_1$--$\Phi_2$ basis and
$\tan\beta\equiv v_2/v_1$ to physical parameters of the model.\footnote{Since the existence of a real Higgs basis implies no spontaneous nor explicit CP-violation in the scalar sector, there exists a $\Phi_1$--$\Phi_2$ basis in which the $\lambda_i$ of \eq{genpot}, $v_1$ and $v_2$ (and hence $\tan\beta$) are simultaneously real.}   The  $\mathbb{Z}_2$  symmetry can be extended to the Higgs-fermion interactions in four inequivalent ways.  In the notation of the Higgs-fermion Yukawa couplings given in \eq{YUK3}, the $\rho^F_{R,I}$ are given by\footnote{As defined here, the parameter $\tan\beta$ flips sign under the redefinition of the Higgs basis field $H_2\to -H_2$, in contrast to the more common convention where $\tan\beta$ is positive (by redefining $H_2\to -H_2$ if necessary).  With this latter definition, the two cases of
$\epsilon_6=\pm 1$ [or equivalently the two cases of ${\rm sgn}(\sbma\cbma)=\mp 1$] represent non-equivalent points of the Type-I, II, X or Y 2HDM parameter space.   However, we do \textit{not} adopt this latter convention in the present work.\label{fnconv}}
\begin{enumerate}
\item 
Type-I: For $F=U,D,E$, $\rho_R^F=\epsilon_6\cot\beta\mathds{1}$ and $\rho_I^F=0$.
\item
Type-II: $\rho_R^U=\epsilon_6\cot\beta\mathds{1}$ and $\rho_I^U=0$.  For $F=D,E$,
$\rho_R^F=-\epsilon_6\tan\beta\mathds{1}$ and $\rho_I^F=0$.
\item
Type-X: $\rho_R^E=-\epsilon_6\tan\beta\mathds{1}$ and $\rho_I^E=0$.  For $F=U,D$, 
$\rho_R^F=\epsilon_6\cot\beta\mathds{1}$ and $\rho_I^F=0$.
\item
Type-Y: $\rho_R^D=-\epsilon_6\tan\beta\mathds{1}$ and $\rho_I^D=0$. For $F=U,E$, $\rho_R^F=\epsilon_6\cot\beta\mathds{1}$ and $\rho_I^F=0$.
\end{enumerate}
Inserting these values for the $\rho_R^F$ and $\rho_I^F$ into \eq{YUK3}, the resulting neutral Higgs--fermion Yukawa couplings are flavor diagonal as advertised.

From a purely phenomenological point of view, one can simply avoid tree-level Higgs-mediated
FCNCs by declaring that the $\rho_R^F$ and $\rho_I^F$ are diagonal matrices.  In the simplest generalization of the Type I, II, X and Y Yukawa interactions, one asserts that both the $\rho_R^F$ and the $\rho_I^F$ are proportional to the identity matrix (where the constants of proportionality can depend on $F$).  This is called the flavor-aligned 2HDM, which we shall discuss in the next subsection.

\subsection{The flavor-aligned 2HDM}

The flavor-aligned 2HDM posits that the Yukawa matrices $\kappa^F$ and $\rho^F$ [cf.~\eq{lyuk}] are proportional.   When written in terms of fermion mass-eigenstates,
$\kappa^F=\sqrt{2}M_F/v$ is diagonal.  Thus in the A2HDM, the $\rho^F$ are likewise diagonal,
which implies that tree-level Higgs-mediated FCNCs are absent.
We define the \textit{alignment parameters} $a^F$ via,
\beq \label{aligned}
\rho^F=e^{-i\theta_{23}}a^F \kappa^F\,,\qquad\quad \text{for $F=U,D,E$},
\eeq
where the (potentially) complex numbers $a^F$ are invariant under the rephasing of the Higgs basis field $H_2\to e^{i\chi}H_2$. It follows from \eqs{rhoR}{rhoI} that
\beq
\rho_R^F=(\Re a^F)\mathds{1}\,,\qquad\quad \rho_I^F=(\Im a^F)\mathds{1}\,.
\eeq
Inserting the above results into \eq{YUK}, the Yukawa couplings take the following form:
\beqa
\!\!\!\!\!\!\!\!\!\!
-\mathscr{L}_Y &=& \frac{1}{v}\,\overline U \sum_{k=1}^3 M_U\biggl\{q_{k1}
+ q_{k2}^* a^U P_R+q_{k2} a^{U*}P_L\biggr\} Uh_k
 \nonumber \\
 &&
+\frac{1}{v}\,\overline D \sum_{k=1}^3 M_D\biggl\{q_{k1}
+ q_{k2} a^{D*} P_R+q^*_{k2}a^{D}P_L\biggr\} Dh_k
 \nonumber \\
  &&
+\frac{1}{v}\,\overline E \sum_{k=1}^3 M_E\biggl\{q_{k1}
+ q_{k2}a^{E*} P_R+q^*_{k2}a^{E}P_L\biggr\} Eh_k
 \nonumber \\
&& +\frac{\sqrt{2}}{v}\biggl\{\overline U\bigl[a^{D*}KM_DP_R-a^{U*}M_U KP_L\bigr] DH^+ +  a^{E*}\,\overline{N}M_EP_R EH^+
+{\rm
h.c.}\biggr\}. \label{YUK4}
\eeqa
This form simplifies further if the neutral Higgs mass-eigenstates are also states of definite CP.  In this case, the corresponding Yukawa couplings are given by
\beqa
\!\!\!\!\!\!\!\!\!\!
-\mathscr{L}_Y &=& \frac{1}{v}\sum_{F=U,D,E} \overline F  M_F\biggl\{\sbma
+\epsilon_6\cbma\bigl[\Re a^F+i\epsilon^F\Im a^F \gamma\ls{5}\bigr]\biggr\} Fh
 \nonumber \\
 &&
+\frac{1}{v}\sum_{F=U,D,E} \overline F  M_F\biggl\{\cbma
-\epsilon_6\sbma\bigl[\Re a^F+i\epsilon^F\Im a^F \gamma\ls{5}\bigr]\biggr\} FH
 \nonumber \\
  &&
+\frac{1}{v}\sum_{F=U,D,E} \overline F  M_F\biggl\{
\epsilon_6\bigl[\Im a^F-i\epsilon^F\Re a^F \gamma\ls{5}\bigr]\biggr\} FA
 \nonumber \\
&& +\frac{\sqrt{2}}{v}\biggl\{\overline U\bigl[a^{D*}KM_DP_R-a^{U*}M_U KP_L\bigr] DH^+ +  a^{E*}\,\overline{N}M_EP_R EH^+
+{\rm
h.c.}\biggr\}. \label{YUK5}
\eeqa
As noted above \eq{sc}, it is convenient to choose a convention in which $s_{\beta-\alpha}\geq 0$.  It then follows from \eq{sc} that
$\epsilon_6 c_{\beta-\alpha}=-|c_{\beta-\alpha}|$.  That is, the neutral Higgs couplings exhibited in \eq{YUK5} do not depend on the sign of $c_{\beta-\alpha}$ (which can be flipped by redefining the overall sign of the Higgs basis field $H_2$).   Note that in this convention, the signs of the alignment parameters $a^F$ are physical.  

The Type-I, II, X and Y Yukawa couplings are special cases of the A2HDM Yukawa couplings.
Since the $a^F$ ($F=U,D,E$) are independent complex numbers, there is no preferred basis for the scalar fields outside of the Higgs basis.  Thus, a priori, there is no separate meaning to the parameters $\alpha$ and $\beta$ in \eq{YUK5}.
Nevertheless, in the special case of a CP-conserving neutral Higgs-lepton interaction governed by \eq{YUK5} with $\Im a^E=0$, it is convenient to introduce the real parameter $\tan\beta$ via 
\beq \label{lepton}
a^E\equiv -\epsilon_6\tan\beta\,,
\eeq
corresponding to a Type-II or Type-X Yukawa couplings of the charged leptons to the neutral Higgs bosons. The theoretical interpretation of $\tan\beta$ defined by \eq{lepton} is as follows. It is always possible to choose a $\Phi_1$--$\Phi_2$ basis with the property that one of the two Higgs-lepton Yukawa coupling matrices vanishes.  Namely, in the notation of
\eq{ymodeliii0}, we have $\eta_2^{E,0}=0$, which means that only $\Phi_1$ couples to leptons.  In the case of a CP-conserving scalar Higgs potential and Higgs vacuum, we can take the $\Phi_1$--$\Phi_2$ basis to be a real basis and  identify 
$\tan\beta=v_2/v_1$, where $\langle\Phi_i^0\rangle\equiv v_i/\sqrt{2}$ (for $i=1$,~2).  However, in contrast to Type-II or Type-X models, $\eta_2^{E,0}=0$ does not correspond to a discrete $\mathbb{Z}_2$ symmetry of the generic A2HDM Lagrangian, since we do not require any of the Higgs-quark Yukawa coupling matrices and the scalar potential parameters 
$\lambda_6$ and $\lambda_7$ to vanish in the same $\Phi_1$--$\Phi_2$ basis.

Note that the sign of $a^E$ in \eq{lepton} is physical since both $\epsilon_6$ and $\tan\beta$ flip sign under the Higgs basis field $H_2\to -H_2$.  In contrast to the standard conventions employed in the 2HDM with Type-I, II, X or Y Yukawa couplings where $\tan\beta$ is defined to be positive [cf.~footnote~\ref{fnconv}], we shall \textit{not} adopt such a convention here.  In practice, we will rewrite \eq{lepton} as, 
\beq \label{leptonE}
a^E=\epsilon_E|\tan\beta|\,,
\eeq
where $\epsilon_E=\pm 1$ correspond to physically non-equivalent points of the A2HDM parameter space.

One theoretical liability of the A2HDM is that for generic choices of the alignment parameters $a^U$ and $a^D$, the flavor-alignment conditions in the quark sector specified in \eq{aligned} are not stable under the evolution governed by the Yukawa coupling renormalization group equations.  Indeed, as shown in Ref.~\cite{Ferreira:2010xe}, \eq{aligned} is stable under renormalization group running if and only if the parameters $a^U$ and $a^D$  satisfy the conditions of the Type I, II, X or Y 2HDMs specified at the end of section~\ref{Sec:neutralCP}.  In the leptonic sector, since we ignore neutrino masses, the Higgs-lepton Yukawa couplings are flavor-diagonal at all scales.  
We therefore assume that\footnote{Under the assumption of a real Higgs basis, $\epsilon_6=e^{i\theta_{23}}$ is fixed via \eq{lepton}.  This factor, which appears in \eq{aligned}, can then be absorbed into the definition of $a^F$. \label{fn}}
\beq \label{eqn:alignment}
\rho^F(\Lambda)=a^F \kappa^F(\Lambda)\,,\qquad\quad \text{for $F=U,D$},
\eeq
at some very high energy scale $\Lambda$ (such as the grand unification (GUT) scale or the Planck scale).  That is, we assume that the alignment conditions are set by some a priori unknown physics at or above the energy scale $\Lambda$.  We take the complex alignment parameters $a^F$ to be
boundary conditions for the RGEs of the Yukawa coupling matrices, and then determine the low-energy values of the Yukawa coupling matrices by numerically solving the RGEs.
To ensure that the resulting low-energy theory is 
consistent with a SM-like Higgs boson observed at the LHC, we shall take $m_h=125$~GeV, and assume that the masses of $H$, $A$ and $H^\pm$ 
are all of order $\Lambda_H\geq 400$~GeV.   In this approximate decoupling regime, $|\cbma|$ is small enough such that the properties of $h$ are within about 20\% of the SM Higgs boson, as required by the LHC Higgs data~\cite{Khachatryan:2016vau}.  We employ the 2HDM RGEs given in Appendix~\ref{sec:RGEs} from $\Lambda$ down to $\Lambda_H$, and then match onto the RGEs of the Standard Model to generate the Higgs-fermion Yukawa couplings at the electroweak scale, which we take to be $m_t$ or $m_Z$.  Note that the values of $\kappa^Q(\Lambda_H)=\sqrt{2} M_Q(\Lambda_H)/v$ (for $Q=U$, $D$) are determined from the known quark masses via Standard Model RG running. 

As noted above for the lepton case ($F=E$), if $\rho^E(\Lambda)$ is proportional to $\kappa^E(\Lambda)$, then $\rho$ is proportional to $\kappa$ at all energy scales.  Thus, we identify the leptonic alignment parameter at low energies by $\tan\beta$.  More precisely [cf.~\eqs{mumd}{leptonE}],
\beq \label{leptonbc}
\rho^E(\Lambda_H)=\sqrt{2}\epsilon_E|\tan\beta|M_E(\Lambda_H)/v\,.
\eeq
Then, $M_E(\Lambda_H)$ is determined by the diagonal lepton mass matrix via Standard Model RG running.

\subsection{Higgs-mediated FCNCs from high scale alignment}

To explore the Higgs-mediated FCNCs that can be generated in the A2HDM at the electroweak scale, we establish flavor-alignment at some high energy scale, $\Lambda$, as for example at the GUT or Planck scale, and run the one-loop RGEs from the high scale to the electroweak scale. Thus, we impose the following boundary conditions for the running of the one-loop 2HDM Yukawa couplings,
\begin{eqnarray}
\kappa^Q(\Lambda_H) &=& \sqrt{2} M_Q(\Lambda_H)/v, \label{eqn:kappalow} \\
\rho^Q(\Lambda) &=& a^Q \kappa^Q(\Lambda), \label{eqn:rhohigh}
\end{eqnarray}
where the $M_Q$ ($Q=U$, $D$) are the diagonal quark matrices, and $\Lambda_H$ is the scale of the heavier doublet, taken to be relatively large to guarantee that we are sufficiently in the decoupling limit. For the lepton sector, the corresponding boundary conditions are [cf.~\eq{leptonE}],
\begin{eqnarray}
\kappa^E(\Lambda_H)&=&\sqrt 2 M_E(\Lambda_H)/v,\\
\rho^E(\Lambda_H)&=&\epsilon_E|\tan\beta| \kappa^E(\Lambda_H).
\end{eqnarray}
Satisfying the two boundary conditions for the quark sector [\eqs{eqn:kappalow}{eqn:rhohigh}] is not trivial, since they are imposed at opposite ends of the RG running.
For example, to set flavor-alignment at the high energy scale, we must know the values of $\kappa^Q(\Lambda)$. This involves running up $\kappa^Q(\Lambda_H)$ to the high scale, but since the one-loop RGEs are strongly coupled to the $\rho^Q$ matrices, we must supply values for $\rho^Q(\Lambda_H)$ to begin the running.

With no {\it{a priori}} knowledge of which values of $\rho^Q(\Lambda_H)$ lead to flavor-alignment at the high scale, we begin the iterative process by assuming flavor-alignment at $\Lambda_H$ via a low-scale alignment parameter $a^{\prime Q}$,
\beq \label{aprime}
\rho^Q(\Lambda_H) = a^{\prime Q} \kappa^Q(\Lambda_H).
\eeq 
This flavor-alignment will be broken during RGE evolution to the high scale, and a procedure is needed to reestablish flavor-alignment at the high scale. To accomplish this, we decompose $\rho^Q(\Lambda)$ into parts that are aligned and misaligned with $\kappa^Q(\Lambda)$, respectively,
\beq\label{eqn:misaligned}
\rho^Q(\Lambda) = a^Q \kappa^Q(\Lambda) + \delta\rho^Q,
\eeq  
where $a^Q$ represents the aligned part (in general, different from $a^{\prime Q})$, and $\delta\rho^Q$ the corresponding degree of misalignment at the high scale. 

To minimize the misaligned part of $\rho^Q(\Lambda)$, we implement the cost function,
\beq
\Delta^Q \equiv \sum_{i,j=1}^3 |\delta\rho_{ij}^Q|^2 = \sum_{i,j=1}^3 |\rho_{ij}^Q(\Lambda)-a^Q\kappa_{ij}^Q(\Lambda)|^2,
\eeq
which, once minimized, provides the optimal value of the complex parameter $a^Q$ for flavor-alignment at the high scale,
\beq
a^Q \equiv \frac{{\sum_{i,j=1}^3}{\kappa^{Q*}_{ij}(\Lambda)\rho^Q_{ij}(\Lambda)}}{{\sum_{i,j=1}^3}{\kappa^{Q*}_{ij}(\Lambda)\kappa^Q_{ij}(\Lambda)}}.
\eeq 
We subsequently impose flavor-alignment at the high scale using this optimized alignment parameter,
\beq
\rho^Q(\Lambda) = a^Q \kappa^Q(\Lambda),
\eeq
and evolve the one-loop RGEs back down to $\Lambda_H$. In principle, further running of
$\kappa^U$ and $\kappa^D$ below $\Lambda_H$ can regenerate off-diagonal terms.  However, these effects are extremely small and can be ignored in practice. At $\Lambda_H$, we use (\ref{eqn:kappalow}) to match the boundary conditions for the 2HDM and SM.  At this point, the matrices $\kappa^U$ and $\kappa^D$ at the scale $\Lambda_H$ are no longer diagonal, so
we must rediagonalize $\kappa^U$ and $\kappa^D$ in analogy with \eq{kappas} [while respectively transforming
$\rho^U$ and $\rho^D$ (at the scale $\Lambda_H$) in analogy with \eq{rhos}].  We can now evolve 
$\kappa^U$ and $\kappa^D$ down to the electroweak scale to check the accuracy of the resulting quark masses.
If any of the quark masses differ from their experimental values by more than 3\%, we reestablish the correct quark masses at the electroweak scale,\footnote{Starting the RG evolution at $m_Z$, we use a five flavor scheme to run up to $m_t$ and a six flavor scheme above $m_t$.  Running quark mass masses at $m_Z$ and $m_t$ are obtained from the
RunDec Mathematica software package~\cite{Chetyrkin:2000yt,Herren:2017osy}, based on quark masses provided in Ref.~\cite{Olive:2016xmw}.  We fix the initial value of the top Yukawa coupling $y_t(m_t)=0.94$, corresponding to an $\overline{\rm MS}$ top quark mass of $m_t(m_t)=163.64$~GeV~\cite{Marquard:2015qpa}. For simplicity, the effects of the lepton masses are ignored, as these contribute very little to the running.} run back up to $\Lambda_H$, and then rerun this procedure repeatedly until the two boundary conditions are satisfied. The result is flavor-alignment between $\kappa^Q(\Lambda)$ and $\rho^Q(\Lambda)$, and a set of $\rho^Q$ matrices at the electroweak scale that provide a source of FCNCs. 

\begin{figure}[t!]
\centering
\includegraphics[width=0.62\textwidth]{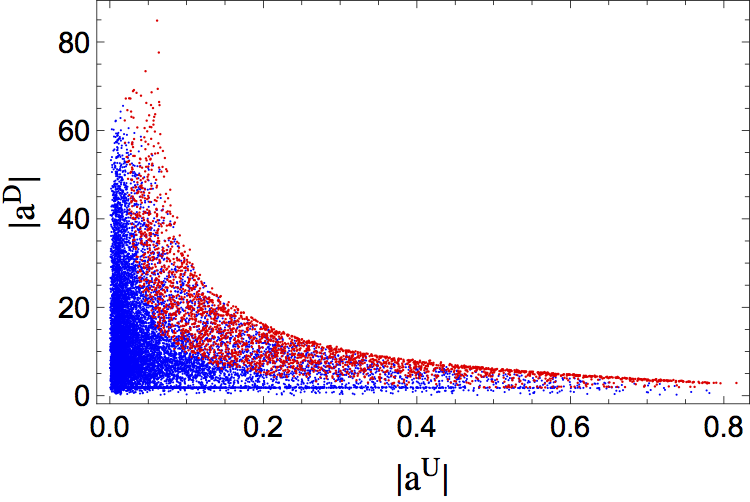}
\caption{The allowed values of $a^U$ and $a^D$ consistent with the absence of Landau poles below $\Lambda=M_{\rm P}$ are exhibited.
The blue points occupy the region of the A2HDM parameter space where the prediction for all entries of the $\rho^Q$ matrices lie within a factor of 3 from the results obtained with the full running. The red points occupy the region where the leading log approximation yields results quite different from the full RG running.} 
\label{fig:auabRd}
\end{figure}

In our iterative procedure, we demand that all scale-dependent Yukawa couplings remain finite from the  electroweak scale to the Planck scale (i.e., Landau poles are absent below 
$\Lambda=M_{\rm P}$).  This restricts the range of the possible seed values, $a^{\prime Q}$, used in \eq{aprime} to initialize the iteration.  Consequently, 
the alignment parameters $a^U$ and $a^D$ cannot be too large in absolute value.
Constraints on the alignment parameters due to Landau pole considerations during one-loop RG running have been given in Ref~ \cite{Bijnens:2011gd}. In our analysis, the allowed values of $a^U$ and $a^D$ consistent with the absence of Landau poles at all scales below $\Lambda$ are exhibited in 
Fig.~\ref{fig:auabRd}.\footnote{If a Landau pole in one of the Yukawa coupling matrices arises at the scale $\Lambda$, then both the corresponding $\rho^Q(\Lambda)$ and $\kappa^Q(\Lambda)$ diverge, whereas their ratio, $a^Q$, remains finite.}   
Assuming $\Lambda_H=400~{\rm{GeV}}$,
these considerations lead to bounds on the alignment parameters evaluated at the Planck scale, $\Lambda=M_{\rm P}$,
\beq\label{eq:Landau}
|a^U|\lesssim 0.8~~{\rm{and}}~~ |a^D|\lesssim 80\,,
\eeq
which are consistent with the results previously obtained in Ref~ \cite{Bijnens:2011gd}.

\subsection{Leading logarithm approximation}\label{Sec:LeadingLog}

In the limit of small alignment parameters, it is possible to obtain approximate analytic solutions to the one-loop RGEs provided in Appendix \ref{sec:RGEs}. 
One can express the $\rho^Q$ matrices at the low scale as
\begin{eqnarray}
\rho^U(\Lambda_H)&\simeq&a^U \kappa^U(\Lambda_H)+\frac{1}{16\pi^2}\log\left(\frac{\Lambda_H}{\Lambda}\right)(\mathcal D\rho^U-a^U\mathcal D\kappa^U),\\
\rho^D(\Lambda_H)&\simeq&a^D \kappa^D(\Lambda_H)+\frac{1}{16\pi^2}\log\left(\frac{\Lambda_H}{\Lambda}\right)(\mathcal D\rho^D-a^D\mathcal D\kappa^D),
\end{eqnarray}
where $D\kappa^D,~D\kappa^U,~D\rho^D,~D\rho^U$ are the $\beta$-functions defined in Eqs. (\ref{betaf})--(\ref{rge4}) and $\kappa^U(\Lambda_H)$ and $\kappa^D(\Lambda_H)$ are proportional to the diagonal quark mass matrices, $M_U$ and $M_D$ respectively, at the scale $\Lambda_H$, according to \eq{mumd}.  Working to one loop order and neglecting higher order terms, it is consistent
to set $\rho^F=a^F\kappa^F=a^F\sqrt{2}M_F/v$ (for $F=U,D,E$) in the corresponding $\beta$-functions,\footnote{The misalignment contributions exhibited in \eqs{eq:rhouLeading}{eq:rhodLeading} were computed for the first time in Ref.~\cite{Jung:2010ik}.}
\beqa
\rho^U(\Lambda_H)_{ij}&\simeq& a^U\delta_{ij}\frac{\sqrt 2(M_U)_{jj}}{v}+\frac{(M_U)_{jj}}{4\sqrt 2 \pi^2v^3}\log\left(\frac{\Lambda_H}{\Lambda}\right)\biggl\{(a^E-a^U)\bigl[1+a^U(a^E)^*\bigr] \delta_{ij}\Tr(M_E^2)
\nonumber \\
&&\qquad\,\,\, +(a^D-a^U)\bigl[1+a^U(a^D)^*\bigr]\biggl[3\delta_{ij} \Tr(M_D^2)-2\sum_k (M_D^2)_{kk} K_{ik}K^*_{jk}\biggr]\biggr\}\,,\label{eq:rhouLeading} \\
\rho^D(\Lambda_H)_{ij}&\simeq& a^D\delta_{ij}\frac{\sqrt 2(M_D)_{ii}}{v}+\frac{(M_D)_{ii}}{4\sqrt 2 \pi^2v^3}\log\left(\frac{\Lambda_H}{\Lambda}\right)\biggl\{(a^E-a^D)\bigl[1+a^D(a^E)^*\bigr] \delta_{ij}\Tr (M_E^2)
\nonumber \\
&&\qquad\,\,\, +(a^U-a^D)\bigl[1+a^D(a^U)^*\bigr] \biggl[3\delta_{ij} \Tr(M_U^2)-2\sum_k(M_U^2)_{kk} K^*_{ki} K_{kj}\biggr]\biggr\}\,.\label{eq:rhodLeading}
\eeqa
It follows that there is a large hierarchy among the several off-diagonal terms of the $\rho^Q$ matrices,
\beqa
\left|\frac{\rho^D(\Lambda_H)_{ij}}{\rho^D(\Lambda_H)_{ji}}\right| &\sim& \frac{(M_D)_{ii}}{(M_D)_{jj}}\ll 1,~~~\text{for $i<j$}\,,
\label{DD}\\
\left|\frac{\rho^U(\Lambda_H)_{ij}}{\rho^U(\Lambda_H)_{ji}}\right| &\sim& \frac{(M_U)_{jj}}{(M_U)_{ii}}\gg 1,~~~\text{for $i<j$}.\label{UU}
\eeqa 
The inequality given in \eq{DD} was previously noted in Ref.~\cite{Braeuninger:2010td},
and provides the justification for ignoring $\rho^{D}_{ij}$ relative to $\rho^{D}_{ji}$, for $i<j$.\footnote{To make contact with the Higgs basis Yukawa couplings $\Delta_u$ and $\Delta_d$ employed by Ref.~\cite{Braeuninger:2010td}, we note the relations $\rho^U=\sqrt{2}\Delta_u$ and
$\rho^D=\sqrt{2}\Delta_d^\dagger$.} 
This hierarchy of Yukawa couplings is reversed for~$\rho^U_{ij}$. This reversal can be traced back to the fact that $\rho^U$ is undaggered in \eq{lyuk} whereas $\rho^D$ is daggered.

It is noteworthy that the leading log results for the off-diagonal terms of the $\rho^Q$ matrices obtained in  \eqs{eq:rhouLeading}{eq:rhodLeading} and the corresponding full numerical calculation are typically within a factor of a few. Even for small alignment parameters, there can be some small discrepancies between the two approaches that can be traced back to the higher order terms that were neglected in \eqs{eq:rhouLeading}{eq:rhodLeading}. These higher order terms are not negligible due to the running performed between the electroweak scale and the high energy scale $\Lambda$. The leading log approximation describes less and less accurately the numerical results at larger and larger alignment parameters. This is shown in Fig.~\ref{fig:auabRd}, where the blue points correspond to the parameter regime in which the leading log approach leads to results within a factor of 3 of the results obtained numerically for all the elements of the $\rho^U$ and $\rho^D$ matrices. In contrast, the red points correspond to the parameter regime in which the leading log approximation leads to results quite different from what is obtained by the full running.

\subsection{A particular type of Minimal Flavor Violation}\label{sec:MFV}
In the quark sector of the A2HDM, only the two Yukawa coupling matrices $\kappa^U$ and $\kappa^D$ break the 
SU(3)$_Q\times$SU(3)$_U\times$SU(3)$_D$ global flavor symmetry of the electroweak Lagrangian involving quarks. For this reason, our model can be thought in terms of a specific realization of a Minimal Flavor Violating (MFV) 2HDM~\cite{Buras:2010mh}.  In particular, in a general 2HDM with MFV one can write the Yukawa Lagrangian as
\beq
-\mathscr{L}_{\rm{Y,MFV}}=\bar Q_L Y_u U_R H_1^\dagger+\bar Q_L Y_d^\dagger D_R H_1+\bar Q_L A_u U_R H_2^\dagger+\bar Q_L A_d^\dagger D_R H_2+{\rm{h.c.}},
\eeq
with $H_1,H_2$ the two Higgs doublets in the Higgs basis as defined in section \ref{sec:2hdm} and $Q_L,U_R,D_R$ flavor eigenstate quarks. In general, $A_u,A_d$ can be expressed by the infinite sum~\cite{D'Ambrosio:2002ex}
\begin{eqnarray}
A_u&=&\sum_{n_1,n_2,n_3} \epsilon^u_{n_1 n_2 n_3}(Y_d Y_d^\dagger)^{n_1}(Y_uY_u^\dagger)^{n_2}(Y_d Y_d^\dagger)^{n_3}Y_u,\\
A_d&=&\sum_{n_1,n_2,n_3} \epsilon^d_{n_1 n_2 n_3}(Y_d Y_d^\dagger)^{n_1}(Y_uY_u^\dagger)^{n_2}(Y_d Y_d^\dagger)^{n_3}Y_d,\end{eqnarray}
with generic $\mathcal O(1)$ complex coefficients $\epsilon^{u,d}_{ni}$.  In order to determine the coefficients $\epsilon^{u,d}_{ni}$ in the A2HDM, we rotate to the quark mass-eigenstate basis: $Y_u\to\kappa^U,A_u\to\rho^U,Y_d\to\kappa^D$, $A_d\to\rho^D$ and compare with the leading log expressions for $\rho^U$ and $\rho^D$ as reported in \eqs{eq:rhouLeading}{eq:rhodLeading}. We find
\begin{eqnarray}
\epsilon^u_{000}&=&a^U-\frac{1}{8\pi^2 v^2}\log\left(\frac{\Lambda_H}{\Lambda}\right)\biggl\{3(a^U-a^D)\bigl[1+a^U(a^D)^*\bigr]\Tr(M_D^2)\nonumber \\[6pt]
&&~~~~~~~~~~~~~~~~~~~~~~~~~~~~~~~~~~~~~~~~~~+(a^U-a^E)\bigl[1+a^U(a^E)^*\bigr]\Tr(M_E^2)\biggr\}\,,\label{eq:epsilonMFVd} \\[6pt]
\epsilon^u_{100}&=&\frac{1}{8\pi^2}\log\left(\frac{\Lambda_H}{\Lambda}\right)(a^U-a^D)\bigl[1+a^U(a^D)^*\bigr]\label{eq:epsilonMFVu}\,,\end{eqnarray}
and all the higher order coefficients equal to zero. The corresponding coefficients for the down sector are obtained from these expressions with the replacement $a^U\to a^D, a^D\to a^U,\kappa^D\to\kappa^U$.
As expected, the leading term in \eq{eq:epsilonMFVd} is given by the alignment parameter at the high scale $a^U$. This coefficient receives one loop corrections. The term in \eq{eq:epsilonMFVu} generates off diagonal terms in the matrix $\rho^U$ and is one loop suppressed.

\section{Predictions of the model for high energy processes}
\label{sec:3hdm}

For our numerical analysis, we use the procedure described in the previous section, taking the A2HDM to be in the decoupling limit, which ensures that the properties of the lightest Higgs boson, $h$, are approximately those of the observed (SM-like) Higgs boson.  As stated below \eq{mtwocp}, we assume that the Higgs scalar potential and the Higgs vacuum are CP-conserving.
In this case, the two heavier neutral scalars, $H$ and $A$, are CP-even and CP-odd mass-eigenstates, respectively. In the decoupling limit, these two scalars are roughly degenerate in mass, i.e., $m_H \approx m_A \approx \Lambda_H\gg m_h$.
The decoupling limit also enforces the condition $|\cos(\beta-\alpha)| \ll 1$, as noted below \eq{sc}.
In this paper, we shall choose a benchmark mass of $m_H=400$~GeV.   Noting that in the case of a SM-like Higgs boson, $m_h^2\simeq Z_1 v^2=(125~{\rm GeV})^2$, which implies that $Z_1\simeq 0.26$, we will furthermore assume that $|Z_6|$ and $Z_1$ are of similar size.  Indeed,  \eq{sc} yields $|\!\cos(\beta-\alpha)|\simeq 0.11$ for $|Z_6|=Z_1$.\footnote{For $m_H=400$~GeV, even a value as large as  $|Z_6|=1$, yields  $|\!\cos(\beta-\alpha)|=0.477$ [cf.~\eq{sc}],  which is (barely) consistent with the measured $WW$ and $ZZ$ couplings of the observed Higgs boson.
}
In particular, if
$\beta - \alpha = \pi/2 - x$, with $|x|\ll 1$, then values $x\neq 0$ imply deviations from SM behavior of the couplings of the 125 GeV Higgs boson to fermions and gauge bosons, as well as the appearance of flavor changing neutral Higgs couplings, the largest of which is the $hbs$ coupling. 

In our analysis, we allow for CP-violating effects to enter in two ways.  
First, CP-violating charged Higgs couplings to fermion pairs are generated via the appearance of the CKM matrix, $K$.  Second, we generically allow for the possibility of complex alignment parameters $a^U$ and $a^D$ at the high energy scale.  Via RG-running, CP-violating neutral Higgs couplings to fermion pairs will be generated.  However, this extra source of CP violation will lead to a loop-suppressed mixing of $H$ and $A$ that is difficult to observe due to the near mass degeneracy of these states in the decoupling limit 
under consideration here.

\subsection{The couplings of the SM-like Higgs boson}\label{sec:SMcoupling}

It is instructive to  examine the $h b\bar b$ coupling, which is the Yukawa coupling that is most affected by New Physics in our framework, and thus plays the leading role in constraining the parameter space.  Following the standard notation of the ATLAS and CMS Collaborations, we denote the coupling of $h$ to bottom quarks normalized to the SM prediction by $\kappa_b$.\footnote{Note that $\kappa_b$ should not be confused with the matrices $\kappa^F$ ($F=U,D,E$) defined in \eqs{lyuk}{mumd}.}
Due to the presence of a CP-violating contribution to the $hb\bar{b}$ coupling when $\Im\rho^D_{33}\neq 0$, both scalar and pseudoscalar contributions to the $hb\bar{b}$ coupling must be considered [see \eq{YUK3}].   In the approximation where $m_b\ll m_h$, 
one can simply replace $\gamma_5$ in the expression for the Yukawa coupling with $\pm 1$, in which case $\kappa_b$ can be expressed by the magnitude of the complex number,
\beq \label{kappab}
\kappa_b=\left|s_{\beta-\alpha}+\frac{v}{m_b\sqrt{2}}c_{\beta-\alpha} \rho^D_{33}\right|\,,
\eeq
and compared to its ATLAS and CMS measurement, extracted from the $h\to b\bar b$ rate.
 In the leading log approximation, \eq{eq:rhodLeading} yields,
\beqa
\kappa_b&=& \biggl|s_{\beta-\alpha}+ a^Dc_{\beta-\alpha}+\frac{1}{8\pi v^2}
\log\left(\frac{\Lambda_H}{\Lambda}\right)\biggl\{(a^E-a^D)\bigl[1+a^D(a^E)^*\bigr] \Tr (M_E^2)
\nonumber \\
&&\qquad\,\,\, +(a^U-a^D)\bigl[1+a^D(a^U)^*\bigr] \biggl[3 \Tr(M_U^2)-2\sum_k(M_U^2)_{kk} K^*_{k3} K_{k3}\biggr]\biggr\}\biggr|\,.
\eeqa

\begin{figure}[t!]
\centering
\includegraphics[width=0.47\textwidth]{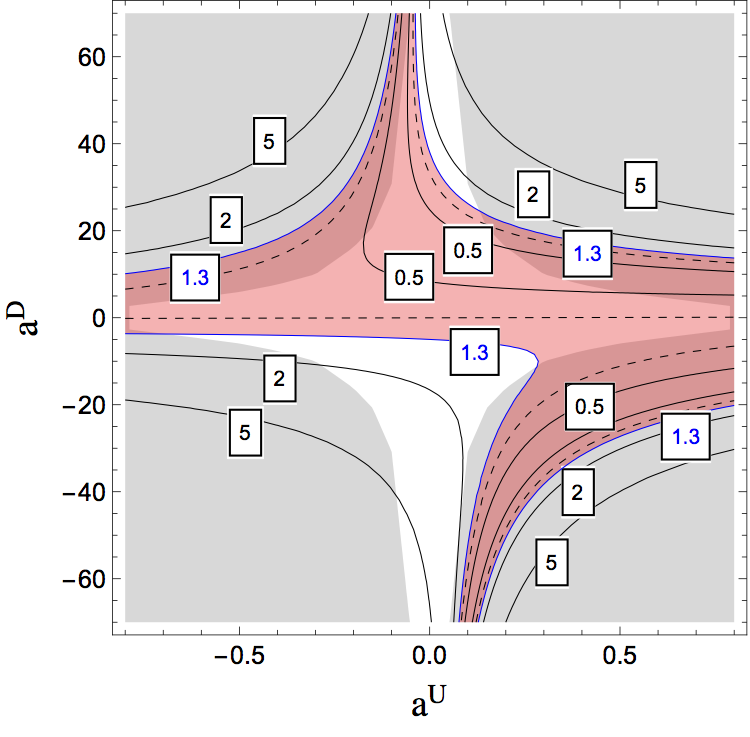}~~~~~
\raisebox{0.03in}{\includegraphics[width=0.47\textwidth]{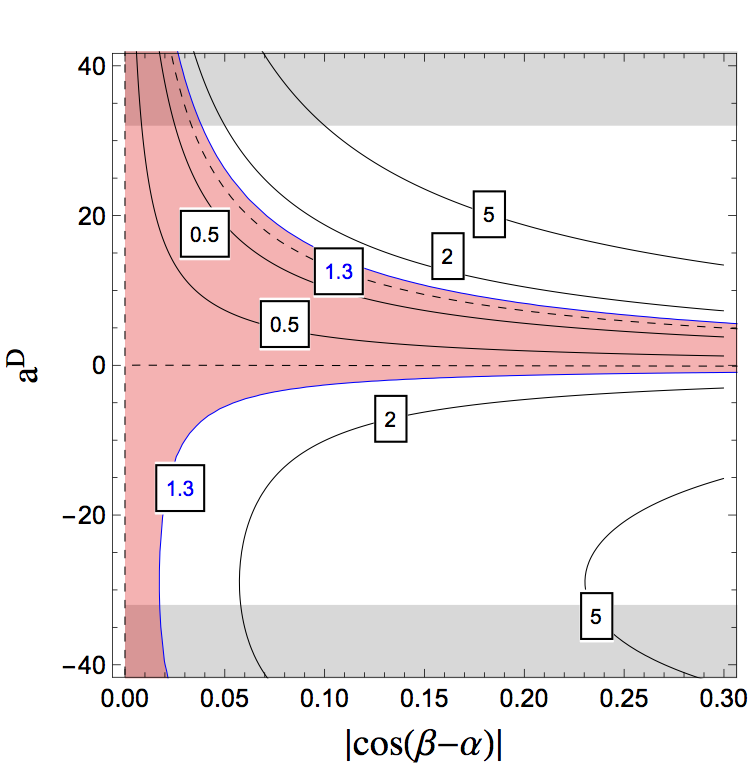}}
\vspace{0.3cm}
~~~~\raisebox{0.38in}{\includegraphics[width=0.5\textwidth]{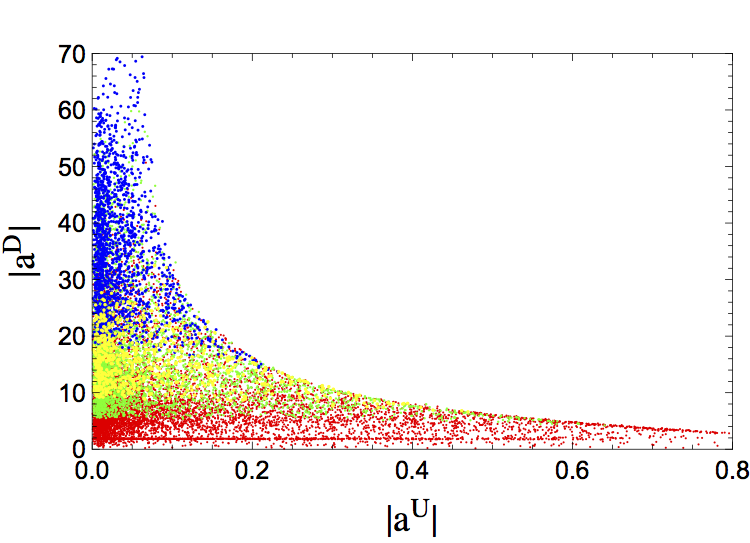}}~~~~~
\includegraphics[width=0.44\textwidth]{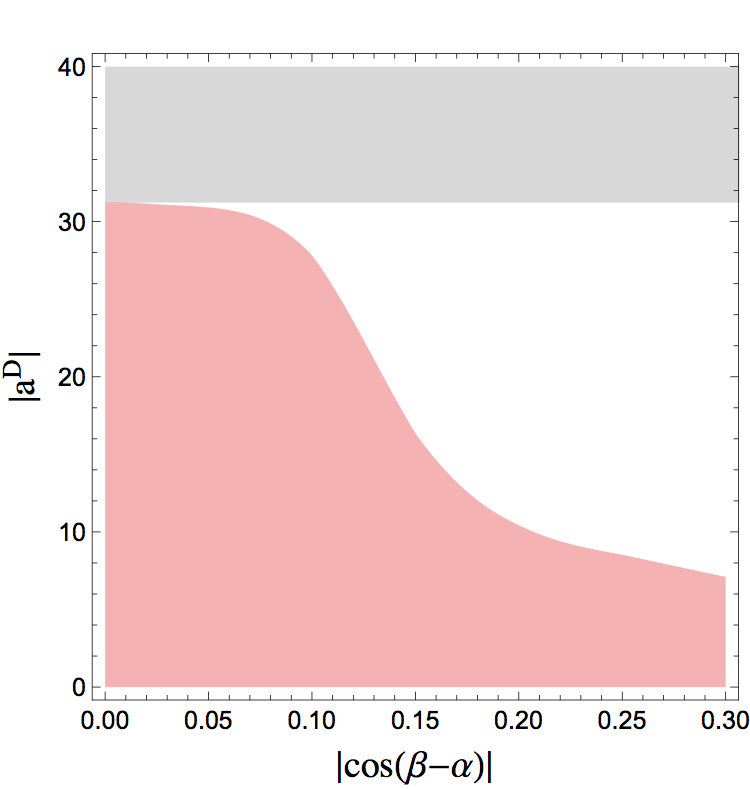}
\caption{Prediction for the SM-like Higgs coupling to bottom quarks, normalized to the SM prediction, $\kappa_b$, as a function of $a^U$ and $a^D$, having fixed $|\cos(\beta-\alpha)|=0.05$ (left panels) and as a function of $|\cos(\beta-\alpha)|$ and $a^D$, having fixed $a^U=0.1$ (right panels). The top panels exhibit the leading log predictions. The dotted line corresponds to the SM value, $\kappa_b=1$. The gray shaded regions produce Landau poles in the Yukawa couplings below $M_{\rm P}$. The pink shaded region is favored by the LHC measurements of $\kappa_b$. The bottom panels show the corresponding results obtained via scanning the parameter space and using the full RG running. In the bottom left panel, yellow, red, green and blue colors correspond to values of $\kappa_b$ in the ranges $<0.5,~[0.5,{\bf{1.3}}],~[{\bf{1.3}},2]$ and $~>2$, respectively. Here, the boldfaced number represents the $3\sigma$ experimental upper bound of $\kappa_b$. All points shown correspond to parameter regimes where Landau poles are absent [cf.~Fig.~\ref{fig:auabRd}]. In the bottom right panel, the gray shaded region produces Landau poles in the Yukawa couplings below $M_{\rm P}$; the pink shaded region contains points favored by the LHC measurements of $\kappa_b$.}
\label{fig:hbb}
\end{figure}

In Fig.~\ref{fig:hbb}, we show the reduced coupling, $\kappa_b$, in the leading log approximation as a function of the free parameters $a^U$, $a^D$ and $|\!\cos(\beta-\alpha)|$. We extend the plots up to $|\!\cos(\beta-\alpha)|\sim 0.3$, consistent with the present measurement of the Higgs couplings to $WW$ and $ZZ$. The two upper panels are obtained using the leading log approximation; the two lower ones using the full RG running. We take real values for $a^U$, $a^D$ to present the leading log results. Generic complex coefficients are employed in parameter scans obtained with the full running. In the left upper panel, we show the reduced coupling as a function of $a^U$ and $a^D$, having fixed $|\!\cos(\beta-\alpha)|=0.05$. In the right upper panel, we show the reduced coupling as a function of $|\!\cos(\beta-\alpha)|$ and $a^D$, having fixed $a^U=0.1$. In the two panels, we show in blue the contour $\kappa_b=1.3$, that roughly corresponds to the present $3\sigma$ bound, as measured by the LHC combining ATLAS and CMS Run I data \cite{Khachatryan:2016vau}.\footnote{Under the assumption of no decay modes of the Higgs boson beyond the SM and no non-SM particles in the loop, Ref~\cite{Khachatryan:2016vau} obtains $\kappa_b=0.67^{+0.22}_{-0.20}$.} The pink regions of Fig.~\ref{fig:hbb} illustrate that values for $|\cos(\beta-\alpha)|\sim\mathcal O(0.1)$ are still allowed for sizable values of $a^D$ of $\mathcal O(20)$. Furthermore, the shape of the constraint is quite different, if compared to the shape obtained for the $(\cos(\beta-\alpha)\,,\,\tan\beta)$ plane in the Type I and II 2HDM \cite{Aad:2015pla,CMS:2016qbe}. The corresponding results obtained using the full RG running are shown in the lower panels. Note that the bounds on the parameter spaces $(|a^U|\,,\,|a^D|)$ and $(|\cos(\beta-\alpha)|\,,\,|a^D|)$ are slightly weaker as compared to the leading log results. 

It is interesting to investigate the Higgs flavor violating couplings in the regions of parameter space favored by the LHC measurements of the SM Higgs rates. The decay to a bottom and a strange quarks is the dominant flavor violating Higgs decay in our model. However, we have checked that the corresponding branching ratio can be at most at the few per-mille level.
Numerically, this is similar to the result for BR($h\to \bar{b}s+b\bar{s})$
obtained by Ref.~\cite{Arhrib:2004xu} in a Type~I and Type II 2HDM due to charged Higgs loop contributions to the decay amplitude.

\subsection{Flavor-changing top decays}\label{sec:top}

We calculate the branching ratios for the decays $t \rightarrow u_ih$ ($u_i=u,c$) arising from misalignment generated via radiative corrections during RG running. This is in contrast to the analysis of Ref.~\cite{Abbas:2015cua} where flavor alignment is assumed to hold at the electroweak scale, in which case only charged Higgs loop diagrams contribute to the top flavor changing decays, 
leading to a  ${\rm BR}(t\to u_i h)$ that depends strongly on the value of the charged Higgs mass. {In this subsection, we show how the charged Higgs contributions compare to the those arising in our model due to tree-level flavor changing top couplings.

Following Ref. \cite{Greljo:2014dka}, we employ the leading order formulae for both $t \rightarrow W b$ and $t \rightarrow u_ih$ decay rates, assuming the top quark decay width is dominated by the SM value of $\Gamma(t\rightarrow W b)$.  In addition, we include the NLO QCD correction to the branching ratio,
\beq\label{eq:topFC}
{\rm BR}(t\rightarrow u_ih) = \cos^2(\beta-\alpha)(|\rho^U_{i3}|^2+|\rho^U_{3i}|^2) 
 \,\frac{v^2}{4m^2_t}\frac{(1-m^2_h/m^2_t)^2}{(1-m^2_W/m^2_t)^2(1 + 2 m^2_W/m_t^2 )}\, \eta_{QCD},
\eeq
where $\eta_{QCD} = 1 + 0.97 \alpha_s \sim 1.10$. 
The flavor violating branching ratios scale with the second power of $\cos(\beta-\alpha)$, and thus suppressed in the $\cos(\beta-\alpha)=0$ limit. The couplings $\rho^U_{i3}$ and $\rho^U_{3i}$ can be easily extracted in the leading logarithmic approximation from 
\eq{eq:rhouLeading}. Generically, the decay into a charm and a Higgs boson has a $\mathcal O(10^2)$ larger branching ratio than the decay into an up quark and a Higgs boson since in the leading logarithmic approximation,
\beq
\frac{{\rm BR}(t\rightarrow ch)}{{\rm BR}(t\rightarrow uh)}=\frac{|\rho^U_{23}|^2+|\rho^U_{32}|^2}{|\rho^U_{13}|^2+|\rho^U_{31}|^2}\sim \frac{|\rho^U_{23}|^2}{|\rho^U_{13}|^2}\sim\left|\frac{K_{cb}}{K_{ub}}\right|^2.
\eeq

\begin{figure}[t!]
\centering
\includegraphics[width=0.47\textwidth]{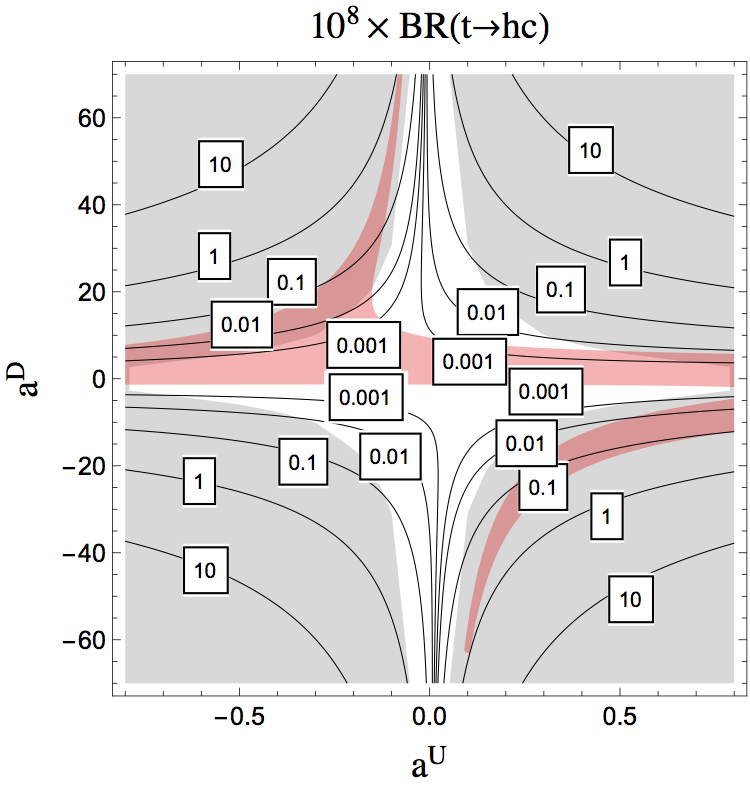}~~~~~
\includegraphics[width=0.47\textwidth]{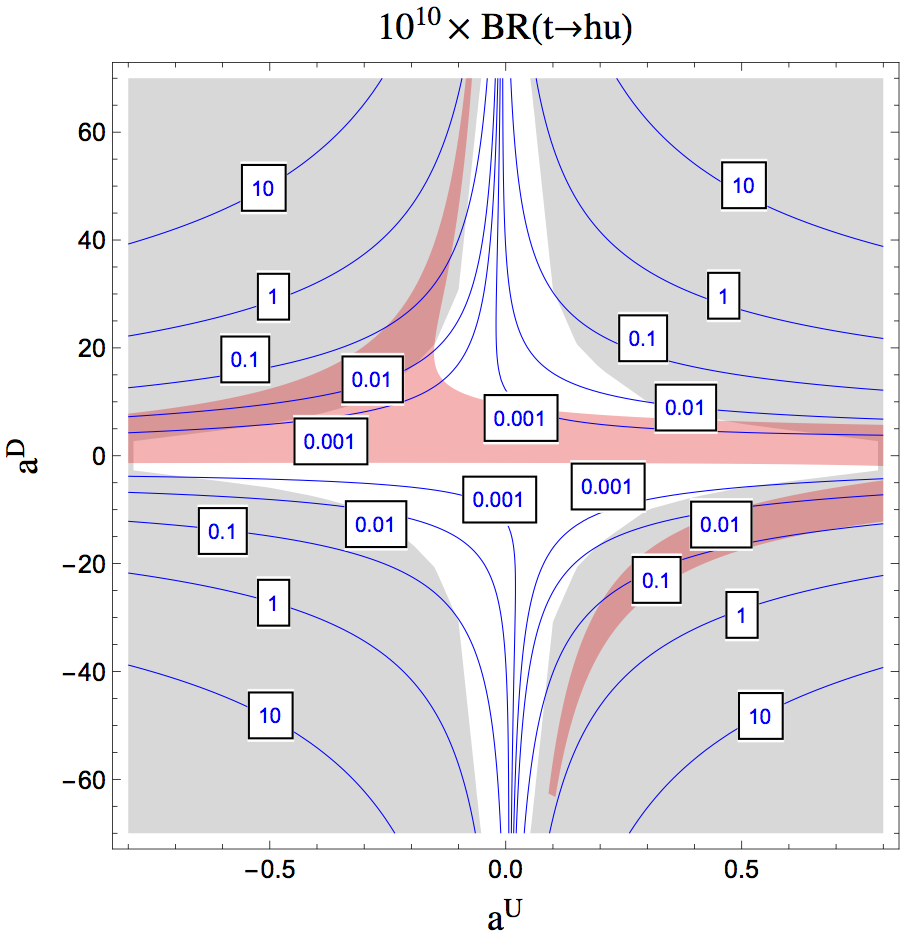}\\
~~~~~\includegraphics[width=0.47\textwidth]{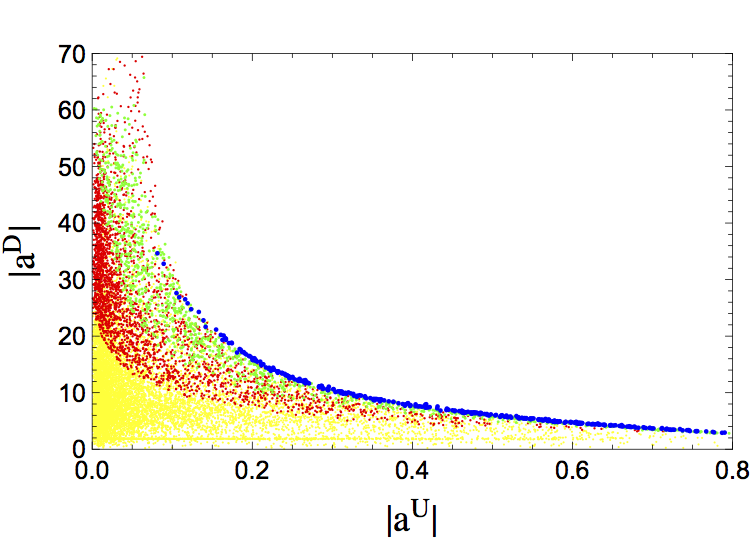} \includegraphics[width=0.47\textwidth]{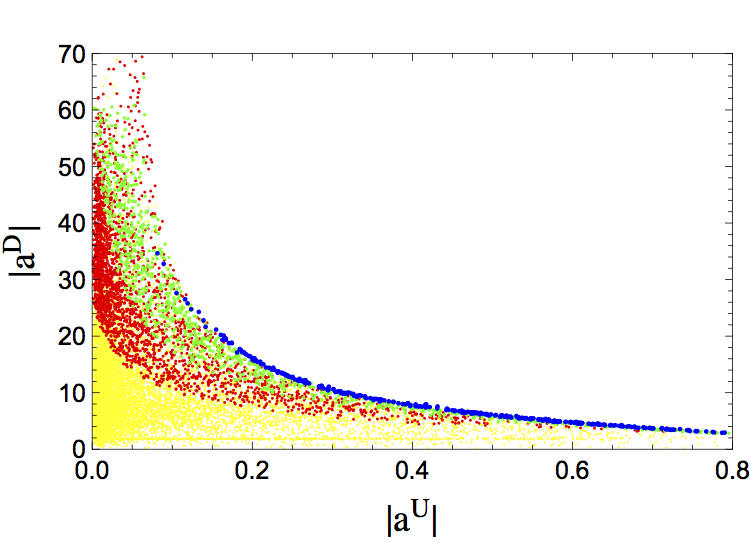}\
\caption{Tree-level contributions to top flavor changing decays assuming that $|\cos(\beta-\alpha)|=0.2$. Top panels: we use the leading log approximation to obtain $10^8\times$ BR($t \rightarrow ch)$ [left panel] and  $10^{10}\times$ BR($t \rightarrow uh)$ [right panel]. The gray shaded region produces Landau poles in the Yukawa couplings below $M_{\rm P}$. The pink region is favored by the LHC measurements of $\kappa_b$ (see section~\ref{sec:SMcoupling}). Bottom panels: we exhibit the corresponding results obtained via the full RG running. Yellow, red, green and blue colors correspond to branching ratios $<10^{-11},~[10^{-11}-10^{-10}],~[10^{-10}-10^{-8}],~>10^{-8}$ for $t\to c h$ [left panel], and to branching ratios $<10^{-13},~[10^{-13}-10^{-12}],~[10^{-12}-10^{-10}],~>10^{-10}$ for $t\to hu$ [right panel]. }
\label{fig:topdecay}
\end{figure}

In the top panel of Fig.~\ref{fig:topdecay}, we show the leading log results for the branching ratios, as a function of the two alignment parameters at the high scale, $a^U$ and $a^D$, having fixed $|\cos(\beta-\alpha)|=0.2$. Gray and pink shaded regions correspond to the region producing Landau poles in the Yukawa couplings below the Planck scale $M_{\rm P}$, and to the region favored by the LHC measurements of $\kappa_b$, respectively. Branching ratios larger than $\sim 10^{-10}$ for $ch$ and $\sim 10^{-12}$ for $uh$ cannot be reached, while being consistent with Higgs coupling measurements and with the requirement of no Landau poles below~$M_{\rm P}$.
For comparison, we also show our results obtained scanning the parameter space and using the full RGEs} (see the bottom panels of 
Fig.~\ref{fig:topdecay}). Comparing the upper and lower panels of  Fig.~\ref{fig:topdecay}, we note that the agreement between the prediction at leading log and the full numerical results is less accurate at larger values of the alignment parameters $|a^U|$ and $|a^D|$, as expected.
Values of the branching ratios as large as $\sim 3\times 10^{-7}$ ($\sim 3\times10^{-9}$) for $t\to c h$ ($t\to u h$) can be reached, while satisfying the condition due to the absence of Landau poles (see the blue points). However, the majority of points with BR$(t \to c h)\gtrsim10^{-8}$ and BR$(t \to u h)\gtrsim10^{-10}$ also produces too large a deviation from SM behavior of the Higgs coupling to bottom quarks. We have checked that the largest branching ratio compatible with Higgs data is at around $10^{-7}$ for $t\to c h$ and at around $10^{-9}$ for $t\to u h$.

These numbers should be compared with the corresponding contributions to flavor-changing top decays from charged Higgs loop diagrams, which are present in all 2HDMs, and are generated by flavor-changing charged Higgs interactions induced by 
CKM mixing~\cite{Arhrib:2005nx,Abbas:2015cua}.  Based on the discussion of Ref.~\cite{Abbas:2015cua}, we see that in the case of light charged Higgs bosons ($m_{H^\pm}\lesssim 200$ GeV) and $hH^+H^-$ couplings as large as allowed by $h\gamma\gamma$ constraints, these latter contributions can be as large as $\mathcal O(10^{-8})$ and therefore comparable to those arising from the tree-level $h\bar{t}u_i$ coupling induced by RG-running in the A2HDM. 

When compared to the ${\rm BR}(t\rightarrow ch)_{\rm{SM}}\sim 3\times 10^{-15},~{\rm BR}(t\rightarrow uh)_{\rm{SM}}\sim 2\times 10^{-17}$, as calculated in the SM by Refs. \cite{Eilam:1990zc,Mele:1998ag,AguilarSaavedra:2004wm,Zhang:2013xya}, the 2HDM in general and the A2HDM in particular exhibit the possibility of a significant enhancement of the branching ratios for flavor-changing $t\to u_i h$ decays. However, both tree-level flavor changing effects and loop-level effects mediated by the charged Higgs boson are generically too small to be probed by the LHC and future colliders.

Searches for top flavor changing decays have been performed by the ATLAS and CMS collaboration using Run I data~\cite{Aad:2015pja,CMS:2015qhe,CMS:2015xqa}, and constrain the branching ratios to ${\rm BR}(t\rightarrow u_ih)\lesssim 0.42\%$ (see also \cite{Gori:2016lga} for a discussion of the most recent experimental results on top flavor changing decays). Projections for the HL-LHC show that the bounds on the branching ratios will likely be at the $10^{-4}$ level \cite{Agashe:2013hma,Selvaggi:2015sdf}.
Hence, it will be very challenging to probe our model at the LHC using top flavor changing decays. FCC estimations show that branching ratios as small as $\sim 10^{-7}$ could be probed with 10 ab$^{-1}$ luminosity \cite{Mangano:2016jyj}. From these numbers, we can conclude that Higgs coupling measurements typically give (and will give) a better probe of our model, since the region of parameter space predicting more sizable top flavor violating branching ratios, also predict large and measurable effects in the Higgs coupling to bottom quarks.

\subsection{Phenomenology of the heavy Higgs bosons}\label{sec:heavyH}

As pointed out in Ref.~\cite{Altmannshofer:2012ar}, the 2HDM with flavor alignment imposed at the electroweak scale predicts a rich and novel phenomenology for the heavy Higgs bosons that is strikingly different than that of the 2HDM with Type I, II, X or Y Higgs-fermion Yukawa couplings. The phenomenology is even more diverse if flavor alignment is imposed at the high scale. For example, the heavy Higgs decay to quarks is flavor non-universal (i.e., the ratios, $y_{Hd_id_i}/m_{d_i}$ and $y_{Hu_iu_i}/m_{u_i}$
are no longer independent of the flavor $i$).   Moreover, flavor changing heavy Higgs decays, which are generated at the loop-level due to the quark flavor-changing charged Higgs interactions~\cite{Arhrib:2004xu,Arhrib:2005nx}, receive an additional contribution from tree-level flavor-changing neutral Higgs interactions. 
In contrast to the flavor-changing top decays discussed in the previous section, these features are not suppressed in the limit of $\cos(\beta-\alpha)=0$, where the couplings of $h$ coincide with those of the SM Higgs boson.
This is exhibited by
the tree-level partial widths of the heavy Higgs bosons to up and down quarks, which are given by
\begin{eqnarray}\nonumber
\Gamma(H\rightarrow \bar f_i f_i)&=&\frac{3G_F}{4\sqrt{2}\pi}m_Hm_{f_i}^2\left[{\rm{Re}}\left(c_{\beta-\alpha}-\epsilon_6s_{\beta-\alpha}\frac{\rho_f^{ii}}{\kappa_f^{ii}}\right)^2\left(1-\frac{4m_{f_i}^2}{m_H^2}\right)^{3/2}\right.\\\label{eq:Hdiag}
&+&\left.{\rm{Im}}\left(c_{\beta-\alpha}-\epsilon_6s_{\beta-\alpha}\frac{\rho_f^{ii}}{\kappa_f^{ii}}\right)^2\left(1-\frac{4m_{f_i}^2}{m_H^2}\right)^{1/2}\right ],
\\\nonumber
\Gamma(H\rightarrow \bar f_i f_j)&=&\Gamma(H\rightarrow \bar f_j f_i)=\frac{3G_F}{8\sqrt{2}\pi}m_H v^2\times s_{\alpha-\beta}^2(|\rho_f^{ij}|^2+|\rho_f^{ji}|^2)\\
&\times&\left[1-\left(\frac{m_{f_i}-m_{f_j}}{m_H}\right)^2\right]\left[\left(1-\frac{m_{f_i}^2+m_{f_j}^2}{m_H^2}\right)^2-\frac{4m_{f_i}^2m_{f_j}^2}{m_H^4}\right]^{1/2}(i\neq j).
\end{eqnarray}
Henceforth, 
we shall set $\cos(\beta-\alpha)=0$, which
automatically avoids constraints from the measured Higgs boson couplings. In the leading log approximation with real values of $a^D$ and $a^U$ assumed,  
the second term of Eq. (\ref{eq:Hdiag}) can be neglected since Im$(\rho_f^{ii})= 0$ [cf.~\eqs{eq:rhouLeading}{eq:rhodLeading}]. 

\begin{figure}[t!]
\centering
\includegraphics[width=0.45\textwidth]{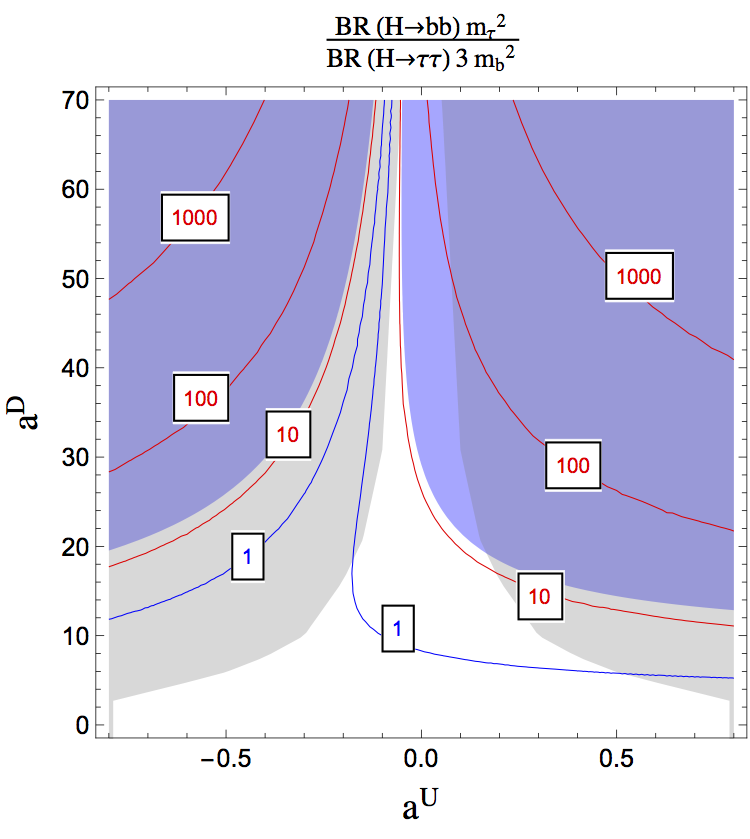}~~~~~
\includegraphics[width=0.45\textwidth]{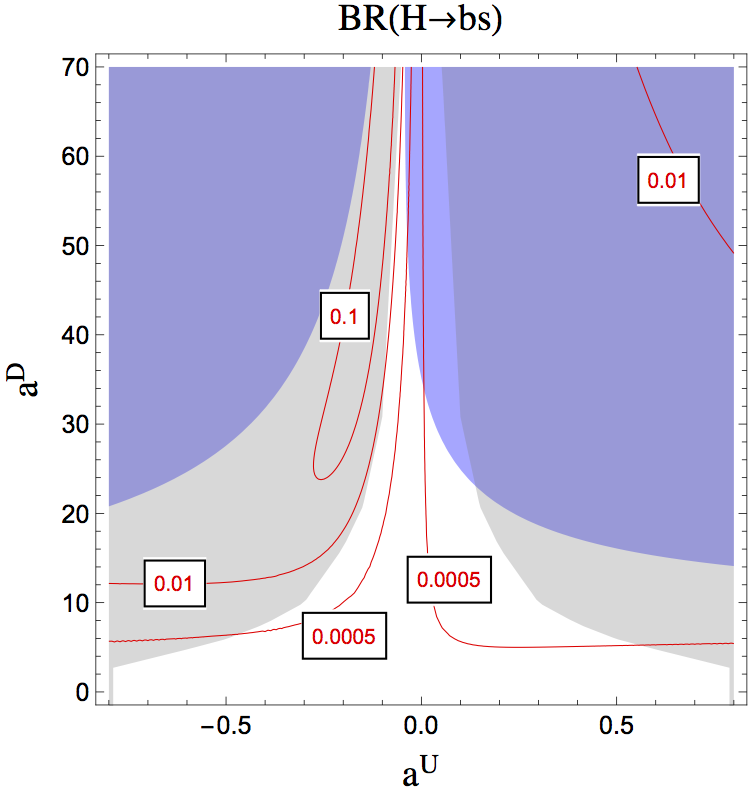}
\vspace{0.3cm}
\includegraphics[width=0.45\textwidth]{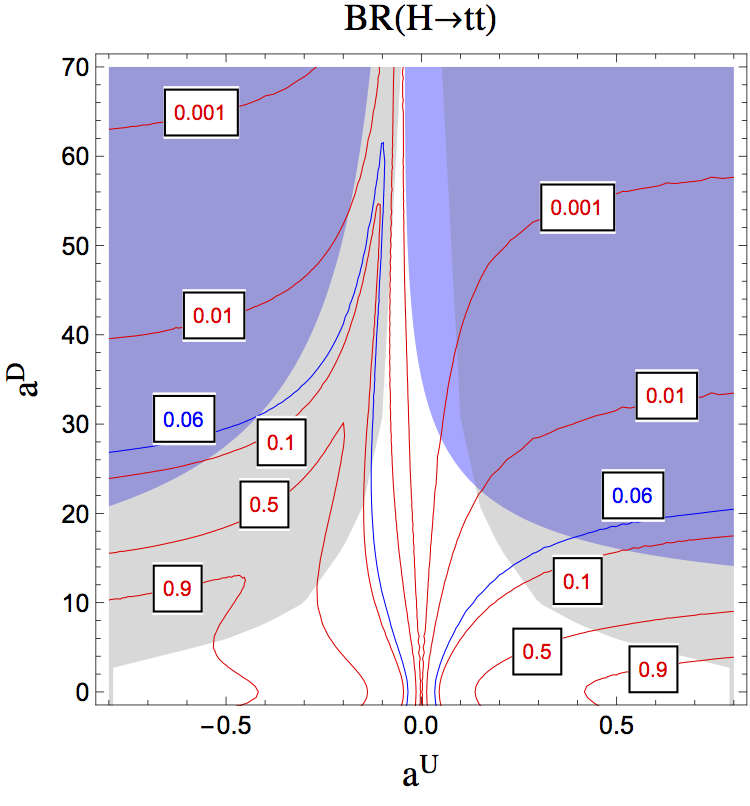}~~~~~
\includegraphics[width=0.45\textwidth]{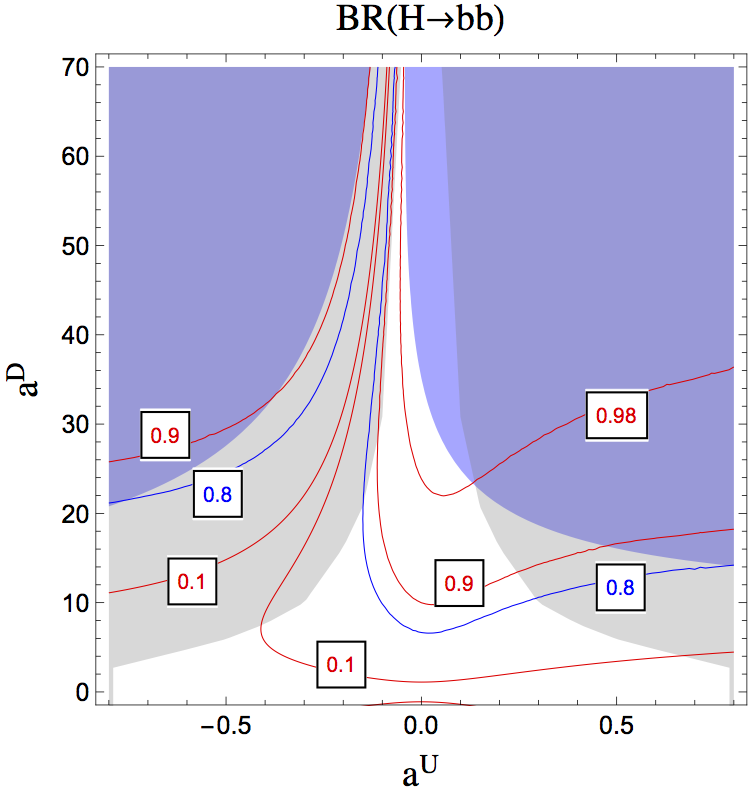}
\vskip -0.2in
\caption{Leading log prediction for the branching ratios of the heavy Higgs boson, $H$, with fixed $\tan\beta=10$, $\cos(\beta-\alpha)=0$, and $m_H=400$ GeV. The blue contours in the upper left and lower panels represent the prediction of a Type II 2HDM. The gray shaded regions produce Landau poles below the Planck scale $\Lambda=M_{\rm P}$. The blue shaded regions have  already been probed by the LHC searches for heavy scalars.}
\label{fig:HeavyLeadingLog}
\end{figure}

In Fig.~\ref{fig:HeavyLeadingLog}, we show the leading log predictions for the most interesting branching ratios ($\bar bb$, $\bar tt$, $\tau^+\tau^-$, $\bar bs+\bar s b$) as a function of the two alignment parameters $a^U$ and $a^D$, where we have fixed $\tan\beta=10$ and $m_H=400$ GeV. In the two panels, we only show positive values of $a^D$, since the results are symmetric under $(a^D,a^U)\leftrightarrow(-a^D,-a^U)$. 
For the predictions of ${\rm BR}(H\to \bar bs+\bar s b)$, we do not include loop contributions involving the charged Higgs boson.  These latter contributions have been examined in Refs.~\cite{Arhrib:2004xu,Arhrib:2005nx} and have been shown generically to be considerably smaller than the corresponding tree-level flavor violating Higgs couplings.
The left upper panel shows that in our model, especially at sizable values of the alignment parameters, the Type I and II 2HDM relation, BR$(H\to \bar bb)/{\rm{BR}}(H\to \tau^+\tau^-)=3m_b^2/m_\tau^2$, is violated. In particular, our model typically predicts a smaller ratio at small values of $a^D$, and therefore the $\tau^+\tau^-$ mode is expected to be even more sensitive than $b\bar b$ relative to that of the Type I or II 2HDM. For $a^D\gtrsim 5$, the hierarchy is reversed, resulting in a larger BR$(H\to \bar bb)$ as compared to BR$(H\to \tau^+\tau^-)$.
Furthermore, the model can predict a non zero decay rate of the heavy Higgs to a bottom and a strange quark (see the right upper panel of Fig.~\ref{fig:HeavyLeadingLog}).  However, the branching ratio predicted in the leading log approximation is at most of order a few percent at large values of $a^D$ in the regions of the parameter space without Landau poles.  

Note that the branching ratios into third generation quarks are different as compared to the Type II 2HDM.  In the latter, BR$(H\to b\bar b)\sim 80\%$ and BR$(H\to t\bar t)\sim 6\%$, for $\tan\beta=10$. For comparison, we present the branching ratios into $t\bar t$ and $b\bar b$ in the lower left and lower right panels of Fig.~\ref{fig:HeavyLeadingLog}. The behavior of the two plots is similar: at small values of $a^U$ ($a^D$) the $t\bar t$ ($b\bar b$) branching ratio is smaller than the one predicted by the Type II 2HDM (see the blue contours in the two plots); the branching ratio can even vanish for particular choices of the alignment parameters $a^U$ and $a^D$. Larger values of the branching ratio are predicted for sizable values $a^U\gtrsim 0.035$ ($a^D\gtrsim 7$). As a byproduct, the ratio of branching ratios ${\rm{BR}}(H\to b\bar b)/{\rm{BR}}(H\to t\bar t)$ differs from the predicted value of the 2HDM with either Type I, II, III, or IV Yukawa couplings. In particular, the A2HDM generically breaks the relation ${\rm{BR}}(H\to b\bar b)/{\rm{BR}}(H\to t\bar t)\simeq m_b^2\tan^4\beta /m_t^2$, which is valid in the Type II 2HDM in the limit $\cos(\beta-\alpha)=0$.  The branching ratios of a Type II 2HDM are recovered by choosing $a^U=\pm1/\tan\beta=\pm1/10$ and $a^D=\mp\tan\beta=\mp10$, as discussed at the end of section \ref{Sec:neutralCP}.

\begin{figure}[t!]
\centering
\includegraphics[width=0.45\textwidth]{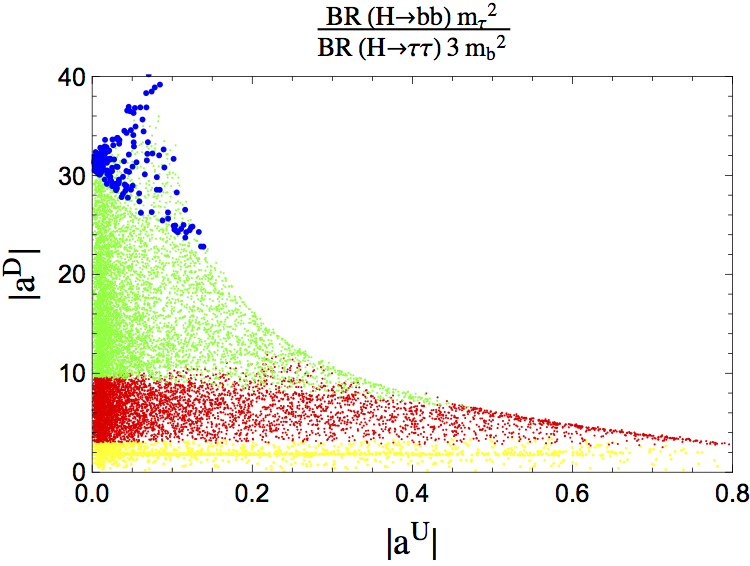}~~~~~
\includegraphics[width=0.45\textwidth]{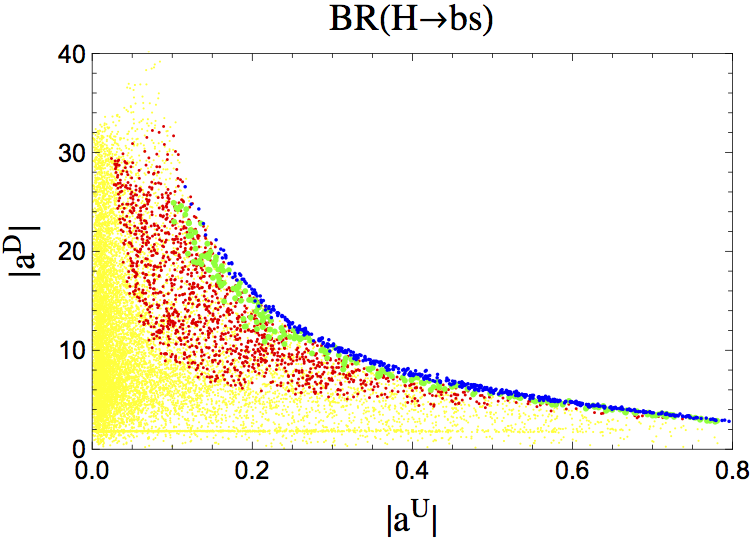}\\
\vspace{0.3cm}
\includegraphics[width=0.45\textwidth]{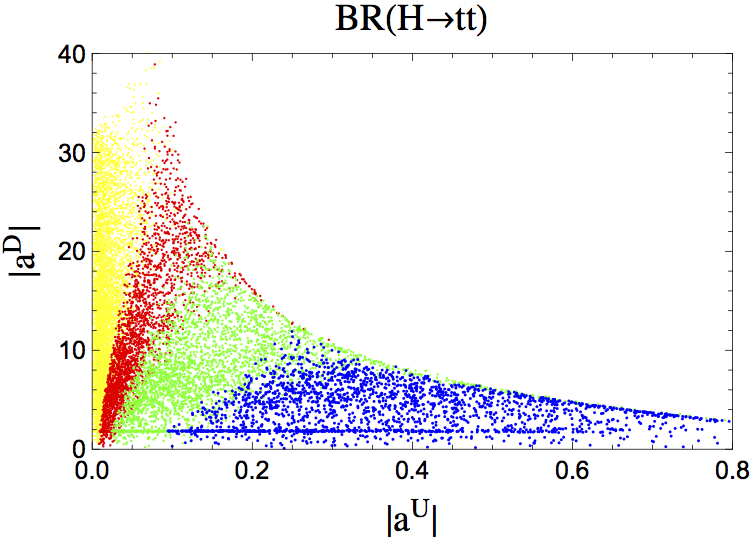}~~~~~
\includegraphics[width=0.45\textwidth]{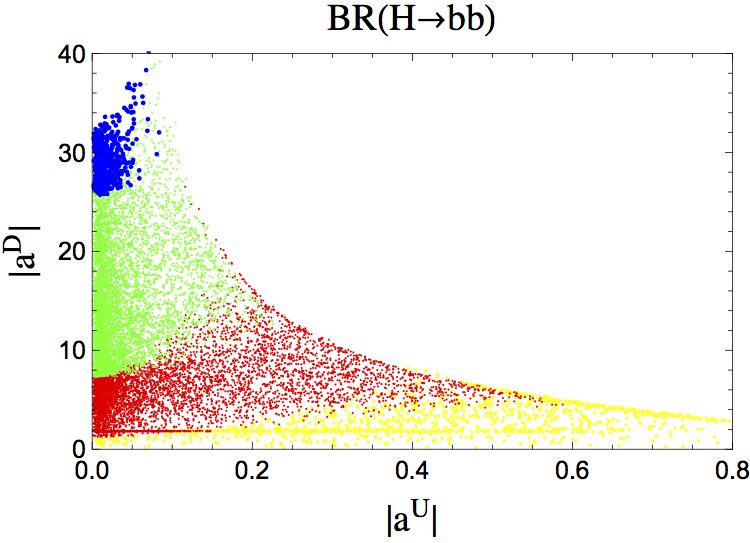}\\
\caption{Branching ratios of the heavy Higgs boson, $H$, obtained by scanning the parameter space and using the full RG running, with fixed $\cos(\beta-\alpha)=0$, $\tan\beta=10$, and $m_H=400$ GeV. The yellow, red, green and blue points correspond to: upper left panel, BR$(H\to \bar bb)m_\tau^2/{\rm{BR}}(H\to \tau^+\tau^-)3m_b^2<{\bf{1}},[{\bf{1}},10],[10,100],>100$; upper right panel, BR$(H\to\bar b s+b\bar s)<0.0005,[0.0005,0.01],[0.01,0.1],>0.1$; lower left panel, BR$(H\to\bar tt)<0.01,[0.01,{\bf{0.06}}],[{\bf{0.06}},0.5],>0.5$; and lower right panel, BR$(H\to\bar bb)<0.1,[0.1,{\bf{0.8}}],[{\bf{0.8}},0.98], >0.98$. In boldface we denote the value of the branching ratios predicted by a Type~II 2HDM with fixed $\tan\beta=10$.  The parameter regime with $|a_D|\gsim 30$--$40$ and $|a_U|\lsim 0.1$ has been eliminated after taking into account the LHC search for heavy Higgs bosons decaying into $\bar{b}b$~\cite{Khachatryan:2015tra}.\label{fig:HeavyHBRScan}}
\end{figure}

The ATLAS and CMS collaborations have performed several searches for heavy Higgs bosons decaying into a fermion pair: $\bar b b$~\cite{Khachatryan:2015tra}, $ \tau^+\tau^-$ \cite{ATLAS:2016fpj,CMS:2016pkt}, $\mu^+\mu^-$ \cite{CMS:2016abv,ATLAS:2016cyf}, and $\bar tt$ \cite{ATLAS:2016pyq}. In a Type II 2HDM, $\tau^+\tau^-$ searches are the most important ones in constraining regions of parameter space at sizable values of $\tan\beta$. Searches for $\bar b b$ can only set weaker bounds in that scenario. However, as discussed e.g. in Ref.~\cite{Carena:2012rw}, 2HDMs with a Yukawa texture different from Type II can be best probed by $\bar bb$ searches. In fact, for $\tan\beta=10$ and $\cos(\beta-\alpha)=0$, only the CMS search for $pp\to b(b)H, H\to\bar bb$, performed with 8 TeV data~\cite{Khachatryan:2015tra}, can probe sizable regions of the parameter space of the A2HDM (see the blue shaded region in Fig.~\ref{fig:HeavyLeadingLog} at large values of $a^D$ and the corresponding parameter regime of Fig.~\ref{fig:HeavyHBRScan}). In the coming years, the LHC will be able to probe complementary regions of parameter space. In addition to the region at large values of $a^D$ best probed by $\bar{b}b$ resonance searches, the region at small values of $a^U$ and $a^D$ will be best probed by searches for $\tau^+\tau^-$ and $\mu^+\mu^-$ resonances; and the region at small values of $a^D$, but sizable values of $a^U$ will be best probed by $\bar{t}t$ resonance searches. 

For comparison, we show in Fig.~\ref{fig:HeavyHBRScan} the corresponding results obtained through the scanning of the parameter space and the running of the full RGEs. Qualitatively, Fig.~\ref{fig:HeavyHBRScan}  shows a similar parameter dependence as the one obtained in the leading log approximation. Numerically, some branching ratios can be quite different, especially in the regime of sizable alignment parameters. In particular, BR$(H\to \bar{b}s+b\bar{s})$ can reach values as large as $\sim 10\%$.

\section{Predictions of the model for low energy processes}
\label{sec:4hdm}
        As we discussed in section \ref{sec:MFV}, the A2HDM is a particular type of 2HDM with Minimal Flavor Violation. As such, it predicts interesting effects in low energy flavor observables, e.g., in meson mixing and in $B$ meson rare decays.  In this section, we shall discuss the predictions of our model for these low energy processes and the corresponding constraints. We shall focus on those observables that receive tree-level Higgs contributions,     
with particular attention to meson mixing, $B\to\mu^+\mu^-$, and $B\to\tau\nu$. 

The lepton universality ratios, ${\rm{BR}}(\overline{B}\to D^{(*)}\tau^-\bar{\nu})/{\rm{BR}}(\overline{B}\to D^{(*)}\ell^-\overline{\nu})$, 
for $\ell=e$, $\mu$, are also notable, especially in light of the early BaBar measurements that yield a combined $3.4\sigma$ deviation from the SM predictions~\cite{Lees:2012xj,Lees:2013uzd}. This anomaly is not inconsistent with subsequent Belle and LHCb measurements, even if with a smaller significance \cite{Aaij:2015yra,Huschle:2015rga,Abdesselam:2016cgx,Abdesselam:2016xqt}. Additional data are required to clarify the implications of 
these measurements and to determine whether new physics beyond the SM is required. 
 If this anomaly persists, New Physics models  need (relatively large) $H^\pm c_L b_R$ and $H^\pm c_R b_L$ couplings of the same order and opposite sign (with $g_{H^\pm cb}^2/m_{H^\pm}^2\sim 1/{\rm{TeV}}^2$), as shown in Ref.~\cite{Freytsis:2015qca}. This is rather challenging to achieve in our model while being consistent with the other flavor bounds.  A more detailed examination of these channels will be left for a future study.
 
In principle, loop induced decays (which typically include contributions from the charged Higgs boson) can also set stringent constraints on the 
allowed regions of the $(m_{H^\pm}\,,\,\tan\beta)$ parameter plane~\cite{Mahmoudi:2009zx}.
For example, in the Type II 2HDM the charged Higgs should be heavier than 580 GeV~\cite{Misiak:2017bgg} to be in agreement with $b \to s\gamma$ measurements (cf.~footnote~\ref{fn2}). Moreover, going beyond the Type II 2HDM, the $b \to s\gamma$ bound depends not only on the charged Higgs mass, but also on the values of $a^U$ and $a^D$, on other non-SM-like Higgs boson masses, as well as on potential contributions of New Physics particles in the loop.   Such constraints merit further investigation.  However, the analysis of this section focuses on parameter regimes in which tree-level Higgs-mediated FCNC effects dominate over competing one-loop contributions.
For this reason, we do not consider further the constraints from $b \to s\gamma$ (which can be avoided for sufficiently heavy Higgs masses)  in this paper.

\subsection{Meson mixing}
Higgs mediated contributions to neutral meson mixing ($B_{d,s}$--$\overline{B}_{d,s}$,~ $K$--$\overline{K}$ and $D$--$\overline{D}$ mixing) arise in our model. Integrating
out the three neutral Higgs bosons, we obtain the following dimension six effective Lagrangian describing $B_s$ meson mixing
\beq
\mathscr{L}_{\rm{eff}}=C_2(\bar b_R s_L)^2+\tilde C_2(\bar b_L s_R)^2+ C_4(\bar b_R s_L)(\bar b_L s_R) +{\rm{h.c.}},\eeq
with Wilson coefficients, 
\begin{eqnarray}\label{eq:C2}
C_2 &=& \frac{(\rho^D_{32})^2}{4}\left(\frac{\sin^2(\beta-\alpha)}{m_H^2}+\frac{\cos^2(\beta-\alpha)}{m_h^2}-\frac{1}{m_A^2}\right),\\\label{eq:C2tilde}
\tilde C_2 &=& \frac{(\rho^{D*}_{23})^2}{4}\left(\frac{\sin^2(\beta-\alpha)}{m_H^2}+\frac{\cos^2(\beta-\alpha)}{m_h^2}-\frac{1}{m_A^2}\right),\\
C_4 &=& \frac{(\rho^D_{32})(\rho^{D*}_{23})}{2}\left(\frac{\sin^2(\beta-\alpha)}{m_H^2}+\frac{\cos^2(\beta-\alpha)}{m_h^2}+\frac{1}{m_A^2}\right),
\end{eqnarray}
and corresponding Wilson coefficients for $B_d$, $K$, and $D$ mixing. 

In the case of degenerate heavy Higgs bosons and in the limit $\cos(\beta-\alpha)=0$, only $C_4$ contributes to meson mixing. In this limit, we expect small Wilson coefficients at leading log, since as discussed in section \ref{Sec:LeadingLog}, $|(\rho^D)_{ij}|\propto m_i/v$ and therefore $|(\rho^D)_{ij}|\ll |(\rho^D)_{ji}|$, for $i<j$. The Wilson coefficients are also relatively small away from the exact $\cos(\beta-\alpha)=0$ and $m_A=m_H$ limit. More specifically, $C_2$ and $\tilde C_2$ will be non zero, but suppressed by the combination of masses and mixing angles shown in \eqs{eq:C2}{eq:C2tilde}, respectively. 
In the following, we will show the numerical results obtained for $\cos(\beta-\alpha)=0$ and $m_A=m_H=400$~GeV. However, we have checked that the constraints on the parameter space do not change considerably by taking small but non-zero values for $\cos(\beta-\alpha)$.

\begin{figure}[b!]
\centering
\includegraphics[width=0.4485\textwidth]{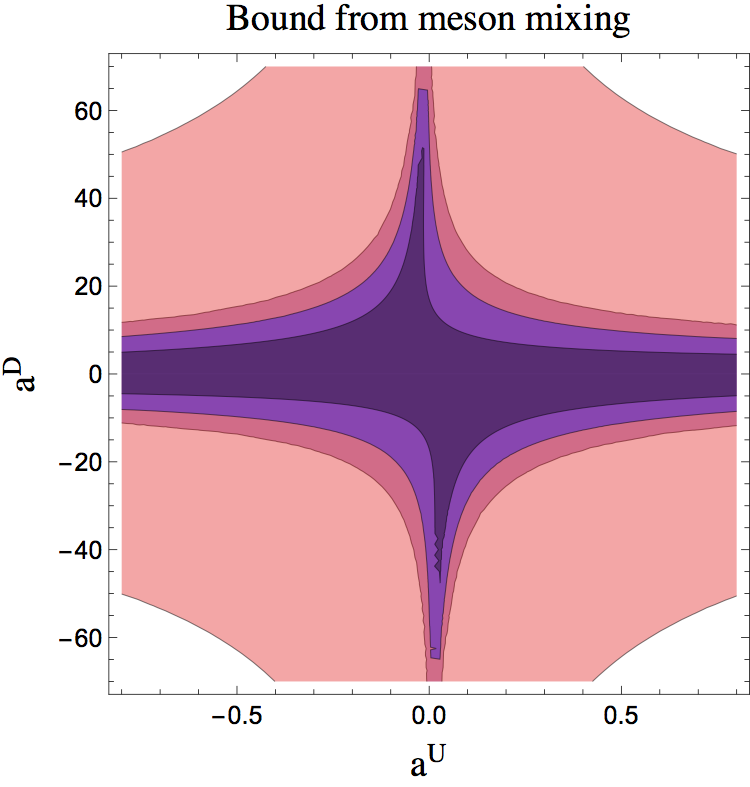}~~~~~
\raisebox{0.6cm}{\includegraphics[width=0.5\textwidth]{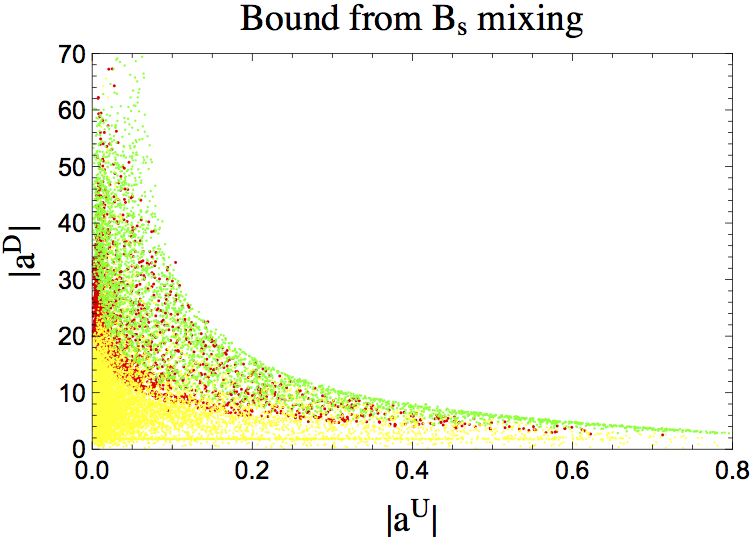}}
\caption{Bounds from meson mixing observables. Left panel: experimentally preferred regions, as computed in our model in the leading logarithmic approximation. The dark purple region is favored by the measurement of $B_s$ mixing, the purple region by $B_d$ mixing, and the dark pink (pink) region by the phase (mass difference) of the Kaon mixing system. $D$ meson mixing does not give any interesting bound on the parameter space and it is not shown in the figure. Right panel: the corresponding bounds from $B_s$ mixing obtained by scanning the parameter space and using the full RG running. The yellow, red, and green points correspond to a Wilson coefficient of $<1/3, [1/3,{\bf{1}}], >{\bf{1}}$ relative to the value that yields the present bound from $B_s$ mixing.} \label{fig:MesonMixing}
\end{figure}

We apply the bounds of Ref.~\cite{Butler:2013kdw} on the $C_4$ Wilson coefficient  (Ref. \cite{Bevan:2014cya} shows slightly stronger constraints). The leading log results for $B_s$, $B_d$, $K$, and $D$ mixing are shown in the left panel 
of~Fig.~\ref{fig:MesonMixing}. 
The dark purple region is favored by the measurement of $B_s$ mixing, the purple region by $B_d$ mixing, the dark pink region by CP violation 
in the Kaon mixing system, and the pink region by the $K$--$\overline{K}$ mass difference. $D$ mixing does not give any interesting bound on the parameter space and is therefore omitted in the figure. $B_s$ mixing leads to the most stringent bound and it constrains $a^D$ to be smaller than $\sim 4.7$ at sizable values of $a^U$. Additionally, the bound from the measurement of CP violation in Kaon mixing (dark pink) is significantly more stringent than the bound from the mass difference  of the Kaon system (in pink). This is due to the fact that
the real and imaginary parts of the Wilson coefficient of the Kaon system have a similar magnitude
 (under the assumption that $a^U$ and $a^D$ are real). In particular, the ratio of the imaginary and real parts of the Wilson coefficient is directly related to the phase of the CKM matrix: ${\rm{Im}}(C_4^K)/{\rm{Re}}(C_4^K) = {\rm{Im}}((K_{32}^*)^2K_{31}^2)/ {\rm{Re}}((K_{32}^*)^2K_{31}^2)$. In contrast, the SM Wilson coefficient has an imaginary part that is much smaller than the real part.
Small differences between the constraints from CP violation and the mass difference also exist in the $B_s$ and $B_d$ systems. In Fig.~\ref{fig:MesonMixing}, we only show the most constraining bound in each system, i.e.~the mass difference in $B_s$ mixing and the phase in $B_d$ mixing.

The right panel of Fig.~\ref{fig:MesonMixing} shows the corresponding results for the $B_s$ mixing system obtained by scanning the parameter space and using the full RG-running. The points in yellow have a Wilson coefficient smaller than 1/3 the present bound on the Wilson coefficient; in red we present the points with a Wilson coefficient smaller than the present bound, and finally in green we present the points that have been already probed by the measurement of the $B_s$ mixing observables. In the limit of sizable $a^U\gtrsim 0.7$, we do not find points with $a^D\gtrsim 4$, in rough agreement with the leading log result.

\subsection{\boldmath $B_{s,d}\to\mu^+\mu^-$ decays}

The $B$-meson rare decays $B_{s,d} \rightarrow \ell^+ \ell^-$ receive contributions from the exchange of the Higgs bosons $H$, $A$ and $h$ at tree-level. This is in contrast to the numerical analysis of Ref.~\cite{Li:2014fea}, where the flavor misalignment at the electroweak scale is set to zero.  The neutral Higgs exchange contributions to the  leptonic decay amplitude are proportional to $m_\ell$ and hence are largest in the case of $B_{s,d} \rightarrow \tau^+ \tau^-$. However, it is more difficult to tag the $\tau$ decay to jets and leptons 
at the LHC and B-factory detectors, as compared to muons. For this reason, the present LHCb bounds \cite{Martini:2016srx},  BR$(B_{s(d)} \rightarrow \tau^+ \tau^-)\lesssim 3\times 10^{-3}~(1.3\times 10^{-3})$, are relatively weak as compared to the SM prediction~\cite{Bobeth:2013uxa}, 
\beq
{\rm BR}(B_{s(d)} \rightarrow \tau^+ \tau^-)_{\rm{SM}}= (7.73\pm 0.49)\times 10^{-7}~\bigl((2.22\pm 0.19)\times 10^{-8}\bigr)\,.
\eeq
At sizable values of $\tan\beta$, the main contributions to $B_{s,d} \rightarrow \mu^+ \mu^-$ are typically due to $H$ and~$A$ exchange, as they are enhanced by the second power of $\tan\beta$. Furthermore, in the $\cos(\beta-\alpha)=0$ limit, the light Higgs ($h$) contribution vanishes at tree-level. For this reason, we shall focus henceforth on the heavy Higgs contributions that are given by~\cite{Altmannshofer:2012az},
\beq
\frac{{\rm BR}(B_{s,d} \rightarrow \mu^+ \mu^-)}{{\rm BR}(B_{s,d} \rightarrow \mu^+ \mu^-)_{\rm{SM}}} \simeq \big(|S_{s,d}|^2 + |P_{s,d}|^2\big) 
\bigg(1 + y_{s,d} \frac{{\rm Re}(P_{s,d}^2) - {\rm Re}(S_{s,d}^2)}{|S_{s,d}|^2 + |P_{s,d}|^2}\bigg)\bigg(\frac{1}{1+y_{s,d}}\bigg), \label{eq:bmumu}
\eeq
where BR($B_{s,d} \rightarrow \mu^+\mu^-)_{\rm{SM}}$ is the SM prediction for the branching ratio extracted from an untagged rate.
In particular, $y_s = (6.1 \pm 0.7)\%$ and $y_d \sim 0 $ have to be taken into account when comparing experimental and theoretical results, and
\beqa
S_{s,d} &\equiv& \frac{m_{B_{s,d}}}{2m_\mu}\frac{(C_{s,d}^S - C_{s,d}^{\prime S})}{C^{SM}_{10\,{s,d}}}\sqrt{1-\frac{4m^2_\mu}{m^2_{B_{s,d}}}}, \\
P_{s,d} &\equiv& \frac{m_{B_{s,d}}}{2m_\mu} \frac{(C_{s,d}^P - C_{s,d}^{\prime P})}{C^{SM}_{10\,{s,d}}} + \frac{(C_{s,d}^{10}-C^\prime_{10\,{s,d}})}{C^{SM}_{10\,{s,d}}}.
\eeqa

\noindent The $C_i$ are the Wilson coefficients corresponding to the Lagrangian
\beq
\mathscr{L}_s= \sum_i (C_i O_i+C_i^\prime O_i^\prime)+\rm{h.c.}\,,
\eeq
with operators
\beqa 
O_s^{(\prime)S} &=& \frac{m_b}{m_{B_s}}(\bar{s}P_{R(L)}b)(\bar{\ell}\ell), \\
O_s^{(\prime) P} &=& \frac{m_b}{m_{B_s}}(\bar{s}P_{R(L)}b)(\bar{\ell}\gamma^5\ell), \\
O^{(\prime)}_{10\,s} &=& (\bar{s}\gamma_\mu P_{L(R)}b)(\bar{\ell}\gamma^\mu\gamma^5\ell), 
\eeqa
and the corresponding ones for the $B_d$ system.  In the limit of $\cos(\beta-\alpha)=0$,
the Wilson coefficients arising from heavy neutral Higgs exchange are given by 
\beqa
C_s^P &=&  -\frac{m_{B_s}}{m_b}\frac{\rho^{D *}_{32}}{\sqrt 2}\frac{m_\mu}{v}\tan\beta\frac{1}{m_A^2}, ~~~~~~~~~~~
C_s^S = - \frac{m_{B_s}}{m_b}\frac{\rho^{D *}_{32}}{\sqrt 2}\frac{m_\mu}{v}\tan\beta\frac{1}{m_H^2}, \label{eq:CSP} \\
C_s^{\prime P} &=&  \frac{m_{B_s}}{m_b}\frac{\rho^{D}_{23}}{\sqrt 2}\frac{m_\mu}{v}\tan\beta\frac{1}{m_A^2}\ll C_s^P, ~~~
C_s^{\prime S} =- \frac{m_{B_s}}{m_b}\frac{\rho^{D}_{23}}{\sqrt 2}\frac{m_\mu}{v}\tan\beta\frac{1}{m_H^2}\ll C_s^S,\label{eq:CSPprime}
\eeqa
 and the analogous results for the $B_d$ system.   There are no tree-level New Physics contributions to the $\mathcal O_{10}^{(\prime)}$ operators.

If $\cos(\beta-\alpha)$ is nonvanishing, then the scalar Wilson coefficients $C_s^S$ and $C_s^{\prime S}$  given in \eqs{eq:CSP}{eq:CSPprime} due to $H$ exchange should be changed accordingly,  $\tan\beta\rightarrow 
 \sin(\beta-\alpha)\tan\beta+\cos(\beta-\alpha)$ and $\rho^{D}\to \rho^{D}\sin(\beta-\alpha)$.
 Moreover, an additional set of contributions arise due to $h$ exchange; the corresponding contributions are obtained from $C_s^S$ and $C_s^{\prime S}$  given in \eqs{eq:CSP}{eq:CSPprime} by making the following replacements,
$\tan\beta\to \sin(\beta-\alpha)-\cos(\beta-\alpha)\tan\beta$,  $\rho^{D}\to -\rho^{D}\cos(\beta-\alpha)$ 
and $m_H\to m_h$.

The SM Wilson coefficient takes the form~\cite{Altmannshofer:2008dz},
\beq
C^{\rm SM}_{10\,{s,d}}=-4.1\frac{e^2}{16\pi^2} \frac{4 G_F}{\sqrt 2}K_{tb}K_{t(s,d)}^*\,,
\eeq
and the predicted branching ratios are given by
\begin{eqnarray}
{\rm{BR}}(B_s \rightarrow \mu^+\mu^-)_{\rm SM} &= &(3.65 \pm 0.23)\times 10^{-9},\\
{\rm{BR}}(B_d \rightarrow \mu^+\mu^-)_{\rm SM} &=& (1.06 \pm 0.09)\times 10^{-10}, 
\end{eqnarray}
as obtained in \cite{Bobeth:2013uxa} with the inclusion of  $\mathcal O(\alpha_{em})$ and $\mathcal O(\alpha_{s}^2)$ corrections. 
These values are in relatively 
\clearpage
\noindent
good agreement with the experimental results. The combination of the LHCb and the CMS measurements at Run I for the $B_s$ and $B_d$ decays to muon pairs are~\cite{CMS:2014xfa}: 
\begin{eqnarray}
{\rm{BR}}(B_s \rightarrow \mu^+\mu^-) =  (2.8^{+ 0.7}_{-0.6})\times 10^{-9},\label{eq:BsmumuCombination} \\
{\rm{BR}}(B_d \rightarrow \mu^+\mu^-) =  (3.9^{+ 1.6}_{-1.4})\times 10^{-10}.\label{eq:BdmumuCombination}
\end{eqnarray}
Note the much larger uncertainty in the latter decay mode.

The ATLAS collaboration has also reported a Run I search for $B_s\to\mu^+\mu^-$, which yielded 
BR($B_s \rightarrow \mu^+\mu^-)=  (0.9^{+ 1.1}_{-0.8})\times 10^{-9}$ \cite{Aaboud:2016ire}, although this measurement is not yet competitive with \eq{eq:BsmumuCombination}.
Very recently, LHCb reported a new measurement for $B_{s,d}\to\mu^+\mu^-$ using Run~II~data~\cite{Aaij:2017vad}. Their result, BR$(B_s \rightarrow \mu^+\mu^-)=(2.8\pm 0.6)\times 10^{-9}$,
agrees very well with the LHCb and CMS combination quoted in \eq{eq:BsmumuCombination}.  In contrast, the new LHCb $B_d$ measurement is closer to the SM prediction, BR$(B_d \rightarrow \mu^+\mu^-)=(1.6^{+1.1}_{-0.9})\times 10^{-10}$. In the following, we will compare the predictions of the A2HDM with the LHCb and CMS combination shown in \eqs{eq:BsmumuCombination}{eq:BdmumuCombination}. In the coming years, the two branching ratios will be measured much more accurately by the LHC. In particular, the $B_s$ and $B_d$ branching fractions will be measured by each experiment with a precision of $\sim 13\%$ and~$\sim 48\%$ at Run-III,  
improving to $\sim 11\%$ and $\sim 18\%$, respectively, at the HL-LHC~\cite{CMS:upgrade}.

 \begin{figure}[t!]
\centering
\includegraphics[width=0.47\textwidth]{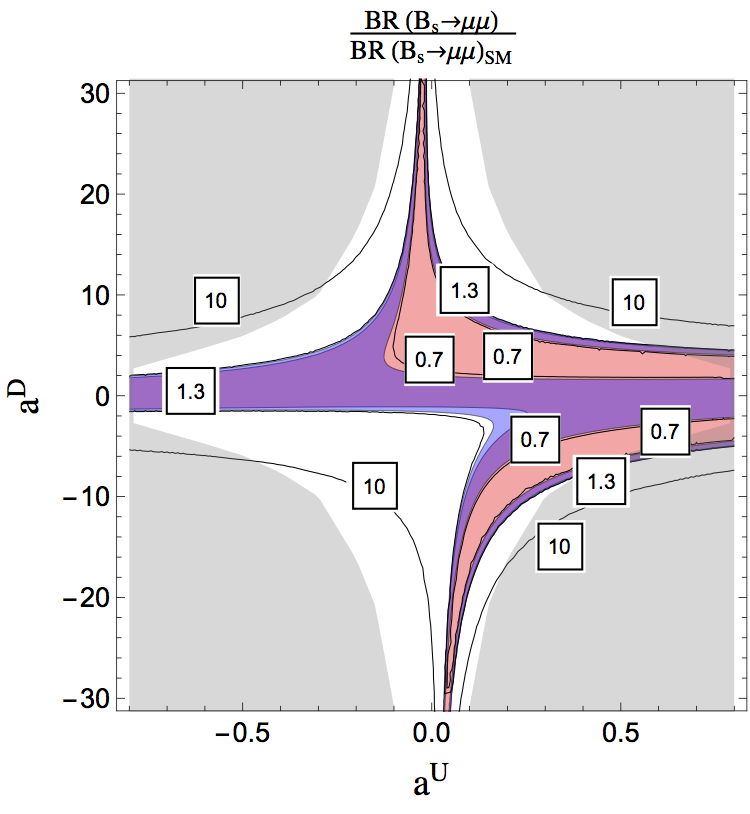}~~~~~
\includegraphics[width=0.47\textwidth]{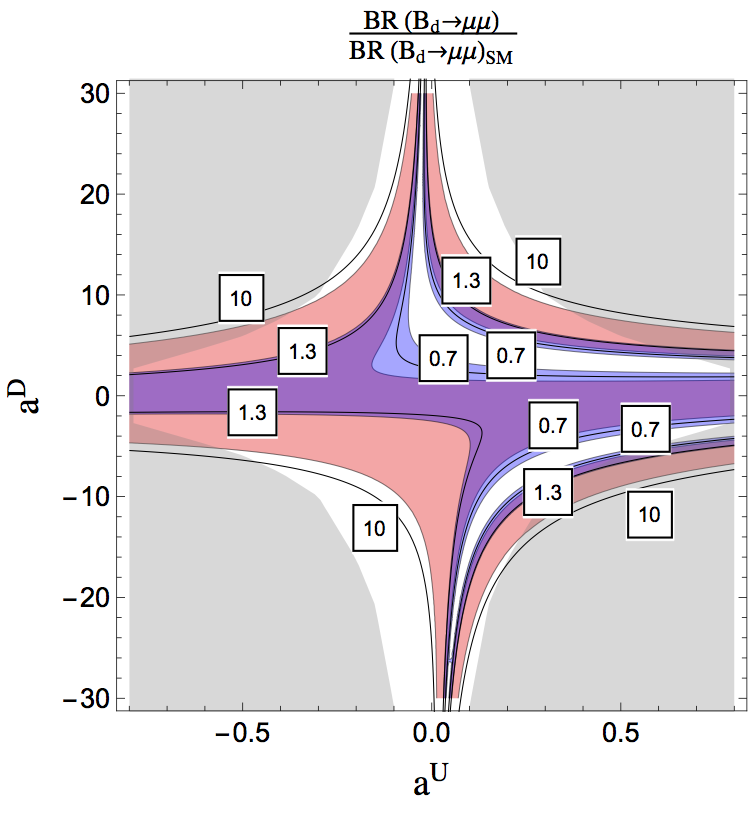}
\caption{Leading log prediction for the branching ratios for $B_s \rightarrow \mu^+\mu^-$ (left panel) and $B_d \rightarrow \mu^+\mu^-$ (right panel) relative the SM, as a function of $a^U$ and $a^D$, with fixed $\tan\beta=10$, $\cos(\beta-\alpha)=0$, and $m_A=m_H=400$ GeV. The regions in pink are allowed at the $2\sigma$ level by the present measurements. The purple shaded regions are anticipated by the more precise HL-LHC measurements, assuming a measured central value equal to the SM prediction.  The gray shaded regions produce Landau poles in the Yukawa couplings below~$M_{\rm P}$.}
\label{fig:BmumuAd}
\end{figure}

In Fig.~\ref{fig:BmumuAd}, we show the constraints
from the measurement of $B_s\to\mu^+\mu^-$ (left panel) and 
$B_d\to\mu^+\mu^-$ (right panel) as functions of $a^U$~and~$a^D$, with fixed $\tan\beta=10$, $\epsilon_E=+1$[see Eq. (\ref{leptonE})],\footnote{Fixing a different sign, $\epsilon_E=-1$, leads to the same results, with the exchange $(a^U,a^D)\rightarrow (-a^U,-a^D)$.}   
 $\cos(\beta-\alpha)=0$, and $m_A=m_H=400$ GeV, based on the leading logarithmic approximation.  The pink shaded region denote the parameter space favored by the CMS and LHCb combined results at the $2\sigma$ level, namely
\beq\label{eq:ranges}
\frac{{\rm{BR}}(B_s \rightarrow \mu^+\mu^-)}{{\rm{BR}}(B_s \rightarrow \mu^+\mu^-)_{SM} }\subset  [0.4, 1.1],~~~~ \frac{{\rm{BR}}(B_d \rightarrow \mu^+\mu^-)}{{\rm{BR}}(B_d \rightarrow \mu^+\mu^-)_{SM} }\subset [0.8, 6.6].
\eeq
The purple shaded region in Fig.~\ref{fig:BmumuAd} is the parameter space favored at $2\sigma$ by the HL-LHC measurement, assuming a measured central value equal to the SM prediction. Comparing the region in pink to the region in purple, one can get a sense of the improvement the HL-LHC can achieve in testing our model. The expected experimental error at the HL-LHC is comparable to the present theory error. For this reason, an additional improvement can be achieved via a more precise calculation of the SM prediction for the two branching ratios, with the benefit of more accurate measurements of the CKM elements that will be obtained at the LHCb and at Belle II in the coming years.

 \begin{figure}[t!]
\centering
\includegraphics[width=0.47\textwidth]{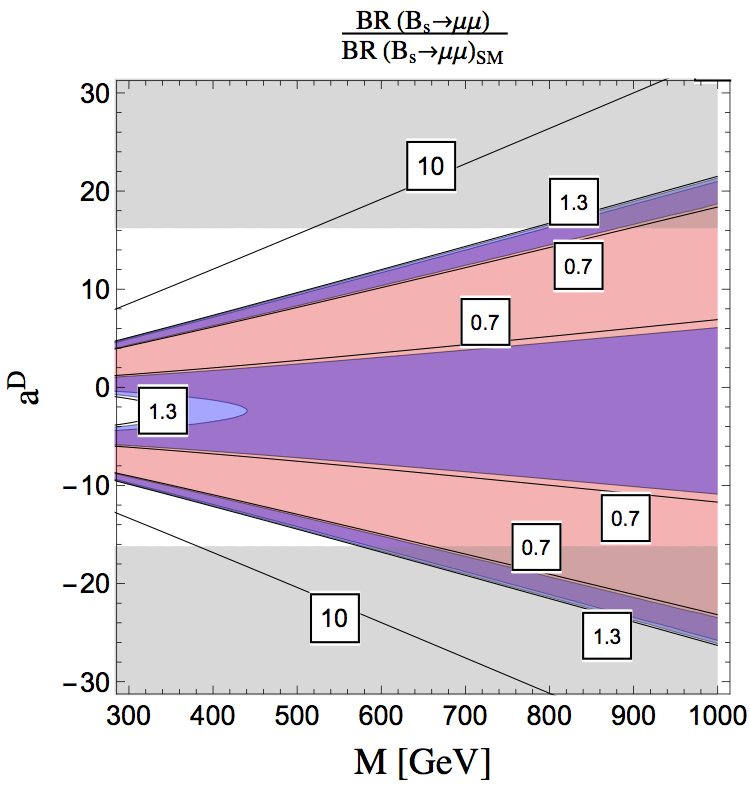}~~~~~
\includegraphics[width=0.47\textwidth]{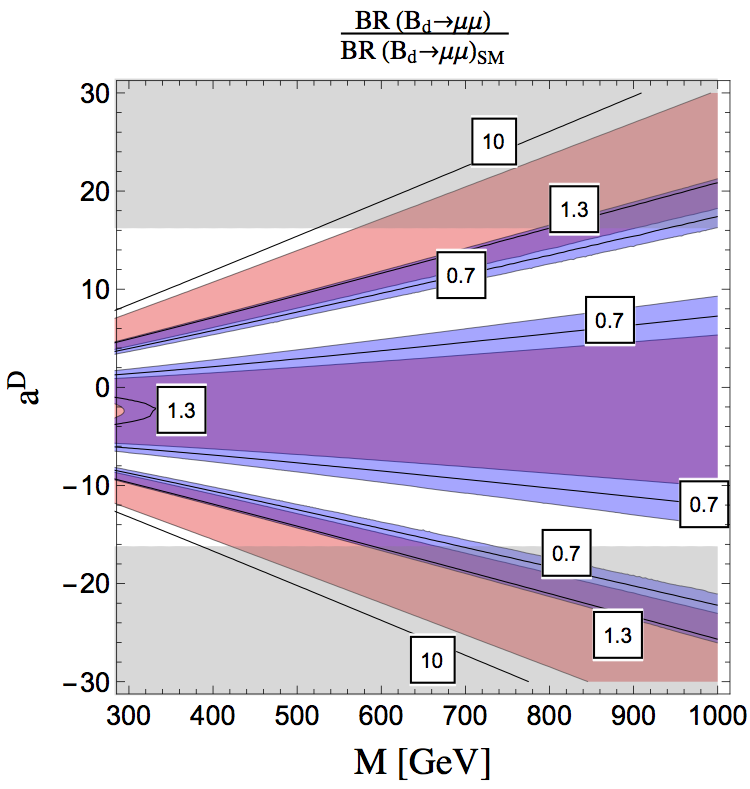}
\caption{Leading log prediction for the branching ratios for $B_s \rightarrow \mu^+\mu^-$ (left panel) and $B_d \rightarrow \mu^+\mu^-$ (right panel) relative the SM, as a function of $M$ (the mass of the heavy scalar and pseudoscalar) and $a^D$. We fix $\tan\beta=10$, $a^U=0.2$, and $\cos(\beta-\alpha)=0$. The pink regions are the regions allowed at the $2\sigma$ level by the present measurements. The purple regions are anticipated by the more precise HL-LHC measurements, assuming a measured central value equal to the SM prediction. The gray shaded regions produce Landau poles in the Yukawa couplings below $M_{\rm P}$.}
\label{fig:BmumuAdMass}
\end{figure}

The present measurement of $B_s\to\mu^+\mu^-$ constrains sizable values of $a^U$ and $a^D$ in our model. The measurement of $B_d\to\mu^+\mu^-$ also sets an interesting constraint at smaller values of $|a^D|$ (cf.~the white region where $|a^D|\sim 3$ and the values of $|a^U|$ are sizable), since the central value of the measurement is larger than the SM prediction: ${\rm{BR}}(B_d\to\mu^+\mu^-)_{\rm{exp}}/{\rm{BR}}(B_d\to\mu^+\mu^-)_{\rm{SM}}\sim 3.7$. However, the deviation from the SM prediction is not yet statistically significant, due to the large experimental uncertainty. Nevertheless, a sizable suppression of the $B_d$ decay mode is presently disfavored. As expected, the contours for ${\rm{BR}}(B_{s,d}\to\mu^+\mu^-)/{\rm{BR}}(B_{s,d}\to\mu^+\mu^-)_{\rm{SM}}$ in the two panels of Fig.~\ref{fig:BmumuAd} are very similar. This is due to the fact that our model is a particular type of MFV model in the leading logarithmic approximation [cf.~section \ref{sec:MFV}].  In particular, MFV models generically predict ${\rm{BR}}(B_d\to\mu^+\mu^-)/{\rm{BR}}(B_s\to\mu^+\mu^-)\sim {\rm{BR}}(B_d\to\mu^+\mu^-)_{\rm{SM}}/{\rm{BR}}(B_s\to\mu^+\mu^-)_{\rm{SM}}$, with corrections arising only from $m_s/m_b$ and $m_d/m_b$ terms. For this reason, it is difficult in our model to enhance one decay mode, while suppressing the other.

 \begin{figure}[t!]
\centering
\includegraphics[width=0.45\textwidth]{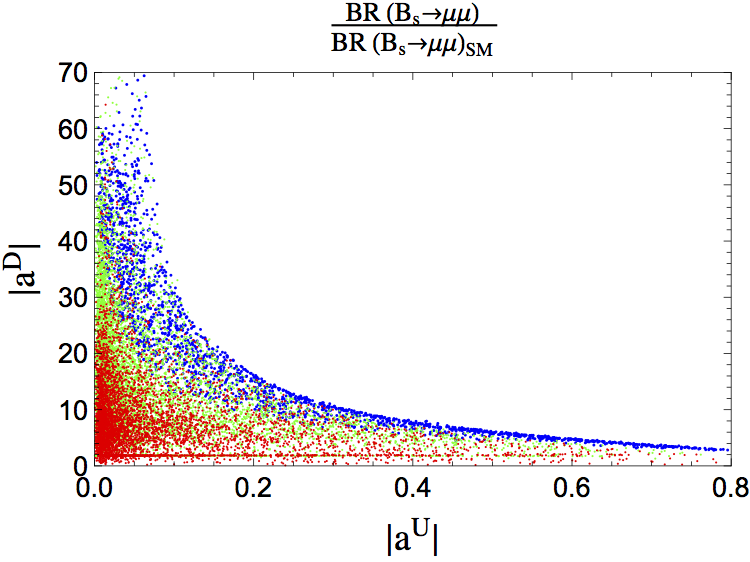}~~~~~
\includegraphics[width=0.45\textwidth]{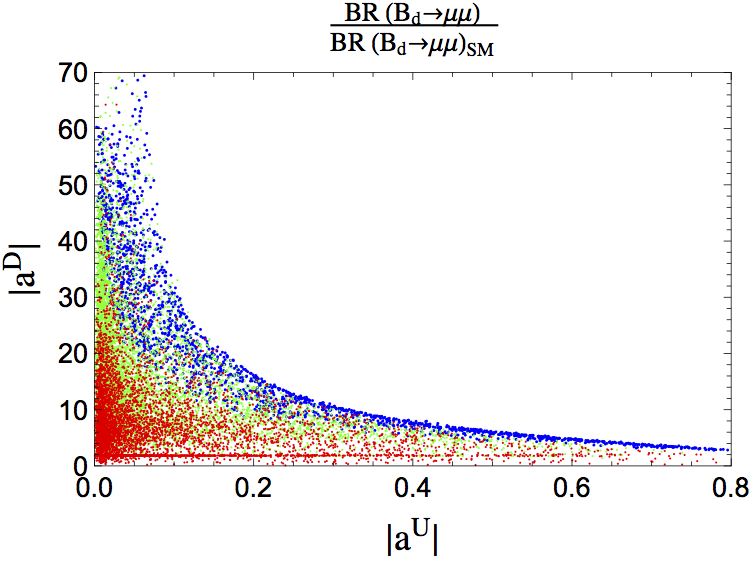}
\caption{The branching ratio for $B_s\to\mu^+\mu^-$ (left panel) and for $B_d\to\mu^+\mu^-$ (right panel) relative to the SM, obtained via scanning the parameter space and using the full RG running, at fixed $\tan\beta=10$, $\cos(\beta-\alpha)=0$, and $m_A=m_H=400$ GeV. The yellow, red, green and blue points corresponds to branching ratios normalized to the SM prediction $<{\bf{0.4}},[{\bf{0.4}},{\bf{1.1}}],[{\bf{1.1}},10],>10$. In boldface we denote the range preferred by the LHCb and ATLAS measurement of $B_s\to\mu^+\mu^-$, as reported in Eq. (\ref{eq:ranges}).}
\label{fig:BmumuScan}
\vskip -0.04in
\end{figure}

It is also interesting to investigate the bounds as a function of the heavy Higgs boson masses. In Fig.~\ref{fig:BmumuAdMass}, we show the same constraints in the $(M,a^D)$ plane, where $M\equiv m_A=m_H$, having fixed $\tan\beta=10$, $a^U=0.2$, and $\cos(\beta-\alpha)=0$. Sizable regions of parameter space are allowed, even for values of $M$ as small as $\sim 300$ GeV.
Finally, in Fig.~\ref{fig:BmumuScan}, we show the results obtained through scanning the parameter space and utilizing the full RG running. These plots are qualitatively similar to the contour plots of Fig.~\ref{fig:BmumuAd} obtained in the leading logarithmic approximation, although the heavy Higgs exchange contributions to the $B_{d,s}\to\mu^+\mu^-$  decay rates computed using the full RG running are somewhat larger (at large alignment parameters) than the corresponding leading log results.

\subsection{\boldmath $B\to\tau\nu$ decays}

The leptonic decays $B \to\ell\nu$ are interesting probes of the Higgs sector of our model and particularly of the charged Higgs couplings, since the charged Higgs boson mediates tree-level New Physics contributions to these decay modes. The $\tau$ channel is the only decay mode of this type observed so far. The present experimental world average is~\cite{Amhis:2014hma}\footnote{Updated results and plots available at: \texttt{http://www.slac.stanford.edu/xorg/hfag}.}
\beq
{\rm{BR}}(B\to\tau\nu)_{\rm{exp}}=(1.06\pm 0.19)\times 10^{-4},\eeq
and is in relatively good agreement with the SM prediction \cite{Charles:2004jd}\footnote{Updated results and plots available at: \texttt{http://ckmfitter.in2p3.fr}.} 
\beq\label{eq:SMBtaunu}
{\rm{BR}}(B\to\tau\nu)_{\rm{SM}}=(0.848^{+0.036}_{-0.055})\times 10^{-4}.\eeq

\begin{figure}[t!]
\centering
\includegraphics[width=0.44\textwidth]{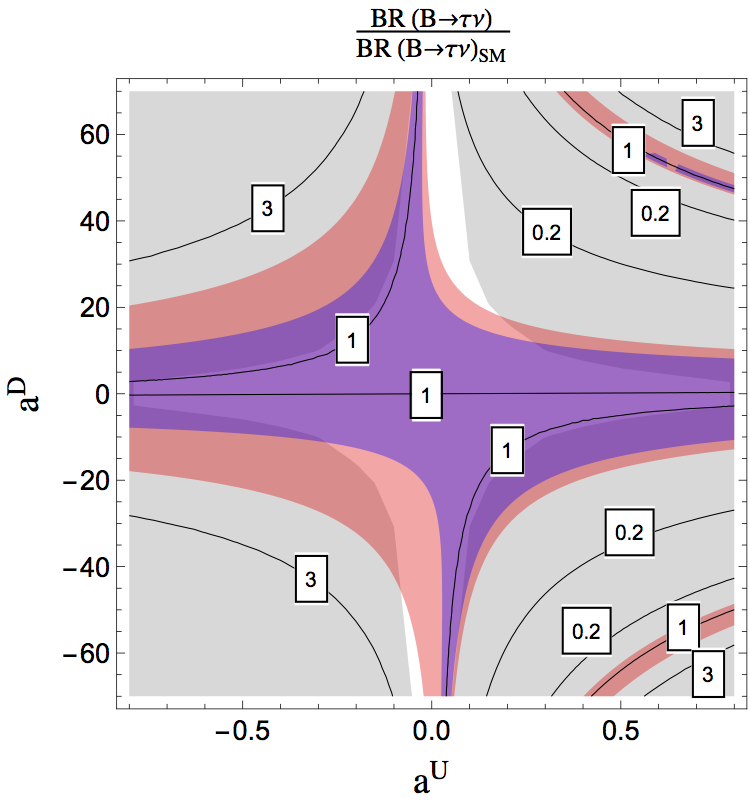}
\raisebox{0.35in}{\includegraphics[width=0.49\textwidth]{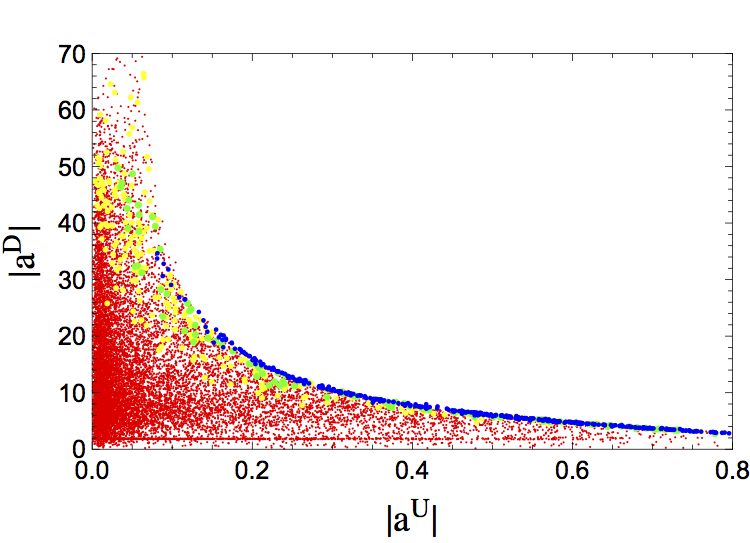}}
\vskip -0.075in
\caption{The ratio ${\rm{BR}}(B\to\tau\nu)/{\rm{BR}}(B\to\tau\nu)_{\rm{SM}}$ 
at fixed $\tan\beta=10$ and $m_{H^\pm}=400$ GeV. Left panel: leading log predictions, where the pink region is favored by the measurement of $B\to\tau\nu$. The purple region is anticipated by future measurement at Belle II, under the assumption that the central value of the measurement is given by the SM prediction. Right panel: result of the parameter space scan, using the full RG running. Yellow, red, green and blue points correspond to the ratios $<0.2,[{\bf{0.79}},{\bf{1.71}}],[{\bf{1.71}},3],>3$, respectively. In boldface we denote the range preferred by the present world average for BR$(B\to\tau\nu)$.}
\label{fig:Btaunu}
\vskip -0.05in
\end{figure}

In our model, the New Physics contribution to this decay reads
\beq
\frac{{\rm{BR}}(B\to\tau\nu)}{{\rm{BR}}(B\to\tau\nu)_{\rm{SM}}}=\left|1+\frac{m_B^2}{m_bm_\tau}\frac{C_L^{ub}-C_R^{ub}}{C_{\rm{SM}}^{ub}}\right|^2,
\eeq
where we have defined the SM Wilson coefficient $C_{\rm{SM}}^{ub}=4 G_F K_{ub}/\sqrt 2$ and $C_{R(L)}^{ub}$ are the Wilson coefficients of the $\mathcal O_{R(L)}^{ub}=(\bar u P_{R(L)} b)(\bar \tau P_L\nu_\tau)$ operators. 
In particular~\cite{Crivellin:2013wna},
\beq
C_{R(L)}^{ub}=\frac{1}{m_{H^\pm}^2}\Gamma_{ub}^{LR(RL)}\frac{\sqrt 2 m_\tau}{v}\tan\beta,
\eeq
with $\Gamma_{ub}^{LR(RL)}$ the two charged Higgs couplings $H^+ \bar u_L b_R$, $H^+ \bar u_R b_L$ given by
\beq
\Gamma_{ub}^{LR}=\sum_i K_{ui}\rho^{D*}_{3i}\,,\qquad\quad
\Gamma_{ub}^{RL}=-\sum_i K_{ib}^*\rho^{U*}_{i1}. \label{up}
\eeq
This leads to the branching ratio,
\beq
\frac{{\rm{BR}}(B\to\tau\nu)}{{\rm{BR}}(B\to\tau\nu)_{\rm{SM}}}=\left|1-\frac{m_B^2}{m_b}\frac{v\tan\beta}{\sqrt 2 K_{ub} m_{H^\pm}^2}\sum_i \left[K_{ui}\rho^{D*}_{3i}+K_{ib}^*\rho^{U*}_{i1}\right]\right|^2.
\eeq
In the leading logarithmic approximation, the most important contributions come from the second term of the above expression ($\propto \rho^{D*}_{3i}$), as one can easily deduce from \eqs{eq:rhouLeading}{eq:rhodLeading}.

In Fig.~\ref{fig:Btaunu}, we show our numerical results as obtained using the leading log approximation (left 
panel) and the scan of the parameter space using the full RGEs, having fixed $m_{H^\pm}=400$ GeV and $\tan\beta=10$. A very large region of parameter is still allowed by the measurement of $B\to\tau\nu$. 
In particular, in the leading logarithmic approximation, every value $|a^D|\lesssim 17$ is allowed, irrespective of the value of the other alignment parameter, $a^U$.  Indeed, in the pink region shown in the left panel of Fig.~\ref{fig:Btaunu}, ${\rm{BR}}(B \rightarrow \tau\nu)/{\rm{BR}}(B \rightarrow \tau\nu)_{\rm SM} \subset  [0.79, 1.71]$, consistent with the current measurements. This is no longer the case when we consider the scan based on the full RG-running.  In this case, a few points at large values of $|a^U|$ are excluded by the measurement of BR$(B \rightarrow \tau\nu)$ (see the blue points in the right panel of the figure). In the left panel of Fig.~\ref{fig:Btaunu}, we also exhibit the purple shaded region of parameter space that would be favored by the future Belle II measurement, under the assumption that the central value of the measurement is given by the SM prediction for this branching ratio [cf.~\eq{eq:SMBtaunu}]. The allowed region of parameter space is expected to shrink considerably, thanks to the anticipated accuracy of the Belle II measurement with a total error of the order of $\sim 5\%$ \cite{Inguglia:2017nxe}, leading to an allowed range, ${\rm{BR}}(B \rightarrow \tau\nu)/{\rm{BR}}(B \rightarrow \tau\nu)_{\rm SM}\subset  [0.86, 1.14]$, where we have assumed no improvement in the SM prediction of this $B$ meson decay mode. 

\section{Conclusions}\label{sec:conclusions}

We have explored the consequences of flavor-alignment at a very high energy scale on flavor observables in the two Higgs doublet Model (2HDM). Flavor alignment at the electroweak scale generically requires an unnatural fine-tuning of the matrix Yukawa couplings.  
If flavor alignment is instead imposed at a higher energy scale such as the Planck scale, perhaps enforced by some new dynamics beyond the SM, then the flavor misalignment at the electroweak scale due to RG running will generate new sources of FCNCs. The resulting tree-level Higgs-mediated FCNCs are somewhat suppressed and relatively mildly constrained by experimental measurements of flavor-changing observables.

\begin{figure}[t!]
\centering
\includegraphics[width=0.445\textwidth]{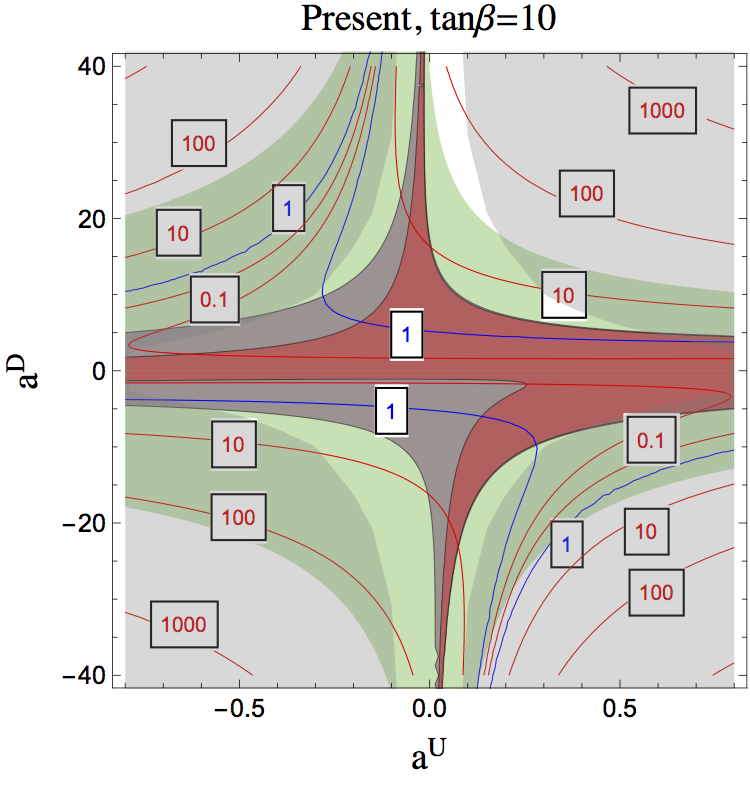}
\includegraphics[width=0.445\textwidth]{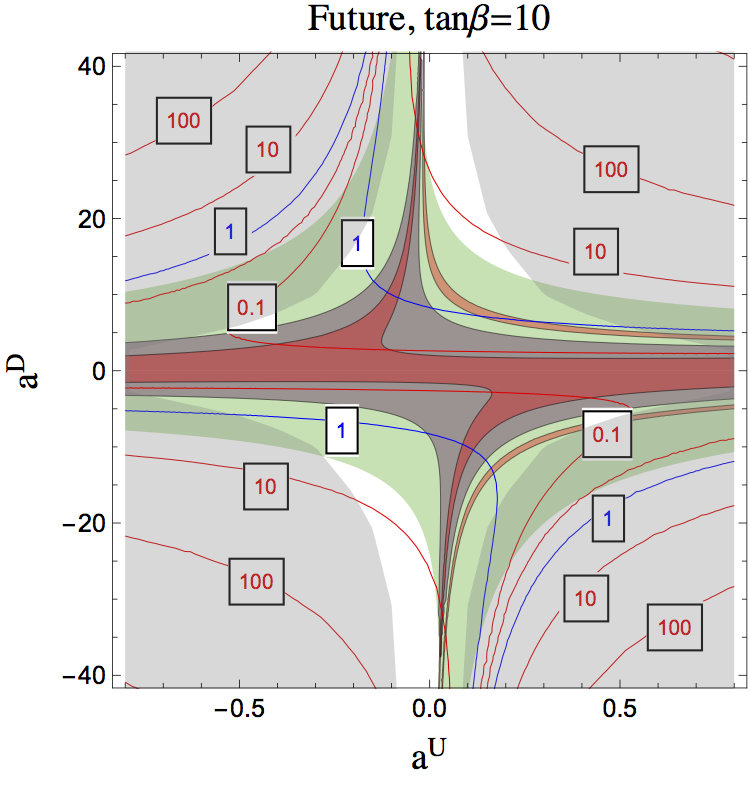}
\includegraphics[width=0.445\textwidth]{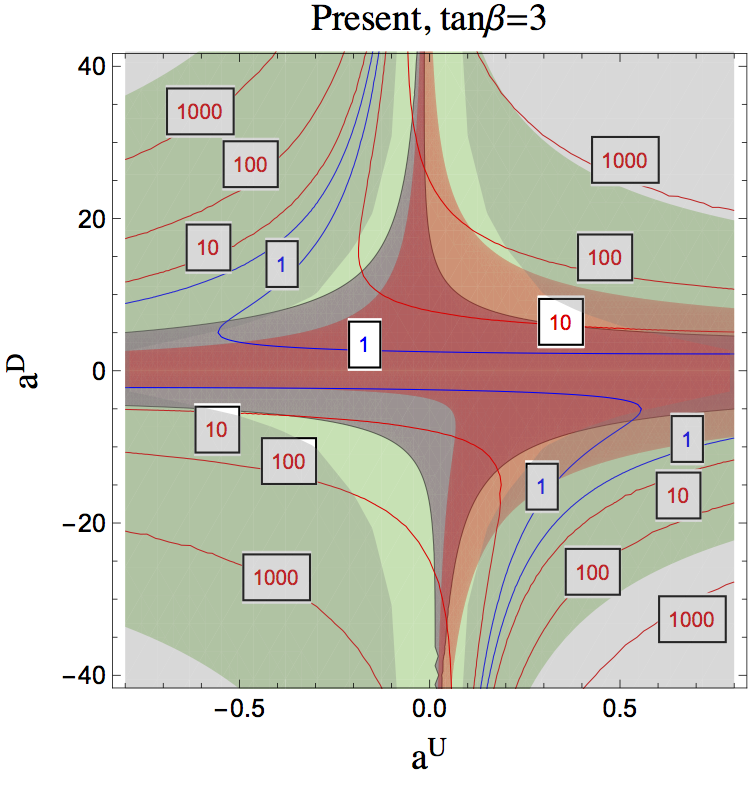}
\includegraphics[width=0.445\textwidth]{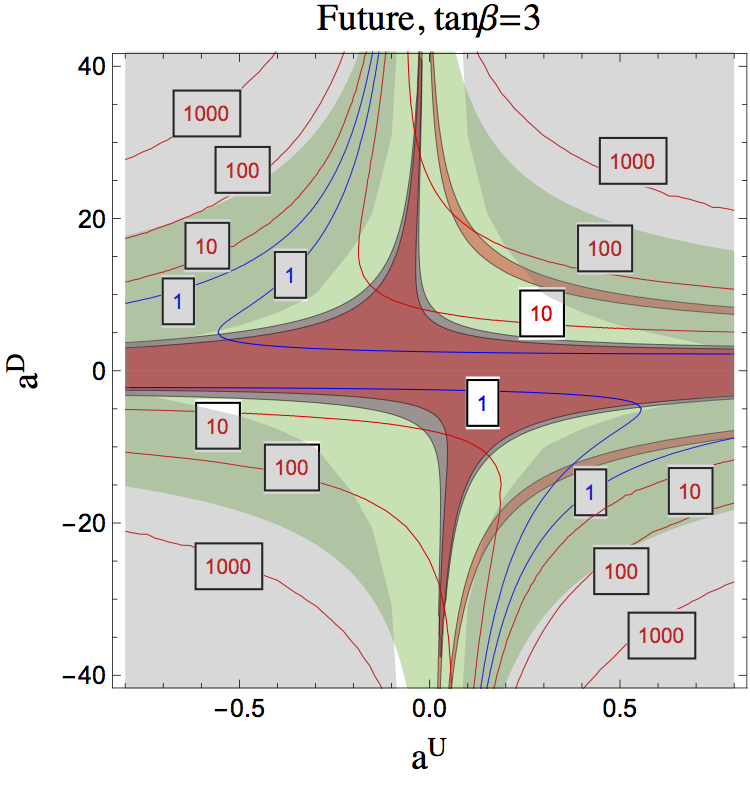}
\vskip -0.1in
\caption{Summary of the constraints and predictions for the heavy Higgs phenomenology, as computed in the leading log approximation. We fix $\cos(\beta-\alpha)=0$, $m_A=m_H=m_{H^\pm}=400$ GeV, $\tan\beta=10$ (upper panels), and $\tan\beta=3$ (lower panels). The contours represent the ratio ${\rm{BR}}(H\to b\bar b)m_\tau^2/[{\rm{BR}}(H\to  \tau^+ \tau^-)3m_b^2]$, where ${\bf{1}}$ is the Type I and Type II 2HDM prediction. The reddish-brown regions are favored by all flavor constraints. The green region is favored by the measurement of $B\to\tau\nu$. Blue-gray and tan regions are favored by $B_s$ mixing and $B_s\to\mu^+\mu^-$, respectively. The gray shaded regions produce Landau poles in the Yukawa couplings below $M_{\rm P}$. The left and right panels represent the bounds as they are now and as projected for the coming years, as detailed in section \ref{sec:4hdm}.}
\label{fig:summaryLeadingLog}
\vskip -0.1in
\end{figure}

\begin{figure}[t!]
\centering
\includegraphics[width=0.45\textwidth]{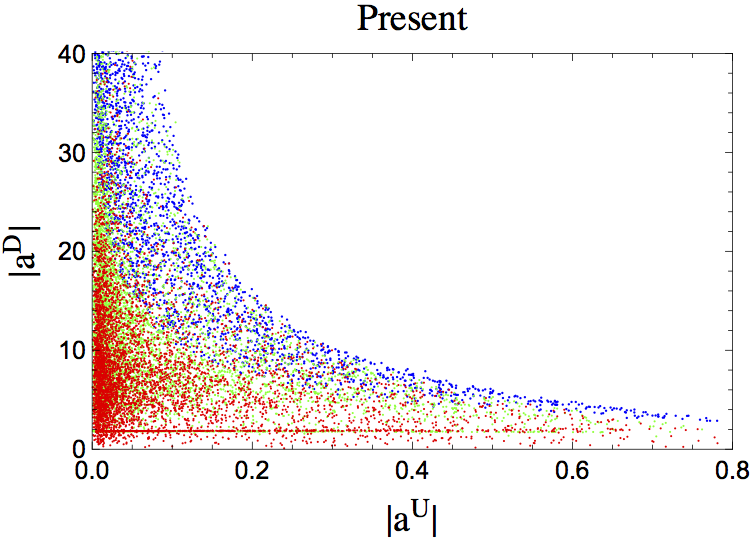}
\includegraphics[width=0.45\textwidth]{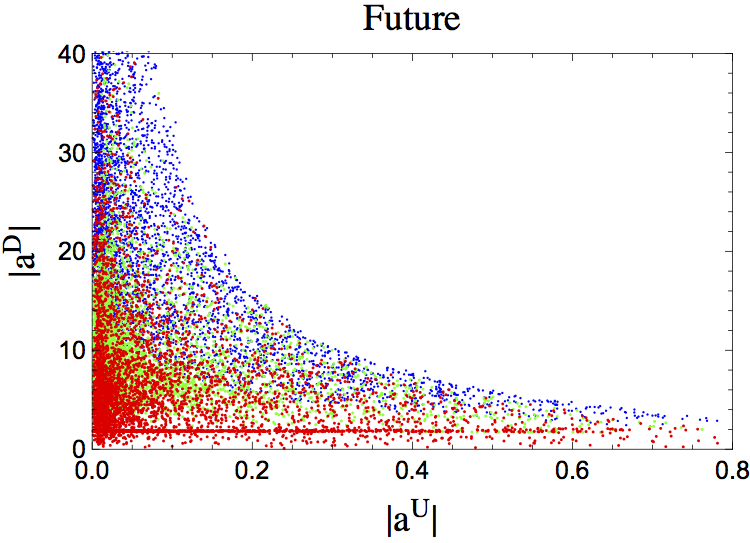}
\caption{Result of the parameter scan using full RG running, with fixed $m_A=m_H=m_{H^\pm}=400$ GeV, $\cos(\beta-\alpha)=0$, and $\tan\beta=10$. Blue points correspond to points allowed by the measurement of $B\to\tau\nu$, but not by the measurement of $B_s$ mixing or $B_s\to\mu^+\mu^-$. Green points are allowed by the measurements of $B\to\tau\nu$ and of meson mixing but not by $B_s\to\mu^+\mu^-$. Red points are allowed by all constraints. The left and right panels represent the bounds as they are now and as projected for the coming years, as detailed in section~\ref{sec:4hdm}.   In the solid white region, Landau poles in the Yukawa couplings are produced below $M_{\rm P}$.}
\label{fig:summaryScan}
\end{figure}

We require that the alignment parameters at the high scale remain perturbative.  In particular, no Landau poles are encountered during RG running. These requirements lead to an upper bound on the values of the alignment parameters at the Planck scale. This in turn provide an upper bound on the size of FCNCs generated at the electroweak scale. The flavor-changing observables considered in this paper that provide the most sensitive probe of the flavor-aligned 2HDM parameter space are meson mixing and rare $B$ decays such as $B_{s,d} \rightarrow \mu^+ \mu^-$ and $B\to\tau\nu$. We also considered constraints from LHC searches of heavy Higgs bosons (the most important of which are searches for $pp\to b(b)H, H\to \bar bb, \tau^+\tau^-$), and measurements of the couplings of the observed (SM-like) Higgs boson. The most stringent constraint on the flavor-aligned 2HDM parameter space arises from the measurement of the rare decay $B_s \rightarrow \mu^+\mu^-$.

We investigated the predictions of the flavor-aligned 2HDM in the regions of the parameter space not yet probed by the measurements listed above. The top rare flavor changing decays, $t\to u h$, $t\to c h$, are generated at tree-level.  However, 
once we impose constraints from Higgs coupling measurements, the predicted
branching ratios for these neutral flavor changing top decays are beyond the LHC reach. Furthermore, the model predicts a novel phenomenology for the heavy Higgs bosons. In particular, the heavy Higgs bosons can have a sizable branching ratios into a bottom and a strange quark, and the ratios, ${\rm{BR}}(H\to \bar tt):{\rm{BR}}(H\to \bar bb):{\rm{BR}}(H\to  \tau^+\tau^-)$, can be very different, if compared to the predictions of the more common Type I and II 2HDMs. These features are exhibited in our summary plots in Figs.~\ref{fig:summaryLeadingLog} and \ref{fig:summaryScan}.

In Fig.~\ref{fig:summaryLeadingLog}, we summarize the constraints on the $(a^U, a^D)$ parameter space, with fixed $\tan\beta=10$ (upper panels) and $\tan\beta=3$ (lower panels). In both panels, we fix the values $\cos(\beta-\alpha)=0$ and $m_A=m_H=m_{H^\pm}=400$ GeV.  The region favored by all flavor constraints is shown in reddish-brown. At sizable values of 
$a^D$, the most relevant constraint comes from the measurement of $B_s\to\mu^+\mu^-$ (tan region). $B_s$ meson mixing also sets an interesting bound on the parameter space (blue-gray region). It offers some complementary with $B_s\to\mu^+\mu^-$, as it does not depend on the particular value of $\tan\beta$.  Moreover, it will be able to probe the small region of parameter space with $a^U>0$ and sizable values of $a^D$ favored by the measurement of $B_s\to\mu^+\mu^-$ in the case of a future measurement with a central value in agreement with the SM prediction.\footnote{We use the results in \cite{Charles:2013aka} for the future prospects in measuring $B_s$ mixing, corresponding to the ``Stage II" scenario.} The measurement of $B\to\tau\nu$ imposes only a relatively weak constraint on the parameter space (green region).  For values of $\tan\beta=10$ (or larger), in the region of parameter space favored by present and future flavor constraints, the ratio $m_\tau^2\,{\rm BR}(H\to \bar bb)/3m_b^2\,{\rm{BR}}(H\to \tau^+\tau^-)$ is smaller than the ratio predicted by Type I and II 2HDM in most of the Aligned 2HDM parameter space. 
The parameter space is somewhat less constrained at lower values of $\tan\beta$, as shown in the lower panels of Fig.~\ref{fig:summaryLeadingLog}. 

In Fig.~\ref{fig:summaryScan}, we present the corresponding results obtained in the numerical scan with full RG running, with fixed $\cos(\beta-\alpha)=0$, $m_A=m_H=m_{H^\pm}=400$ GeV, and $\tan\beta=10$. 
The qualitative features of the leading log approximation continue to hold.  In particular, we again see that  $B_s\to\mu^+\mu^-$ provides the most stringent constraint on the aligned 2HDM parameter space.  Note that in order to emphasize the comparison of the constraints obtained from the different $B$ physics observables in Figs.~\ref{fig:summaryLeadingLog} and \ref{fig:summaryScan}, we do not include the constraints due to the LHC searches for the heavy Higgs bosons decaying into fermion pairs in these figures.   As shown in Figs.~\ref{fig:HeavyLeadingLog} and \ref{fig:HeavyHBRScan} for the heavy Higgs mass values quoted above, in the 
region of the Aligned 2HDM parameter space consistent with no Landau poles below $M_{\rm P}$, the
current LHC limits on $H$ and $A$ production eliminate the parameter regime with $|a_D|\gsim 30$--$40$ and $|a_U|\lsim 0.1$.

In considering the phenomenological implications of extended Higgs sectors, the most conservative approach is to impose only those constraints that are required by the current experimental data.  In most 2HDM studies in the literature, the Yukawa couplings are assumed to be of Type I, II, X or~Y.  In this paper, we have argued that the current experimental data allows for a broader approach in which the Yukawa couplings are approximately aligned in flavor at the electroweak scale.   
The resulting phenomenology can yield some unexpected surprises.
We hope that the search strategies of future Higgs studies at the LHC will be expanded to accommodate the broader phenomenological framework of the (approximately) flavor-aligned extended Higgs sector.

\section*{Acknowledgments}
H.E.H. gratefully acknowledges Paula Tuzon for numerous interactions during her two month long visit to Santa Cruz in 2010--2011.  Her work on the aligned 2HDM provided inspiration for this work. S.G. thanks Wolfgang Altmannshofer for discussions. H.E.H. and E.S. are supported in part by the U.S. Department of Energy grant number DE-SC0010107. S.G. acknowledges support from the University of Cincinnati.
S.G. and H.E.H. are grateful to the hospitality and the inspiring working atmosphere of both
the Kavli Institute for Theoretical Physics in Santa Barbara, CA, supported in part by the National Science Foundation under Grant No. NSF PHY11-25915, and 
the Aspen Center for Physics, supported by the National Science Foundation Grant No.\ PHY-1066293, where some of the research reported in this work was carried out. 

\appendix

\section{Review of the Higgs-fermion Yukawa couplings in the Higgs basis}
\label{appA}

In a general 2HDM,
the Higgs fermion
interactions are governed by the following interaction Lagrangian:\footnote{We follow the conventions of Ref.~\cite{Haber:2006ue}, in which covariance is manifest with respect to U(2) flavor transformations, $\Phi_a\to U_{a\bar{b}}\Phi_b$ [where $U\in$~U(2)], by implicitly summing over barred/unbarred index pairs of the same letter.} 
\beq \label{ymodeliii0}
-\mathscr{L}_{\rm Y}
=\anti \qlo\, \wtil\Phi_{\abar}\eiuoa\,  \uro +\anti Q_L^0\,\Phi_a(\eidoab)^\dagger \dro
+\anti E_L^0\,\Phi_a(\eieoab)^\dagger\,\ero
+{\rm h.c.}\,,
\eeq
summed over $a=\abar=1,2$,
where $\Phi_{1,2}$ are the Higgs doublets, $\wtil\Phi_{\abar}\equiv
i\sigma_2 \Phi^*_{\abar}$,
$\qlo $ and $\elo$ are the weak isospin quark and lepton doublets,
and $\uro$, $\dro$, $\ero$ are weak isospin quark and lepton singlets.\footnote{The right and left-handed fermion fields are defined as usual:
$\psi_{R,L}\equiv P_{R,L}\psi$, where $P_{R,L}\equiv \half(1\pm\gamma_5)$.}
Here, $\qlo $, $\elo$, $\uro $,
$\dro$, $\ero$ denote the interaction basis states, which
are vectors in the quark and lepton
flavor spaces, and $\eiuoa,\eidoa,\eieoa$ are $3\times 3$
matrices in quark and lepton flavor spaces.  

Note that $\eta_a^{U,0}$ appears undaggered in \eq{ymodeliii0}, whereas the corresponding Yukawa coupling matrices for down-type fermions ($D$ and $E$) appear daggered.  In this convention, the transformation of the Yukawa coupling matrices under a scalar field basis change is the same for both up-type and down-type fermions.  That is, under a change of basis, $\Phi_a\to U_{a\bar{b}}\Phi_b$ (which implies
that $\wtil\Phi_{\abar}\to \wtil\Phi_{\bbar}U^\dagger_{b\abar}$), the Yukawa coupling matrices transform as
$\eta_a^F\to U_{a\bbar}\eta_b^F$ and $\eta^{F\,\dagger}_{\abar}\to \eta^{F\,\dagger}_{\bbar}U^\dagger_{b\abar}$ (for $F=U$, $D$ and~$E$), which reflects the form-invariance of $\mathscr{L}_{\rm Y}$ under the basis change.

The neutral Higgs states acquire vacuum expectation values,
\beq
\langle\Phi^0_a\rangle=\frac{v\hat{v}_a}{\sqrt{2}}\,,
\eeq
where $\hat v_a \hat v^*_{\abar}=1$ and $v=246$~GeV.  It is also convenient to define
\beq
\hat w_b\equiv \hat v^*_{\abar}\epsilon_{ab}\,,
\eeq
where $\epsilon_{12}=-\epsilon_{21}=1$ and $\epsilon_{11}=\epsilon_{22}=0$.

Following Refs.~\cite{Davidson:2005cw,Haber:2006ue}, we define invariant and pseudo-invariant matrix Yukawa couplings,
\beq \label{kapparho}
\kappa^{F,0}\equiv \hat v^*_\abar\eta^{F,0}_a\,,\qquad\qquad
\rho^{F,0}\equiv \hat w^*_\abar\eta^{F,0}_a\,,
\eeq
where $F=U$, $D$ or $E$.  
Inverting these equations yields
\beq \label{inverted}
\eta^{F,0}_a=\kappa^{F,0}\hat v_a+\rho^{F,0}\hat w_a\,.
\eeq
Note that under the U(2) transformation, $\Phi_a\to U_{a\bar{b}}\Phi_b$,
\beq \label{rhotrans}
\kappa^{F,0}~~\hbox{is invariant and}~~\rho^{F,0}\to (\det U)\rho^{F,0}\,.
\eeq

The Higgs fields in the Higgs basis are defined by
$H_1\equiv \hat v^*_{\abar}\Phi_a$
and $H_2\equiv \hat w^*_{\abar}\Phi_a$,
which can be inverted to yield
$
\Phi_a=H_1 \hat v_a+H_2 w_a
$~\cite{Haber:2006ue}.
Rewriting \eq{ymodeliii0} in terms of the Higgs basis fields,
\beqa \label{yukhbasis}
-\mathscr{L}_{\rm Y}
&=&\anti \qlo\, (\wtil H_1\kappa^{U,0}+\wtil H_2 \rho^{U,0})\,  \uro +\anti Q_L^0\,(H_1\kappa^{D,0\,\dagger}+H_1\rho^{D,0\,\dagger})\, \dro \nonumber \\
&&\qquad\qquad +\anti E_L^0\,(H_1\kappa^{E,0\,\dagger}+H_1\rho^{E,0\,\dagger})\,\ero
+{\rm h.c.}
\eeqa

The next step is to identify the quark and lepton mass-eigenstates.  This is accomplished by
replacing $H_1\to  (0\,,\,v/\sqrt{2})$ and
performing unitary transformations of the left and right-handed up-type and
down-type fermion multiplets such that the resulting quark and charged lepton mass matrices are
diagonal with non-negative entries.  In more detail, we define:
\beqa \label{biunitary}
&& P_L U=V_L^U P_L U^0\,,\qquad P_R U=V_R^U P_R U^0\,,\qquad
P_L D=V_L^D P_L D^0\,,\qquad P_R D=V_R^D P_R D^0\,,\nonumber \\
&& P_L E=V_L^E P_L E^0\,,\qquad P_R E=V_R^D P_R E^0\,,\qquad
 P_L N=V_L^E P_L N^0\,,
\eeqa
and the Cabibbo-Kobayashi-Maskawa (CKM) matrix is defined as
$
K\equiv V_L^U V_L^{D\,\dagger}\,.
$
Note that for the neutrino fields, we are free to choose
$V_L^N=V_L^E$ since neutrinos are exactly massless in this analysis.\footnote{Here we
are ignoring the right-handed neutrino sector, which gives mass to neutrinos via the seesaw mechanism.}
In particular, the unitary matrices $V_L^F$ and $V_R^F$ (for $F=U$, $D$ and $E$)
are chosen such that
\beqa
M_U&=&\frac{v}{\sqrt{2}}V_L^U \kappa^{U,0} V_R^{U\,\dagger}={\rm diag}(m_u\,,\,m_c\,,\,m_t)\,,\label{MU}\\[8pt]
M_D&=&\frac{v}{\sqrt{2}}V_L^D \kappa^{D,0\,\dagger} V_R^{D\,\dagger}={\rm
diag}(m_d\,,\,m_s\,,\,m_b) \,,\label{MD}\\[8pt]
M_E&=&\frac{v}{\sqrt{2}}V_L^E \kappa^{E,0\,\dagger} V_R^{E\,\dagger}={\rm
diag}(m_e\,,\,m_\mu\,,\,m_\tau) \label{ME}\,.
\eeqa

It is convenient to define
\beqa
\kappa^{U}&=& V_L^U \kappa^{U,0} V_R^{U\,\dagger}\,,\qquad \kappa^{D}= V_R^D \kappa^{D,0} V_L^{D\,\dagger}\,,\qquad
\kappa^{E}= V_R^D \kappa^{E,0} V_L^{E\,\dagger}\,,\label{kappas} \\
\rho^{U}&=& V_L^U \rho^{U,0} V_R^{U\,\dagger}\,,\qquad \rho^{D}= V_R^D \rho^{D,0} V_L^{D\,\dagger}\,,\qquad
\rho^{E}= V_R^D \rho^{E,0} V_L^{E\,\dagger}\,.\label{rhos}
\eeqa
\Eq{rhotrans} implies that under the U(2) transformation, $\Phi_a\to U_{a\bar{b}}\Phi_b$,
\beq
\kappa^{F}~~\hbox{is invariant and}~~\rho^{F}\to (\det U)\rho^{F}\,,
\eeq
for $F=U$, $D$ and $E$.   Indeed, $\kappa^F$ is invariant since \eqst{MU}{ME} imply
that
\beq \label{MF}
M_F=\frac{v}{\sqrt{2}}\kappa^F\,,
\eeq
which is a physical observable.
The matrices $\rho^U$, $\rho^D$ and $\rho^E$ are independent pseudo-invariant complex $3\times 3$ matrices.
The Higgs-fermion interactions
given in \eq{yukhbasis} can be rewritten in terms of the quark and lepton mass-eigenstates,
\beqa \label{yukhbasis2}
-\mathscr{L}_{\rm Y}&=&\anti U_L (\kappa^U H_1^{0\,\dagger}
+\rho^U H_2^{0\,\dagger})\ur
-\anti D_L K^\dagger(\kappa^U H_1^{-}+\rho^U H_2^{-})\ur \nonumber \\[6pt]
&& +\anti U_L K (\kappa^{D\,\dagger}H_1^++\rho^{D\,\dagger}H_2^+)\dr
+\anti D_L (\kappa^{D\,\dagger}H_1^0+\rho^{D\,\dagger}H_2^0)\dr \nonumber \\[6pt]
&& +\anti N_L  (\kappa^{E\,\dagger}H_1^++\rho^{E\,\dagger}H_2^+)\er
+\anti E_L (\kappa^{E\,\dagger}H_1^0+\rho^{E\,\dagger}H_2^0)\er
+{\rm h.c.}
\eeqa

\section{Renormalization group equations for the Yukawa matrices}\label{sec:RGEs}

We first write down the renormalization group equations (RGEs) for the Yukawa matrices $\eiuoa$, $\eidoa$ and $\eieoa$.
Defining $\mathcal{D} \equiv 16 \pi^2 \mu (d/d\mu)=16 \pi^2 (d/dt)$, the RGEs are given by~\cite{Ferreira:2010xe}:
\beqa
\mathcal{D}\eiuoa&=& -\bigl(8g_s^2+\tfrac{9}{4}g^2+\tfrac{17}{12}g^{\prime\,2}\bigr) \eiuoa
+\biggl\{3{\rm Tr}\bigl[\eiuoa({\eiuobb})^\dagger+\eidoa({\eidobb})^\dagger\bigr]
+{\rm Tr}\bigl[\eieoa({\eieobb})^\dagger \bigr]\biggr\}\eiuob \nonumber \\[8pt]
&&-2(\eidobb)^\dagger\eidoa\eiuob+\eiuoa(\eiuobb)^\dagger\eiuob+\half (\eidobb)^\dagger\eidob\eiuoa+\half \eiuob(\eiuobb)^\dagger\eiuoa\,,\label{rged}\\[6pt]
\mathcal{D}\eidoa&=& -\bigl(8g_s^2+\tfrac{9}{4}g^2+\tfrac{5}{12}g^{\prime\,2}\bigr)\eidoa
+\biggl\{3{\rm Tr}\bigl[({\eidobb})^\dagger \eidoa+({\eiuobb})^\dagger \eiuoa\bigr]
+{\rm Tr}\bigl[({\eieobb})^\dagger \eieoa\bigr]\biggr\}\eidob \nonumber \\[8pt]
&&-2\eidob\eiuoa(\eiuobb)^\dagger+\eidob(\eidobb)^\dagger\eidoa+\half \eidoa\eiuob(\eiuobb)^\dagger
+\half \eidoa(\eidobb)^\dagger\eidob\,,\label{rgeu}  \\[8pt]
\mathcal{D}\eieoa&=& -\bigl(\tfrac{9}{4}g^2+\tfrac{15}{4}g^{\prime\,2}\bigr)\eieoa
+\biggl\{3{\rm Tr}\bigl[({\eidobb})^\dagger \eidoa+({\eiuobb})^\dagger \eiuoa\bigr]
+{\rm Tr}\bigl[({\eieobb})^\dagger \eieoa\bigr]\biggr\}\eieob \nonumber \\[8pt]
&&+\eieob(\eieobb)^\dagger\eieoa+\half \eieoa(\eieobb)^\dagger\eieob\,.\label{rgee}
\eeqa

The RGEs above are true for any basis choice.  Thus, they must also be true in the Higgs basis in which $\hat v=(1,0)$ and $\hat w=(0,1)$.
In this case, we can simply choose $\eta_1^{F,0}=\kappa^{F,0}$ and $\eta_2^{F,0}=\rho^{F,0}$ to obtain the RGEs
for the $\kappa^{F,0}$ and $\rho^{F,0}$.
Alternatively, we can multiply \eqst{rged}{rgee} first by $\hat v_a^*$ and then by $\hat w_a^*$.  Expanding $\eta_{\bar a}^\dagger$,
which appears on the right-hand sides of \eqst{rged}{rgee},
in terms of $\kappa^\dagger$ and $\rho^\dagger$ using \eq{inverted}, we again obtain the RGEs
for the $\kappa^{F,0}$ and $\rho^{F,0}$.  Of course, both methods yield the same result,
since the diagonalization matrices employed in \eqst{MU}{ME} are defined as
those that bring the mass matrices to their diagonal form at the electroweak scale.
No scale dependence is assumed in the diagonalization matrices, and as such they are
not affected by the operators~$\mathcal{D}$.
\beqa\nonumber
\mathcal{D}\kappa^{U,0}&=& -\bigl(8g_s^2+\tfrac{9}{4}g^2+\tfrac{17}{12}g^{\prime\,2}\bigr) \kappa^{U,0}
+\biggl\{3{\rm Tr}\bigl[\kappa^{U,0}\kappa^{U,0\,\dagger}+\kappa^{D,0}\kappa^{D,0\,\dagger}\bigr]
+{\rm Tr}\bigl[\kappa^{E,0}\kappa^{E,0\,\dagger }\bigr]\biggr\}\kappa^{U,0} \nonumber\\[8pt]
&+&\biggl\{3{\rm Tr}\bigl[\kappa^{U,0}\rho^{U,0\,\dagger}+\kappa^{D,0}\rho^{D,0\,\dagger}\bigr]
+{\rm Tr}\bigl[\kappa^{E,0}\rho^{E,0\,\dagger }\bigr]\biggr\}\rho^{U,0}
-2\bigl(\kappa^{D,0\,\dagger}\kappa^{D,0}\kappa^{U,0}+\rho^{D,0\,\dagger}\kappa^{D,0}\rho^{U,0}\bigr) \nonumber\\[8pt]
&+&\kappa^{U,0}(\kappa^{U,0\,\dagger}\kappa^{U,0}+\rho^{U,0\,\dagger}\rho^{U,0})
+\half(\kappa^{D,0\,\dagger}\kappa^{D,0}+\rho^{D,0\,\dagger}\rho^{D,0})\kappa^{U,0}  \nonumber\\[8pt]
&+&\half(\kappa^{U,0}\kappa^{U,0\,\dagger}+\rho^{U,0}\rho^{U,0\,\dagger})\kappa^{U,0} \,,\\[12pt]
\mathcal{D}\rho^{U,0}&=& -\bigl(8g_s^2+\tfrac{9}{4}g^2+\tfrac{17}{12}g^{\prime\,2}\bigr) \rho^{U,0}
+\biggl\{3{\rm Tr}\bigl[\rho^{U,0}\kappa^{U,0\,\dagger}+\rho^{D,0}\kappa^{D,0\,\dagger}\bigr]
+{\rm Tr}\bigl[\rho^{E,0}\kappa^{E,0\,\dagger }\bigr]\biggr\}\kappa^{U,0} \nonumber\\[8pt]
&+&\biggl\{3{\rm Tr}\bigl[\rho^{U,0}\rho^{U,0\,\dagger}+\rho^{D,0}\rho^{D,0\,\dagger}\bigr]
+{\rm Tr}\bigl[\rho^{E,0}\rho^{E,0\,\dagger }\bigr]\biggr\}\rho^{U,0}
-2\bigl(\kappa^{D,0\,\dagger}\rho^{D,0}\kappa^{U,0}
+\rho^{D,0\,\dagger}\rho^{D,0}\rho^{U,0}\bigr) \nonumber\\[8pt]
&+&\rho^{U,0}(\kappa^{U,0\,\dagger}\kappa^{U,0}+\rho^{U,0\,\dagger}\rho^{U,0})
+\half(\kappa^{D,0\,\dagger}\kappa^{D,0}+\rho^{D,0\,\dagger}\rho^{D,0})\rho^{U,0} \nonumber \nonumber\\[8pt]
&+&\half(\kappa^{U,0}\kappa^{U,0\,\dagger}+\rho^{U,0}\rho^{U,0\,\dagger})\rho^{U,0} \,,\\[12pt]
\mathcal{D}\kappa^{D,0}&=& -\bigl(8g_s^2+\tfrac{9}{4}g^2+\tfrac{5}{12}g^{\prime\,2}\bigr) \kappa^{D,0}
+\biggl\{3{\rm Tr}\bigl[\kappa^{D,0\,\dagger}\kappa^{D,0}+\kappa^{U,0\,\dagger}\kappa^{U,0}\bigr]
+{\rm Tr}\bigl[\kappa^{E,0\,\dagger}\kappa^{E,0}]\biggr\}\kappa^{D,0} \nonumber \\[8pt]
&+&\biggl\{3{\rm Tr}\bigl[\rho^{D,0\,\dagger}\kappa^{D,0}+\rho^{U,0\,\dagger}\kappa^{U,0}\bigr]
+{\rm Tr}\bigl[\rho^{E,0\,\dagger}\kappa^{E,0}]\biggr\}\rho^{D,0}-2(\kappa^{D,0}\kappa^{U,0}\kappa^{U,0\,\dagger} \nonumber \\[8pt]
&+&\rho^{D,0}\kappa^{U,0}\rho^{U,0\,\dagger}) +(\kappa^{D,0}\kappa^{D,0\,\dagger}+\rho^{D,0}\rho^{D,0\,\dagger})\kappa^{D,0}
+\half\kappa^{D,0}(\kappa^{U,0}\kappa^{U,0\,\dagger}+\rho^{U,0}\rho^{U,0\,\dagger}) \nonumber \\[8pt]
&+&\half\kappa^{D,0}(\kappa^{D,0\,\dagger}\kappa^{D,0}+\rho^{D,0\,\dagger}\rho^{D,0})\,,
\\[12pt]
\mathcal{D}\rho^{D,0}&=& -\bigl(8g_s^2+\tfrac{9}{4}g^2+\tfrac{5}{12}g^{\prime\,2}\bigr) \rho^{D,0}
+\biggl\{3{\rm Tr}\bigl[\kappa^{D,0\,\dagger}\rho^{D,0}+\kappa^{U,0\,\dagger}\rho^{U,0}\bigr]
+{\rm Tr}\bigl[\kappa^{E,0\,\dagger}\rho^{E,0}]\biggr\}\kappa^{D,0} \nonumber \\[8pt]
&+&\biggl\{3{\rm Tr}\bigl[\rho^{D,0\,\dagger}\rho^{D,0}+\rho^{U,0\,\dagger}\rho^{U,0}\bigr]
+{\rm Tr}\bigl[\rho^{E,0\,\dagger}\rho^{E,0}]\biggr\}\rho^{D,0}-2(\kappa^{D,0}\rho^{U,0}\kappa^{U,0\,\dagger}+\rho^{D,0}\rho^{U,0}\rho^{U,0\,\dagger}) \nonumber \\[8pt]
&+&(\kappa^{D,0}\kappa^{D,0\,\dagger}+\rho^{D,0}\rho^{D,0\,\dagger})\rho^{D,0}+\half\rho^{D,0}(\kappa^{U,0}\kappa^{U,0\,\dagger}+\rho^{U,0}\rho^{U,0\,\dagger}) \nonumber \\[8pt]
&+&\half\rho^{D,0}(\kappa^{D,0\,\dagger}\kappa^{D,0}+\rho^{D,0\,\dagger}\rho^{D,0})\,,
\eeqa
\beqa
\mathcal{D}\kappa^{E,0}&=& -\bigl(\tfrac{9}{4}g^2+\tfrac{15}{4}g^{\prime\,2}\bigr) \kappa^{E,0}
+\biggl\{3{\rm Tr}\bigl[\kappa^{D,0\,\dagger}\kappa^{D,0}+\kappa^{U,0\,\dagger}\kappa^{U,0}\bigr]
+{\rm Tr}\bigl[\kappa^{E,0\,^\dagger}\kappa^{E,0}\bigr]\biggr\}\kappa^{E,0} \nonumber \\[8pt]
&+&\biggl\{3{\rm Tr}\bigl[\rho^{D,0\,\dagger}\kappa^{D,0}+\rho^{U,0\,\dagger}\kappa^{U,0}\bigr]
+{\rm Tr}\bigl[\rho^{E,0\,^\dagger}\kappa^{E,0}\bigr]\biggr\}\rho^{E,0} \nonumber \\[8pt]
&+&(\kappa^{E,0}\kappa^{E,0\,\dagger}+\rho^{E,0}\rho^{E,0\,\dagger})\kappa^{E,0}+\half\kappa^{E,0}(\kappa^{E,0\,\dagger}\kappa^{E,0}+\rho^{E,0\,\dagger}\rho^{E,0})\,,\\[12pt]
\mathcal{D}\rho^{E,0}&=& -\bigl(\tfrac{9}{4}g^2+\tfrac{15}{4}g^{\prime\,2}\bigr) \rho^{E,0}
+\biggl\{3{\rm Tr}\bigl[\kappa^{D,0\,\dagger}\rho^{D,0}+\kappa^{U,0\,\dagger}\rho^{U,0}\bigr]
+{\rm Tr}\bigl[\kappa^{E,0\,^\dagger}\rho^{E,0}\bigr]\biggr\}\kappa^{E,0} \nonumber \\[8pt]
&+&\biggl\{3{\rm Tr}\bigl[\rho^{D,0\,\dagger}\rho^{D,0}+\rho^{U,0\,\dagger}\rho^{U,0}\bigr]
+{\rm Tr}\bigl[\rho^{E,0\,^\dagger}\rho^{E,0}\bigr]\biggr\}\rho^{E,0} \nonumber \\[8pt]
&+&(\kappa^E\kappa^{E,0\,\dagger}+\rho^{E,0}\rho^{E,0\,\dagger})\rho^{E,0}+\half\rho^{E,0}(\kappa^{E,0\,\dagger}\kappa^{E,0}+\rho^{E,0\,\dagger}\rho^{E,0})\,.
\eeqa

Using \eqs{kappas}{rhos}, we immediately obtain the RGEs for the 
$\kappa^F$ and $\rho^F$.
Schematically, we shall write,
\beq \label{betaf}
\mathcal{D}\kappa^F= \beta_{\kappa^F}\,,\qquad\quad \mathcal{D}\rho^F= \beta_{\rho^F}\,,
\eeq
for $F=U$, $D$ and $E$.  Explicitly, the corresponding $\beta$-functions at one-loop order are given by,
\beqa\label{rge0}
\mathcal{D}\kappa^U&=& -\bigl(8g_s^2+\tfrac{9}{4}g^2+\tfrac{17}{12}g^{\prime\,2}\bigr) \kappa^U
+\biggl\{3{\rm Tr}\bigl[\kappa^U\kappa^{U\,\dagger}+\kappa^D\kappa^{D\,\dagger}\bigr]
+{\rm Tr}\bigl[\kappa^E\kappa^{E\,\dagger }\bigr]\biggr\}\kappa^U  \label{rge1} \\[8pt]
&+&\biggl\{3{\rm Tr}\bigl[\kappa^U\rho^{U\,\dagger}+\kappa^D\rho^{D\,\dagger}\bigr]
+{\rm Tr}\bigl[\kappa^E\rho^{E\,\dagger }\bigr]\biggr\}\rho^U
-2K\bigl(\kappa^{D\,\dagger}\kappa^D K^\dagger\kappa^U+\rho^{D\,\dagger}\kappa^D K^\dagger\rho^U\bigr)\nonumber \\[8pt]
&+&\kappa^U(\kappa^{U\,\dagger}\kappa^U+\rho^{U\,\dagger}\rho^U)
+\half K (\kappa^{D\,\dagger}\kappa^D+\rho^{D\,\dagger}\rho^D)K^\dagger\kappa^U 
+\half(\kappa^U\kappa^{U\,\dagger}+\rho^U\rho^{U\,\dagger})\kappa^U \,, \nonumber\\[12pt]
\mathcal{D}\rho^U&=& -\bigl(8g_s^2+\tfrac{9}{4}g^2+\tfrac{17}{12}g^{\prime\,2}\bigr) \rho^U
+\biggl\{3{\rm Tr}\bigl[\rho^U\kappa^{U\,\dagger}+\rho^D\kappa^{D\,\dagger}\bigr]
+{\rm Tr}\bigl[\rho^E\kappa^{E\,\dagger }\bigr]\biggr\}\kappa^U  \label{rge2} \\[8pt]
&+&\biggl\{3{\rm Tr}\bigl[\rho^U\rho^{U\,\dagger}+\rho^D\rho^{D\,\dagger}\bigr]
+{\rm Tr}\bigl[\rho^E\rho^{E\,\dagger }\bigr]\biggr\}\rho^U
-2K\bigl(\kappa^{D\,\dagger}\rho^D K^\dagger\kappa^U+\rho^{D\,\dagger}\rho^D K^\dagger\rho^U\bigr)\nonumber \\[8pt]\nonumber
&+&\rho^U(\kappa^{U\,\dagger}\kappa^U+\rho^{U\,\dagger}\rho^U)
+\half K (\kappa^{D\,\dagger}\kappa^D+\rho^{D\,\dagger}\rho^D)K^\dagger\rho^U 
+\half(\kappa^U\kappa^{U\,\dagger}+\rho^U\rho^{U\,\dagger})\rho^U \,,\nonumber \\[8pt] 
\mathcal{D}\kappa^D&=& -\bigl(8g_s^2+\tfrac{9}{4}g^2+\tfrac{5}{12}g^{\prime\,2}\bigr) \kappa^D
+\biggl\{3{\rm Tr}\bigl[\kappa^{D\,\dagger}\kappa^D+\kappa^{U\,\dagger}\kappa^U\bigr]
+{\rm Tr}\bigl[\kappa^{E\,\dagger}\kappa^E]\biggr\}\kappa^D \label{rge3}  \\[8pt]
&+&\biggl\{3{\rm Tr}\bigl[\rho^{D\,\dagger}\kappa^D+\rho^{U\,\dagger}\kappa^U\bigr]
+{\rm Tr}\bigl[\rho^{E\,\dagger}\kappa^E]\biggr\}\rho^D-2(\kappa^D K^\dagger \kappa^U\kappa^{U\,\dagger}+\rho^D K^\dagger \kappa^U\rho^{U\,\dagger})K \nonumber \\[8pt]
&+&(\kappa^D\kappa^{D\,\dagger}+\rho^D\rho^{D\,\dagger})\kappa^D+\half\kappa^D K^\dagger(\kappa^U\kappa^{U\,\dagger}+\rho^U\rho^{U\,\dagger})K
+\half\kappa^D(\kappa^{D\,\dagger}\kappa^D+\rho^{D\,\dagger}\rho^D)\,,\nonumber 
\eeqa
\beqa
\mathcal{D}\rho^D&=& -\bigl(8g_s^2+\tfrac{9}{4}g^2+\tfrac{5}{12}g^{\prime\,2}\bigr) \rho^D
+\biggl\{3{\rm Tr}\bigl[\kappa^{D\,\dagger}\rho^D+\kappa^{U\,\dagger}\rho^U\bigr]
+{\rm Tr}\bigl[\kappa^{E\,\dagger}\rho^E]\biggr\}\kappa^D \label{rge4}\ \\[8pt]
&+&\biggl\{3{\rm Tr}\bigl[\rho^{D\,\dagger}\rho^D+\rho^{U\,\dagger}\rho^U\bigr]
+{\rm Tr}\bigl[\rho^{E\,\dagger}\rho^E]\biggr\}\rho^D-2(\kappa^D K^\dagger \rho^U\kappa^{U\,\dagger}+\rho^D K^\dagger \rho^U\rho^{U\,\dagger})K  \nonumber\\[8pt]
&+&(\kappa^D\kappa^{D\,\dagger}+\rho^D\rho^{D\,\dagger})\rho^D+\half\rho^D K^\dagger(\kappa^U\kappa^{U\,\dagger}+\rho^U\rho^{U\,\dagger})K 
+\half\rho^D(\kappa^{D\,\dagger}\kappa^D+\rho^{D\,\dagger}\rho^D)\,,\nonumber\\[12pt]
\mathcal{D}\kappa^E&=& -\bigl(\tfrac{9}{4}g^2+\tfrac{15}{4}g^{\prime\,2}\bigr) \kappa^E
+\biggl\{3{\rm Tr}\bigl[\kappa^{D\,\dagger}\kappa^D+\kappa^{U\,\dagger}\kappa^U\bigr]
+{\rm Tr}\bigl[\kappa^{E\,^\dagger}\kappa^E\bigr]\biggr\}\kappa^E \label{rge5} \\[8pt]
&+&\biggl\{3{\rm Tr}\bigl[\rho^{D\,\dagger}\kappa^D 
+\rho^{U\,\dagger}\kappa^U\bigr]
+{\rm Tr}\bigl[\rho^{E\,^\dagger}\kappa^E\bigr]\biggr\}\rho^E \nonumber \\[8pt]
&+&(\kappa^E\kappa^{E\,\dagger}+\rho^E\rho^{E\,\dagger})\kappa^E 
+\half\kappa^E(\kappa^{E\,\dagger}\kappa^E+\rho^{E\,\dagger}\rho^E)\,,\nonumber \\[12pt]
\mathcal{D}\rho^E&=& -\bigl(\tfrac{9}{4}g^2+\tfrac{15}{4}g^{\prime\,2}\bigr) \rho^E 
+\biggl\{3{\rm Tr}\bigl[\kappa^{D\,\dagger}\rho^D+\kappa^{U\,\dagger}\rho^U\bigr]
+{\rm Tr}\bigl[\kappa^{E\,^\dagger}\rho^E\bigr]\biggr\}\kappa^E \label{rge6}  \\[8pt]
 &+&\biggl\{3{\rm Tr}\bigl[\rho^{D\,\dagger}\rho^D +\rho^{U\,\dagger}\rho^U\bigr]
+{\rm Tr}\bigl[\rho^{E\,^\dagger}\rho^E\bigr]\biggr\}\rho^E \nonumber  \\[8pt]
&+&(\kappa^E\kappa^{E\,\dagger}+\rho^E\rho^{E\,\dagger})\rho^E +\half\rho^E(\kappa^{E\,\dagger}\kappa^E+\rho^{E\,\dagger}\rho^E)\,.\nonumber
\eeqa

For the numerical analysis of the RGEs, it is convenient to define 
\beq
\widetilde{\kappa}^D\equiv \kappa^D K^\dagger\,,\qquad\quad
\widetilde{\rho}^D\equiv \rho^D K^\dagger\,,
\eeq
keeping in mind that the (unitary) CKM matrix $K$ is defined at the electroweak scale and thus is not taken to be a running quantity.  The RGEs given in \eqst{rge1}{rge6} can now be rewritten by taking $\kappa^D\to \widetilde{\kappa}^D$, $\rho^D\to\widetilde{\rho}^D$ and $K\to \mathds{1}$.  The advantage of the RGEs written in this latter form is that the CKM matrix $K$ no longer appears explicitly in the differential equations, and enters
only in the initial condition of $\widetilde{\kappa}^D$ at the low scale [cf.~\eq{eqn:kappalow}], 
\beq
\widetilde{\kappa}^D(\Lambda_H)=\sqrt{2}M_D(\Lambda_H)K^\dagger/v\,.
\eeq
In particular, the high scale boundary condition given by \eq{eqn:rhohigh} also applies to $\widetilde{\kappa}^D$ and $\widetilde{\rho}^D$, i.e.,
\beq
\widetilde{\rho}^D(\Lambda)=a^D\widetilde{\kappa}^D(\Lambda)\,.
\eeq

\bibliography{Align.bib}
\bibliographystyle{JHEP}


\end{document}